\documentclass[twocolumn]{aastex631}
\usepackage{amsmath}
\usepackage{float} 
\usepackage{url}

\shorttitle{Mass Through Splashback}
\shortauthors{Gabriel-Silva \& Sodr\'{e}}

\begin{document}

\title{Galaxy Cluster Mass Estimation Through The Splashback Radius}

\author[0000-0002-8881-0694]{Lucas Gabriel-Silva}
\affiliation{Instituto de Astronomia, Geof\'{i}sica e Ci\^{e}ncias Atmosf\'{e}ricas da Universidade de S\~{a}o Paulo (IAG/USP), \\
Rua do Mat\~{a}o 1226, Cidade Universit\'{a}ria, S\~{a}o Paulo, 05508-090, Brazil}
\email{lucasgabriel@usp.br}

\author[0000-0002-3876-268X]{Laerte Sodr\'{e} Jr.}
\affiliation{Instituto de Astronomia, Geof\'{i}sica e Ci\^{e}ncias Atmosf\'{e}ricas da Universidade de S\~{a}o Paulo (IAG/USP), \\
Rua do Mat\~{a}o 1226, Cidade Universit\'{a}ria, S\~{a}o Paulo, 05508-090, Brazil}
\email{laerte.sodre@iag.usp.br}

\begin{abstract}
We present an analysis of the splashback radius ($R_{\text{sp}}$) and the associated splashback mass ($M_{\text{sp}}$) for a sample of galaxy clusters using SDSS spectroscopic data and mock simulations. $R_{\text{sp}}$ marks a physical boundary between the virialized core and the outer infall regions of clusters, providing a robust measure of cluster mass accretion history without being affected by pseudo-evolution. We model the cumulative galaxy number profile of clusters, testing different halo density models and considering the impact of cluster properties, such as center definitions, magnitude limits, galaxy colors, and field contamination, on the estimation of splashback features. Our results show that observed splashback radii, measured in projection (2D), are consistently smaller than predicted by dark matter simulations, with $R_\text{sp}/R_{200m} \approx 1$, supporting previous discrepancies in the literature. We also explore the relationship between $M_{\text{sp}}$ and $R_{\text{sp}}$, proposing a new scaling relation for future cosmological studies, as $R_{\text{sp}}$ is easily observable. Our findings indicate that splashback masses strongly correlate with radii, with a dispersion of $\approx 0.15$ dex, competitive with other mass-observable relations. However, the fitted relation diverges from the constant density expectations of galaxy clusters around $R_\text{sp}$. Additionally, the $M_{\text{sp}} \textendash R_{\text{sp}}$ relation shows significant redshift evolution, though the predominantly low-redshift range of our sample limits our ability to confirm this trend conclusively. The approach developed here may play a key role in cluster characterization and cosmology in the era of large galaxy surveys.
\end{abstract}


\section{Introduction} \label{sec:intro}

Galaxy clusters form a bridge between the smallest galactic scales and the vastness of the cosmic web, capturing the complexities of structure formation and evolution in a single, dynamic environment. Due to their nature, these systems serve as valuable laboratories for advancing our understanding of cosmology, galaxy evolution, and the large-scale structure of the universe \citep[e.g.][]{allen+11, kravtsov+12}. The study of galaxy cluster evolution across cosmic time is closely tied to various cosmological parameters, including the growth rates of primordial density fluctuations and the cosmic volume-redshift relation \citep{peebles80}. These relations arise because galaxy cluster halos occupy the exponential tail of the cosmic mass function \citep[e.g.][]{haiman+01}. By measuring a large number of galaxy clusters spanning a wide range of masses and redshifts, we can place strong constraints on cosmological models \citep[e.g.][]{allen+04, vikhlinin+09, pratt+19}.

However, accurately estimating galaxy cluster masses typically requires expensive instrumentation or complex methods, such as spectroscopic measurements using the virial theorem \citep[e.g.][]{carlberg+97}, X-ray observations mapping the gas to infer the cluster's gravitational potential \citep[e.g.][]{arnaud+05}, or weak-lensing analyses, which can introduce significant biases and systematics if not carefully performed \citep{umetsu+20}.

Traditionally, the characteristic mass of galaxy clusters, and thus of dark matter halos, has been defined based on spherical overdensity. These criteria assign mass within an overdensity relative to a reference density, such as the critical (c) or mean (m) density of the universe, leading to common mass definitions like $M_{200c}$, $M_{200m}$, and $M_{vir}$, with associated radii $R_{200c}$, $R_{200m}$, and $R_{vir}$ \citep[e.g.][]{gunn+72, lacey+93}. However, these definitions have limitations. For instance, many satellites may orbit outside the virial radius \citep[e.g.][]{wetzel+14}, and infalling subhalos begin to experience stripping well outside $R_{vir}$ \citep{behroozi+14}. Moreover, as shown by \cite{diemer+13}, these boundary definitions undergo pseudo-evolution: as the universe expands and the reference density decreases, the overdensity threshold is lowered, leading to an apparent increase in halo radius and mass even when the actual density profile remains unchanged. This pseudo-evolution can cause a non-physical increase in halo mass by up to a factor of two since $z = 1.0$. To address these issues, the concept of the splashback radius ($R_{\text{sp}}$) was proposed.

\cite{fillmore+84} and \cite{bertschinger85} demonstrated that models of self-similar secondary infall of matter onto a spherical overdensity predict a density jump where recently infalling material reaches its first apocenter, corresponding to the last density caustic. This location, initially referred to as the second turnaround radius, has been identified in numerical simulations \citep[e.g.][]{diemer+14, adhikari+14, more+15}. These simulations reveal a sharp transition in the outer density profile of clusters, marked by a steepening of the slope at a certain radius. This transition, which is not captured by traditional profiles such as Navarro, Frenk \& White \citep[NFW;][]{navarro+96} or Einasto \citep{merritt+05}, defines what is now known as the splashback radius, a key tracer of a cluster’s dynamical history.

The splashback radius is closely linked to mass accretion rates and delineates the boundary between the virialized core of the cluster and the surrounding infalling material. A rapidly growing potential well reduces the apocenters of particles' orbits, leading to a smaller splashback radius \citep{diemer+14}. Unlike conventional radii, $R_{\text{sp}}$ is a more physically motivated boundary, as it includes all the matter that has been accreted by the cluster by a given redshift. Crucially, it does not undergo pseudo-evolution, as it is not defined based on a reference density, making it a reliable marker of a halo’s growth history. Furthermore, $R_{\text{sp}}$ can provide insights into galaxy evolution, as galaxies within this region have experienced intracluster processes, such as dynamical friction and ram pressure, which can influence their star formation rates, colors, and morphologies.

The first observational evidence of the splashback radius was presented by \citet{more+16}, who detected the feature in the stacked surface density profiles of galaxies from the Sloan Digital Sky Survey (SDSS). Subsequent works, such as \citet{baxter+17}, confirmed these results, identifying truncation in the halo model using similar SDSS data. However, their findings suggested discrepancies with the predicted values from simulations, later confirmed by other studies \citep[e.g.][]{oneil+21, oneil+22, rana+23}. Observational measurements of $R_{\text{sp}}$ typically yield values near $R_{200m}$, whereas simulations predict larger ratios, reflecting the lower accretion rates expected in galaxy clusters \citep[e.g.][]{valles-parez+20}.

Since these initial detections, numerous efforts have been made to understand these discrepancies, proposing that effects such as dynamical friction, miscentering, or projection may play a role \citep[e.g.][]{more+16, murata+20, shin+19}. In particular, \citet{zurcher+19} demonstrated that clusters selected via the Sunyaev-Zel'dovich (SZ) effect \citep{sunyaev+72} show splashback radii consistent with expectations from dark matter simulations. Additionally, new techniques have been developed to reliably estimate $R_{\text{sp}}$, such as directly modeling the logarithmic slope of the surface density \citep{oneil+22} or using visual inspection with the cumulative number profile \citep{kopylova+20}. The splashback radius has even been explored in the context of modified gravity theories and cosmological parameter constraints \citep[e.g.][]{adhikari+18, contigiani+19, haggar+24}.

\cite{adhikari+21} further explored the relationship between galaxy colors and $R_{\text{sp}}$, finding that blue galaxies, which are more recently accreted, have not yet reached their apocenters, while red galaxies exhibit larger splashback radii. This color dependence offers insights into the timing of galaxy infall and its relationship with environmental processes within clusters.

Given the robustness of $R_{\text{sp}}$, we can also define a mass associated with it, the splashback mass ($M_{\text{sp}}$), which, like the radius, is free from pseudo-evolution. $M_{\text{sp}}$ offers a more representative mass estimate for the cluster, accounting for all the matter accreted by a given redshift, and is thus an excellent tracer for studying the growth of structure over cosmic time, with strong implications for cosmological constraints. Nonetheless, these mass definitions are not widely used or well understood, as there has been relatively little intuitive development around them \citep[e.g.][]{ryu+21, diemer20}.

In this work, we explore the possibility of estimating the splashback radii and masses of galaxy clusters using SDSS spectroscopic data and mock catalogs. We test halo density profiles from the literature and model the cumulative number profile of these systems. We also investigate how cluster properties such as center definition, magnitude limit, galaxy colors and field contamination influence the splashback feature, and explore the potential discrepancies between observations and dark matter simulations. Finally, we discuss the establishment of a scaling relation between $R_{\text{sp}}$ and $M_{\text{sp}}$, leveraging the fact that $R_{\text{sp}}$ is an easily observable feature. Throughout this paper, we adopt a flat $\Lambda$CDM cosmology with parameters $\Omega_M = 0.28$ and $\Omega_\Lambda = 0.72$, and a Hubble constant of $H_0 = 70 h_{70} \quad \text{km}~\text{s}^{-1}~\text{Mpc}^{-1}$, with $h_{70} = 1.0$. The outline of this work is as follows: in Section \ref{sec:data}, we describe the observational and mock data; in Section \ref{sec:methods}, we explain our methods for cumulative profile modeling and model selection; in Section \ref{sec:results}, we present our results and compare them with previous findings; and in Section \ref{sec:conclusion}, we summarize our results and discuss their broader implications for future research.

\section{Data} \label{sec:data}

In this section, we describe the galaxy cluster catalogs used in our work. Section \ref{sec:data:sdss} details the publicly available catalogs of weak-lensing (WL) mass measurements for galaxy clusters and the selection criteria for Sloan Digital Sky Survey galaxies. In Section \ref{sec:data:mock}, we describe the mock catalog of simulated galaxies used for comparison in our analysis.

\subsection{SDSS clusters}\label{sec:data:sdss}

We use a publicly available compilation of galaxy clusters with weak-lensing mass measurements from \citet{sereno15}. This dataset includes at least 822 WL mass measurements reported for overdensities of 2500c, 500c, 200c, and virial masses. We selected only clusters not identified as part of complex structures. Additionally, we combined this dataset with WL mass measurements ($M_{500\text{c}}$ and $M_{200\text{c}}$) from \citet{herbonnet20}, which includes 48 clusters from the Multi Epoch Nearby Cluster Survey (MENeaCS) and 52 clusters from the Canadian Cluster Comparison Project (CCCP). Their analysis leverages high-precision photometric redshift catalogs and advanced shape measurements, achieving low systematic uncertainties in the lensing signal. Duplicate entries were carefully handled during the combination process.

From this combined dataset, we selected clusters within the mass range $14 < \log(M_{200c}/M_{\odot}) < 15.5$ and redshift range $0.01 < z < 0.2$ to minimize the impact of cosmological evolution in our analysis, which will be explored in future work. We also limited our sample to clusters within the SDSS DR17\footnote{\url{skyserver.sdss.org/dr17}} footprint, as we intend to use spectroscopic redshifts to determine cluster membership. SDSS is complete up to a magnitude of $\approx 18$ in the r-band, except in regions affected by fiber collisions \citep{strauss02}.

Due to the challenges of performing accurate weak-lensing analysis at low redshifts, where clusters span large areas of the sky, and the limitations of SDSS imaging for these methods, our final sample consists of 60 clusters that meet our criteria and are covered by SDSS beyond 10 Mpc from the cluster center. We also merged this final sample with an X-ray catalog of galaxy clusters from \citet{wang14} to obtain X-ray centers for our dataset, allowing for a comparison with optical centers (see Section \ref{sec:results}). This merge covered about 87\% of our dataset, corresponding to 54 clusters, ensuring a robust comparison. We refer to this final subset as our SDSS sample.

For each cluster in our final catalog, we excluded galaxies flagged as \verb|SATURATED|, \verb|SATUR_CENTER|, \verb|BRIGHT|, or \verb|DEBLENDED_AS_MOVING| in SDSS. We considered only extinction-corrected magnitudes and applied k-corrections to each SDSS filter using the analytical approximations from \citet{chilingarian10} and \citet{chilingarian11}, although the redshift interval is small (with typical corrections of $\approx 0.09$ mag in the r band). We also included only galaxies brighter than $M_r < -20.13$ (corresponding to $m_r < 18$ at our median redshift of 0.09) to avoid unfair comparisons for low-redshift clusters, which are dominated by fainter objects, although we verify the impact of this limit in our measurements. For membership determination, we considered galaxies to be cluster members if their recession velocities fell within $\pm 3\sigma_v$ of the cluster's mean velocity, where $\sigma_v$ is the velocity dispersion of the cluster. We also explored different velocity intervals to assess the influence of field contamination. This sample is presented in Table \ref{tab:wl}, where we list the cluster names, positions in both optical (R.A.$^{\text{BCG}}$, Dec.$^{\text{BCG}}$) and X-ray (R.A.$^{\text{X-ray}}$, Dec.$^{\text{X-ray}}$) catalogs, redshifts ($z$), mass determinations in terms of critical density ($M_{200c}$), and the references for these mass estimates. From the mass estimates, we also derived the velocity dispersions of the clusters ($\sigma_v$) following the expressions from \citet{carlberg+97}:

\begin{equation}
    \sigma_v = \sqrt{\frac{GM_{200c}}{3R_{200c}}},
\end{equation}

where $R_{200c}$ was also derived from the mass as:

\begin{equation}
    R_{200c} = \left(\frac{3M_{200c}}{4\pi\delta\rho_c}\right)^{1/3}.
\end{equation}

We additionally estimated the parameter $N_{200c}$, representing the cluster richness within $R_{200c}$, constrained by the velocity recession interval and the magnitude limit described above.

\begin{deluxetable*}{cccccccccc}
\tabletypesize{\scriptsize}
\tablewidth{0pt} 
\tablecaption{Properties of clusters in SDSS sample.\label{tab:wl}}
\tablehead{
 \colhead{Cluster} & \colhead{R.A.$^{\text{BCG}}$} & \colhead{Dec.$^{\text{BCG}}$} &  \colhead{R.A.$^{\text{X-ray}}$} & \colhead{Dec.$^{\text{X-ray}}$} & \colhead{$z$} & \colhead{$M_{200c}$} & \colhead{$\sigma_v$} & \colhead{$N_{200c}$} & \colhead{Reference}\\
 \colhead{} & \colhead{(deg)} & \colhead{(deg)} & \colhead{(deg)} & \colhead{(deg)} & \colhead{} & \colhead{($10^{14}$ $h_{70}^{-1} M_{\odot}$)} & \colhead{(kms$^{-1}$)} & \colhead{} & \colhead{}}
\startdata
A1033 & 157.935 & 35.041 & 157.935 & 35.041 & 0.123 & $8.5 \pm 3.3$ & 880.272 & 37 & \citet{herbonnet20}\\
A1066 & 159.778 & 5.21 & 159.883 & 5.167 & 0.069 & $8.98 \pm 4.66$ & 886.511 & 72 & \citet{kubo+09}$^\dagger$\\
A1068 & 160.142 & 40.064 & 160.185 & 39.953 & 0.139 & $5.0 \pm 2.2$ & 740.002 & 15 & \cite{herbonnet20}\\
A1132 & 164.599 & 56.795 & 164.599 & 56.795 & 0.135 & $11.2 \pm 2.3$ & 967.354 & 37 & \citet{herbonnet20}\\
A1139 & 164.715 & 1.651 & -- & -- & 0.04 & $1.19 \pm 0.22$ & 447.136 & 27 & \citet{wojtak+10}\\
A1234 & 170.625 & 21.406 & 170.632 & 21.409 & 0.164 & $6.9 \pm 2.4$ & 828.422 & 23 & \citet{herbonnet20}\\
A1246 & 171.016 & 21.491 & 170.995 & 21.479 & 0.193 & $6.1 \pm 2.3$ & 800.029 & 17 & \citet{herbonnet20}\\
A1314 & 173.497 & 49.062 & 173.705 & 49.078 & 0.033 & $2.93 \pm 0.46$ & 604.823 & 30 & \citet{wojtak+10}$^\dagger$\\
A1413 & 178.743 & 23.422 & 178.825 & 23.405 & 0.141 & $10.8 \pm 2.9$ & 956.807 & 19 & \cite{herbonnet20}\\
A1437 & 180.216 & 3.316 & 180.106 & 3.347 & 0.134 & $3.93 \pm 2.19$ & 682.071 & 15 & \citet{cypriano+04}$^\dagger$\\
A1456 & 180.953 & 4.345 & 180.961 & 4.291 & 0.135 & $11.71 \pm 8.64$ & 981.717 & 27 & \citet{dietrich+09}$^\dagger$\\
A1553 & 187.704 & 10.546 & 187.704 & 10.546 & 0.166 & $7.65 \pm 4.18$ & 857.67 & 9 & \citet{cypriano+04}$^\dagger$\\
A1650 & 194.659 & -1.575 & 194.673 & -1.762 & 0.084 & $10.5 \pm 2.5$ & 936.802 & 59 & \citet{herbonnet20}\\
A1691 & 197.824 & 39.288 & 197.786 & 39.227 & 0.072 & $4.07 \pm 1.02$ & 681.029 & 58 & \citet{wojtak+10}$^\dagger$\\
A1767 & 203.986 & 59.233 & 204.035 & 59.206 & 0.071 & $3.43 \pm 2.53$ & 642.716 & 71 & \citet{kubo+09}$^\dagger$\\
A1773 & 205.623 & 2.201 & 205.54 & 2.227 & 0.077 & $5.07 \pm 1.0$ & 733.824 & 56 & \citet{wojtak+10}$^\dagger$\\
A1781 & 206.219 & 29.771 & 206.239 & 29.737 & 0.062 & $1.5 \pm 1.5$ & 485.841 & 27 & \citet{herbonnet20}\\
A1795 & 207.422 & 26.717 & 207.219 & 26.593 & 0.063 & $13.9 \pm 2.8$ & 1023.759 & 80 & \citet{herbonnet20}\\
A1809 & 208.277 & 5.15 & 208.277 & 5.15 & 0.08 & $4.22 \pm 0.66$ & 690.409 & 69 & \citet{wojtak+10}\\
A1914 & 216.486 & 37.816 & 216.486 & 37.816 & 0.167 & $11.3 \pm 2.5$ & 976.56 & 32 & \citet{herbonnet20}\\
A1927 & 217.644 & 25.647 & 217.778 & 25.634 & 0.095 & $4.4 \pm 1.8$ & 702.491 & 30 & \citet{herbonnet20}\\
A1983 & 223.18 & 16.904 & 223.23 & 16.703 & 0.045 & $1.19 \pm 0.27$ & 447.678 & 47 & \citet{wojtak+10}$^\dagger$\\
A1991 & 223.7 & 18.564 & 223.631 & 18.642 & 0.058 & $3.7 \pm 2.4$ & 657.688 & 56 & \citet{herbonnet20}\\
A2029 & 227.75 & 5.783 & 227.734 & 5.745 & 0.078 & $18.1 \pm 3.4$ & 1120.757 & 143 & \citet{herbonnet20}\\
A2033 & 227.932 & 6.179 & 227.86 & 6.349 & 0.08 & $3.2 \pm 1.8$ & 629.262 & 54 & \cite{herbonnet20}\\
A2034 & 227.584 & 33.486 & 227.549 & 33.486 & 0.113 & $8.09 \pm 4.85$ & 864.078 & 51 & \citet{okabe+08}$^\dagger$\\
A2048 & 228.809 & 4.386 & 228.819 & 4.395 & 0.098 & $9.2 \pm 5.08$ & 899.179 & 63 & \citet{kubo+09}\\
A2050 & 229.075 & 0.089 & 228.714 & 0.304 & 0.12 & $4.6 \pm 2.0$ & 716.755 & 25 & \citet{herbonnet20}\\
A2051 & 229.184 & -0.969 & 229.242 & -1.111 & 0.118 & $4.73 \pm 0.42$ & 723.222 & 33 & \citet{ettori+11}$^\dagger$\\
A2052 & 229.185 & 7.022 & 229.185 & 7.022 & 0.034 & $2.48 \pm 0.42$ & 572.015 & 42 & \citet{wojtak+10}$^\dagger$\\
A2055 & 229.669 & 6.238 & 229.69 & 6.232 & 0.102 & $2.9 \pm 1.8$ & 611.78 & 25 & \citet{herbonnet20}\\
A2063 & 230.772 & 8.609 & 230.772 & 8.609 & 0.034 & $5.21 \pm 0.86$ & 733.942 & 51 & \citet{wojtak+10}$^\dagger$\\
A2065 & 230.6 & 27.714 & 230.622 & 27.708 & 0.072 & $12.0 \pm 2.6$ & 976.976 & 124 & \citet{herbonnet20}\\
A2069 & 231.1 & 30.006 & 231.091 & 30.012 & 0.116 & $3.2 \pm 1.9$ & 634.146 & 34 & \citet{herbonnet20}\\
A21 & 5.012 & 28.75 & -- & -- & 0.094 & $6.1 \pm 2.1$ & 783.396 & 30 & \citet{herbonnet20}\\
A2107 & 234.845 & 21.739 & 234.913 & 21.783 & 0.042 & $2.42 \pm 0.46$ & 568.173 & 39 & \citet{wojtak+10}$^\dagger$\\
A2142 & 239.583 & 27.233 & 239.583 & 27.233 & 0.09 & $14.5 \pm 3.0$ & 1043.928 & 172 & \citet{herbonnet20}\\
A2175 & 245.13 & 29.891 & 245.13 & 29.891 & 0.096 & $5.03 \pm 1.29$ & 734.825 & 47 & \citet{wojtak+10}$^\dagger$\\
A2199 & 247.159 & 39.551 & 247.163 & 39.553 & 0.03 & $6.98 \pm 3.86$ & 808.717 & 92 & \citet{kubo+09}$^\dagger$\\
A2244 & 255.628 & 34.042 & 255.677 & 34.06 & 0.099 & $4.68 \pm 3.54$ & 717.658 & 64 & \citet{kubo+09}$^\dagger$\\
A2259 & 260.045 & 27.704 & 260.04 & 27.669 & 0.16 & $6.7 \pm 2.1$ & 819.68 & 10 & \citet{herbonnet20}\\
A2627 & 354.212 & 23.924 & -- & -- & 0.125 & $3.0 \pm 1.9$ & 621.857 & 16 & \citet{herbonnet20}\\
A2634 & 355.118 & 27.109 & -- & -- & 0.032 & $4.95 \pm 0.68$ & 721.043 & 17 & \citet{wojtak+10}$^\dagger$\\
A2700 & 0.89 & 2.138 & -- & -- & 0.094 & $1.9 \pm 0.23$ & 529.801 & 17 & \citet{ettori+11}$^\dagger$\\
A291 & 30.43 & -2.197 & -- & -- & 0.196 & $9.02 \pm 2.67$ & 911.872 & 24 & \citet{okabe+10}$^\dagger$\\
A383 & 42.014 & -3.529 & -- & -- & 0.189 & $4.6 \pm 2.4$ & 727.524 & 23 & \citet{herbonnet20}\\
A671 & 127.132 & 30.431 & 127.132 & 30.431 & 0.05 & $4.85 \pm 0.78$ & 718.81 & 50 & \citet{wojtak+10}$^\dagger$\\
A7 & 2.939 & 32.416 & -- & -- & 0.104 & $4.4 \pm 1.9$ & 703.839 & 33 & \citet{herbonnet20}\\
A779 & 140.151 & 33.651 & 139.945 & 33.75 & 0.023 & $1.29 \pm 0.2$ & 457.8 & 17 & \citet{wojtak+10}$^\dagger$\\
A795 & 141.064 & 14.128 & 141.022 & 14.173 & 0.139 & $16.0 \pm 3.9$ & 1089.488 & 46 & \citet{herbonnet20}\\
A85 & 10.459 & -9.43 & 10.46 & -9.303 & 0.055 & $8.4 \pm 2.7$ & 864.671 & 76 & \citet{herbonnet20}\\
A954 & 153.496 & -0.082 & 153.9 & 0.013 & 0.094 & $4.32 \pm 1.14$ & 698.039 & 28 & \cite{wojtak+10}$^\dagger$\\
A961 & 154.095 & 33.638 & 154.095 & 33.638 & 0.127 & $7.0 \pm 2.3$ & 825.962 & 32 & \cite{herbonnet20}\\
Coma-Cluster & 195.034 & 27.977 & 194.899 & 27.959 & 0.024 & $12.3 \pm 4.8$ & 975.326 & 136 & \citet{gavazzi+09}$^\dagger$\\
MKW-4 & 181.113 & 1.896 & 181.113 & 1.896 & 0.02 & $1.55 \pm 0.22$ & 486.744 & 18 & \citet{wojtak+10}$^\dagger$\\
MKW-9 & 233.133 & 4.681 & 233.138 & 4.684 & 0.039 & $1.2 \pm 0.3$ & 448.321 & 22 & \citet{pointecouteau+05}$^\dagger$\\
MKW3S & 230.466 & 7.709 & 230.404 & 7.719 & 0.045 & $2.5 \pm 2.1$ & 574.819 & 31 & \cite{herbonnet20}\\
MS0906 & 137.303 & 10.975 & 137.303 & 10.975 & 0.166 & $10.0 \pm 2.1$ & 937.653 & 25 & \cite{herbonnet20}\\
RX-J1720.1+2638 & 260.042 & 26.626 & 260.042 & 26.626 & 0.161 & $7.3 \pm 5.56$ & 843.682 & 17 & \citet{pedersen+07}$^\dagger$\\
ZWCL1215 & 184.421 & 3.656 & 184.421 & 3.656 & 0.077 & $5.1 \pm 2.7$ & 735.243 & 97 & \citet{herbonnet20}\\
\enddata
\tablecomments{$^\dagger$Clusters present in \citet{sereno15} compilation.}
\end{deluxetable*}

\subsection{Mock clusters}\label{sec:data:mock}

As a control sample, we used a mock catalog from a simulated sky light-cone from \citet{araya-araya21}, referred to as our Mock sample. This catalog emulates a projected area of 324 square degrees, using synthetic galaxies generated by the semi-analytical model (SAM) from \citet{henriques15}. The algorithm is based on the Millennium Run simulation \citep{springel05}, scaled to match the Planck 1 cosmological parameters \citep{planck14}.

\begin{figure}
    \centering
    \includegraphics[width=0.85\linewidth]{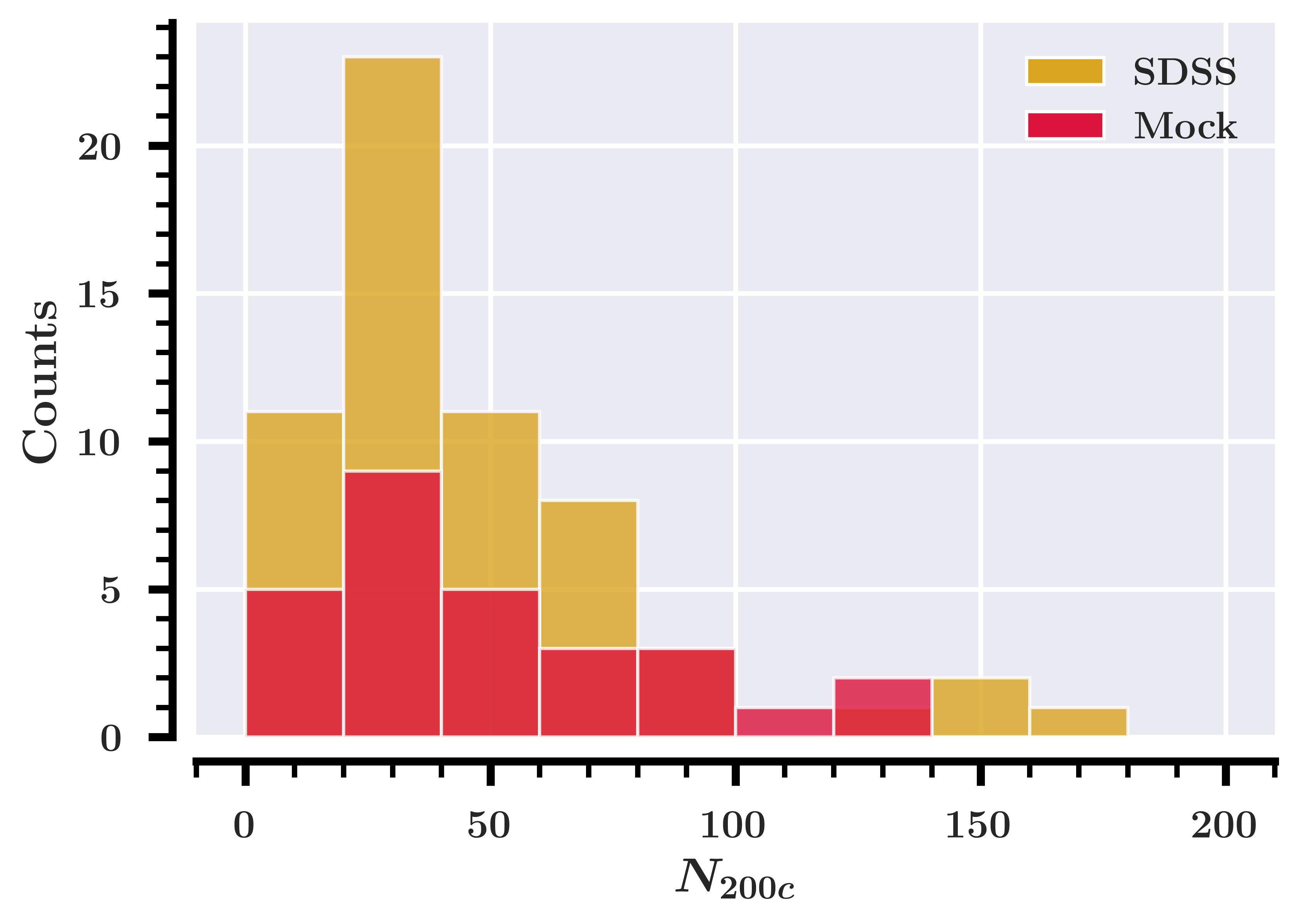}
    \caption{Distribution of $N_{200c}$ richness for the SDSS (yellow) and Mock (pink) cluster samples.}
  \label{fig:abundances}
\end{figure}

For consistency, we applied the same selection criteria as for the SDSS sample, focusing on clusters within the mass range $14 < \log(M_{200c}/M_{\odot}) < 15.5$ and redshift range $0.01 < z < 0.2$. For each cluster, we also only considered galaxies brighter than $M_r < -20.13$, ensuring that we avoid issues related to the mock catalog's stellar mass resolution ($\approx 10^8 M_{\odot} h^{-1}$). Due to the limited volume of the simulation box and the rarity of massive clusters ($\log{(M_{200\text{c}}/M_{\odot})} > 14.5$), where our observational dataset is centered, our Mock sample contains only 30 clusters. Nevertheless, we believe these clusters are representative of the SDSS dataset for comparing the distributions of observables. As a consistency check, we show in Figure \ref{fig:abundances} the distribution of $N_{200c}$ richness for both the SDSS and Mock samples. After applying the same magnitude and velocity cuts, we find that the galaxy abundances in both samples are in close agreement, supporting the representativeness of our Mock sample for comparison with observations.

\begin{figure*}
    \centering
    \includegraphics[width=0.85\linewidth]{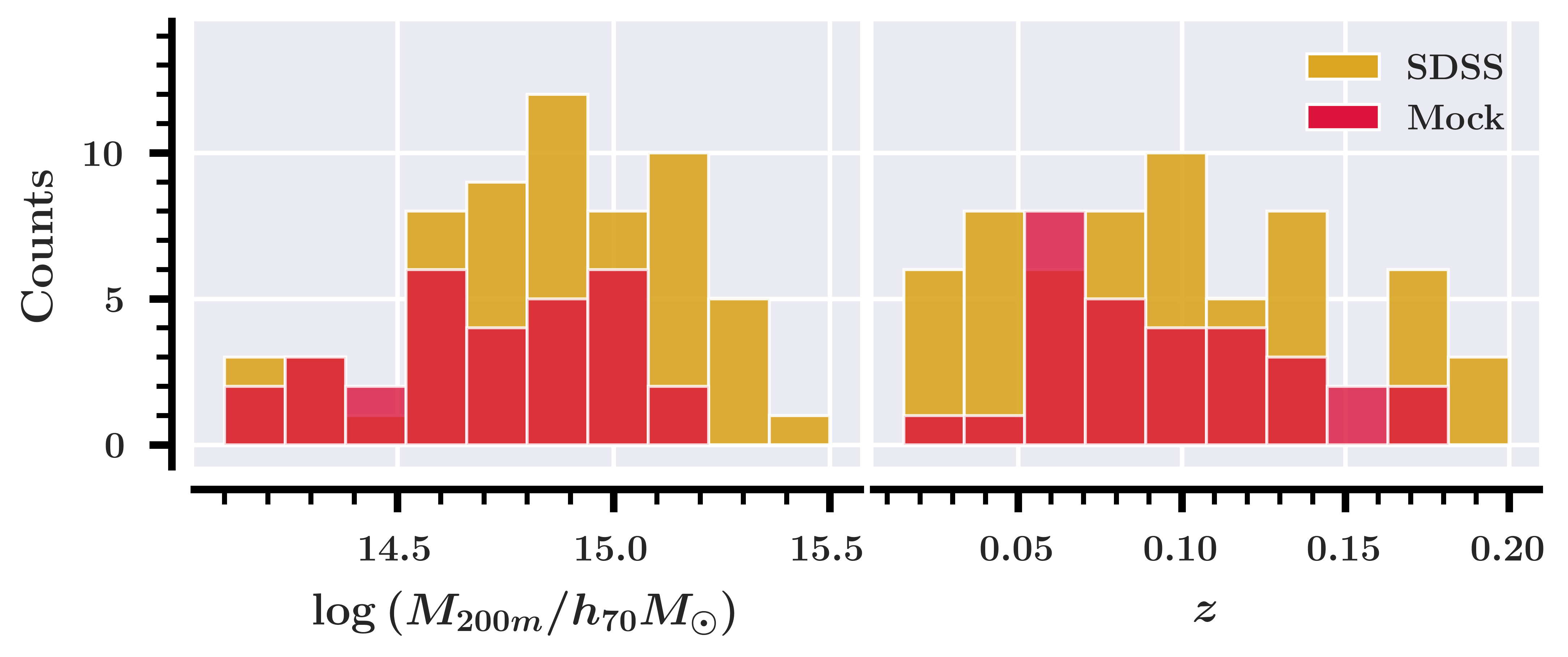}
    \caption{Distributions of $M_{200\text{m}}$ masses (left panel) and redshift (right panel) for our samples. SDSS clusters are shown in yellow, while Mock clusters are in pink.}
  \label{fig:mass_z_dist}
\end{figure*}

For comparison with previous works, where $R_{\text{sp}}$ is often reported relative to $R_{200\text{m}}$ because cluster outskirts are most self-similar when stacked by this quantity \citep[e.g.][]{diemer+14, shi16, umetsu17}, we converted our mass definition from 200c to 200m and derived $R_{200\text{m}}$ estimates assuming the mass-concentration relation from \citet{diemer19}. The distributions of mass and redshift for both the SDSS and Mock samples are shown in Figure \ref{fig:mass_z_dist}, with observational clusters in yellow and simulated ones in pink. It is important to note that the mock sample does not exhibit mass distributions as high as those of the SDSS sample. Additionally, there is a lack of clusters at $z < 0.05$ due to their large sizes within the simulation box, causing even their inner regions to intersect the edges of the projected light cone.

\section{Methods} \label{sec:methods}

In this section, we describe the models we tested and the process used to analyze cumulative and individual cluster profiles. We also outline how we estimate splashback masses from $M_{200c}$ catalogs.

\subsection{Cumulative Model} \label{subsec:model}

Typically, splashback radius estimates are derived from modeling surface number density profiles, which have been shown to approximate the real density of dark matter halos \citep[e.g.][]{lebeau+24}. To improve the signal-to-noise ratio, previous studies have stacked these surface number densities, as individual profiles tend to be noisy, complicating the estimation of $R_{sp}$ for individual clusters. However, the primary objective of this work is to obtain splashback radii for single clusters without relying on stacked profiles. Instead of using surface number density, we chose to employ cumulative number profiles, which are more robust, particularly in systems with low galaxy counts, such as poor or high-redshift clusters. Assuming spherical symmetry, the cumulative profile is easily convertible from the surface number density profile via the relation:

\begin{equation} N(<R) = 2\pi \int_0^R \Sigma(R) R dR, \end{equation}

where the surface density, $\Sigma(R)$, follows the model proposed by \citet{diemer+14}, combining a one-halo and two-halo term with a smooth transition function:

\begin{equation} \Sigma(R) = \Sigma_{\text{1-halo}} f_t + \Sigma_{\text{2-halo}}. \end{equation}

To minimize numerical errors in typical Abel integrals, we directly apply known analytical projected models. For the one-halo term, we test two models. First, the Sérsic profile \citep{sersic63}:

\begin{equation}
\Sigma_{\text{sersic}}(R) = \Sigma_e \exp{\left\{-b_n \left[\left(\frac{R}{R_e}\right)^{\frac{1}{n}} - 1\right]\right\}}.
\end{equation}

where $b_n$ can be approximated analytically using the expression from \citet{ciotti+99}:

\begin{equation}
b_n = 2n - \frac{1}{3} + \frac{4}{405n} + \frac{46}{25515n^2} + \frac{131}{1148175n^3}.
\end{equation}

We also consider the analytical projected Navarro-Frenk-White (NFW) model \citep{navarro+96, merritt+05}, as derived by \citet{lokas+01}:

\begin{align}
&\Sigma_{\text{NFW}}(R) = \frac{2 r_s \rho_s}{x^2 - 1} \left( 1 - \frac{\mathcal{C}(x)}{\sqrt{|x^2 - 1|}} \right), \\
&\mathcal{C}(x) = 
\begin{cases} 
    \cosh^{-1}(1/x) & \text{if } x < 1, \\
    \cos^{-1}(1/x) & \text{if } x > 1,
\end{cases} \\
&x = \frac{R}{r_s}.
\end{align}

The transition function, $f_t$, is modeled using the smooth exponential form proposed by \citet{diemer22}:

\begin{equation} f_t = \exp{\left[ -(R/R_t)^\tau \right]}\label{eq:f_t}, \end{equation}

For the two-halo term, we employ the projected two-point correlation function from \citet{davis+83}:

\begin{equation}
\Sigma_{\text{2-halo}} = \rho_m \left[ \left(\frac{R}{r_{\text{out}}}\right)^{1 - \gamma} B\left(\frac{\gamma-1}{2}, \frac{1}{2}\right) + 1 \right]\label{eq:2-halo},
\end{equation}

where $B(X, Y)$ is the Beta function. We fix $r_{\text{out}} = 1.5$ Mpc (or $1.5R_{200m}$ in the case of stacked profiles) due to its degeneracy with $\rho_m$.

The Sérsic model (hereafter trunc-Sérsic) includes seven free parameters ($\Sigma_e$, $R_e$, $n$, $R_t$, $\tau$, $\rho_m$, $\gamma$), while the NFW model (hereafter trunc-NFW) has six ($\rho_s$, $r_s$, $R_t$, $\tau$, $\rho_m$, $\gamma$). We fit these parameters using a Markov Chain Monte Carlo (MCMC) sampler with a Poisson likelihood.

Nonetheless, the deprojection can be readily achieved using a conversion constant, as noted in previous studies on scaling relations of projected quantities \citep[e.g.][]{tully15} and specifically confirmed for splashback measurements in the results of \citet{more+16}, where the modeling of the 3D splashback radius ($r_\text{sp}$) is converted into the 2D $R_\text{sp}$ using:

\begin{equation} \label{eq:r_sp}
r_\text{sp} = \sqrt{\frac{\pi}{2}} R_{\text{sp}}.
\end{equation}

In modeling individual profiles, we focus on applying priors based on typical values from the literature. These priors are corroborated through our stacked profile analysis, which uses non-informative priors to explore the full parameter space (see the next section for details).

\subsubsection{Stacked Profiles}\label{subsubsec:stacked}

To determine the best-fitting model, we apply a traditional stacking procedure separately for the SDSS and Mock samples. Clusters are sorted into two redshift bins, $0 < z < 0.09$ and $0.09 < z < 0.2$, and three mass bins: $14 < \log{(M_{200\text{m}}/M_{\odot})} < 14.8$, $14.8 < \log{(M_{200\text{m}}/M_{\odot})} < 15$, and $15 < \log{(M_{200\text{m}}/M_{\odot})} < 15.5$. Given the limited sample size in the Mock data, we use only two mass bins for these data: $\log{(M_{200\text{m}}/M_{\odot})} < 14.8$ and $\log{(M_{200\text{m}}/M_{\odot})} > 14.8$. Observed and simulated clusters are not mixed in the stacking. Each bin interval is defined to contain an equal number of clusters, 10 for the SDSS sample and $\approx$ 8 for the Mock one, ensuring balanced representation across bins.

To prevent massive clusters from dominating the stack, we normalize the clustercentric distance by $R_{200m}$ and fit the data on a logarithmic grid of 100 points between 0.1 and 5 $R_{200m}$. We also normalize the cumulative counts to mitigate potential incompleteness at higher redshifts. The background galaxy density, $\Sigma_{bg}$, is estimated in both the SDSS footprint and the Mock field under the same magnitude and redshift conditions for each cluster. We randomly select 100 points within the available area and measure the surface number density over a $9 \times 9$ square degree region, large enough to encompass our low-redshift clusters. However, for the Mock sample, we reduce the area to $2 \times 2$ square degrees due to the limited size of the simulation box. The median of these 100 points is used to define the ``cumulative background", which is unique to each cluster:

\begin{equation} N_{bg}(<R) = \pi R^2 \Sigma_{bg}. \end{equation}

\begin{figure}
\centering
\includegraphics[width=1.0\linewidth]{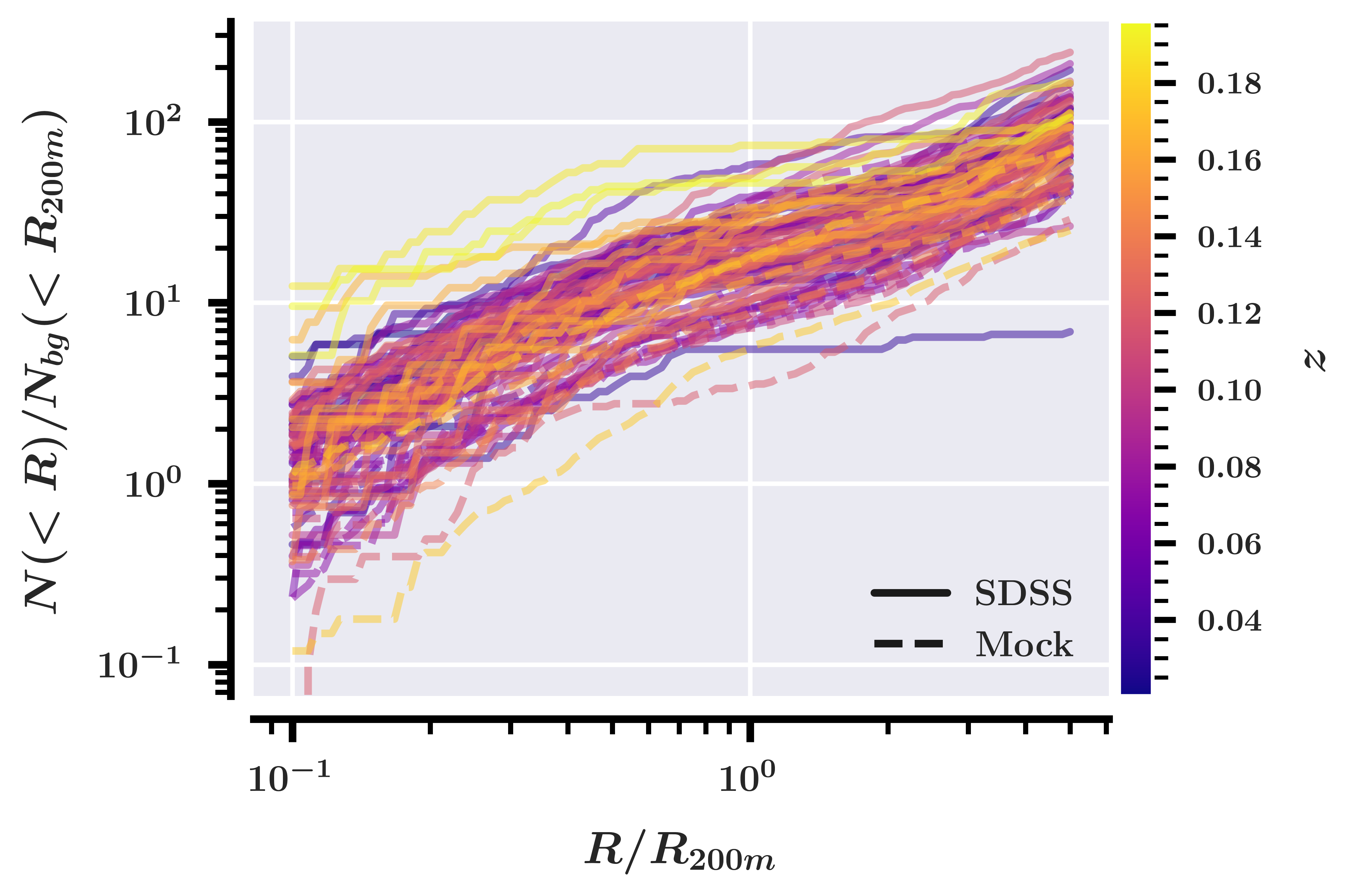} \caption{Normalized cumulative number profiles for SDSS (filled lines) and Mock (dashed lines) clusters, color-coded by redshift.}\label{fig:norm_profiles}
\end{figure}

We arbitrarily choose $R=R_{200m}$ for normalization. The normalized cumulative number profiles for the SDSS and Mock samples are shown in Figure \ref{fig:norm_profiles}, with the SDSS profiles represented by filled lines and the Mock profiles by dashed lines, both color-coded by redshift. Since we observe no significant differences between the normalized profiles of observational and simulated clusters, we conclude that our normalization procedure is effective. For the fitting, we also smooth the profiles using a Savitzky–Golay filter with a third-order polynomial and a five-point window. This smoothing step helps mitigate noise, which primarily affects the smaller radii due to the low galaxy count in those regions.

\subsubsection{Individual Profiles}\label{subsubsec:individual}

For individual clusters, our primary objective, we compute the cumulative profile within a comoving clustercentric distance range of 0.3-10 cMpc. This choice avoids the effects of miscentering on small scales and ensures a proper description of the two-halo term on larger scales. However, for the Mock clusters, we restrict this interval to 0.3–6.5 cMpc to prevent sampling regions that extend beyond the boundaries of the projected light cone. Additionally, by excluding the innermost 0.3 cMpc, we mitigate potential resolution-related issues in the central regions of the simulated clusters. The profiles are not smoothed, and we fit them using 100 points in clustercentric distance.

As demonstrated in Section \ref{subsec:model_sel}, the trunc-NFW model outperforms the trunc-Sérsic model. Therefore, we adopt the trunc-NFW as our primary model for individual clusters. We introduce informative priors based on our stacking results, while ensuring consistency with the literature. For example, $\tau$ in Eq. \ref{eq:f_t} is typically around 4, consistent with the findings of \citet{diemer22}, and $\gamma$ in the two-point correlation function (Eq. \ref{eq:2-halo}) is approximately 1.7, in agreement with previous studies \citep[e.g.][]{wang+13, shi+16}. Table \ref{tab:priors} summarizes the priors used for individual cluster modeling.

We estimate the splashback radius by selecting the minimum in the logarithmic slope of the surface density within a range of $\pm 0.5R_t$ in clustercentric distance, avoiding spurious local minima caused by central core features.

\begin{deluxetable}{ccc}
\tabletypesize{\scriptsize}
\tablewidth{0pt} \tablecaption{Priors used in the trunc-NFW model for fitting individual cumulative number profiles.\label{tab:priors}} \tablehead{
\colhead{Parameter} & \colhead{Value} & \colhead{Type}}
\startdata
$\rho_s$ & 0--500 & Uniform \\
$r_s$ & 0.1--1.5 & Uniform \\
$R_t$ & 0.1--8.0 & Uniform \\
$\tau$ & $4 \pm 0.8$ & Gaussian \\
$\rho_m$ & 0--100 & Uniform \\
$r_{\text{out}}$ & 1.5 & Fixed \\
$\gamma$ & $1.7 \pm 0.4$ & Gaussian \\
\enddata
\end{deluxetable}

\subsection{Splashback Mass} \label{subsec:Msp}

To estimate the splashback mass, we take the $M_{200c}$ values from the catalogs and extend the mass profile of the clusters using the NFW profile expression from \citet{lokas+01}, combined with the same previous truncation function:

\begin{align}\label{eq:M_r}
&M(<r) = 4\pi\int_0^r \frac{r^2 M_{200c} dr}{4\pi r_s^3 g_c} \frac{f_t}{x(x+1)^2}, \\
&g_c = \log{(1 + c_{200c})} - \frac{c_{200c}}{1 + c_{200c}}, \\
&c_{200c} = \frac{R_{200c}}{r_s}.
\end{align}

By utilizing the fitted parameters $r_s$, $R_t$, and $\tau$, together with the tabulated $M_{200c}$, we can estimate the mass at any radius, assuming the cluster follows an NFW profile. For instance, in the splashback radius:

\begin{equation}\label{eq:M_sp}
    M_{\text{sp}} \equiv M(<r_{\text{sp}}).
\end{equation}

In practice, the truncation function does not significantly alter other characteristic mass estimates, as noted by \citet{giocoli+24}. Since $R_t$ is of the order of $R_{\text{sp}}$, which is typically much larger than $R_{200c}$, the truncation does not affect mass estimates within radii smaller than $R_{\text{sp}}$.

However, it is important to note that we are combining 2D and 3D profiles, which may introduce bias in our mass estimates. Nonetheless, our tests show that the uncertainties in these mass estimates are large enough that any potential bias does not significantly impact our results. Thus, this approach provides a reasonable estimate for the splashback mass, especially since a direct estimation remains challenging. Still, to minimize inconsistencies, the splashback mass determination is made within the 3D splashback radius, as defined in Equation \ref{eq:r_sp}.

Furthermore, the scale radius estimated in our model fits reflects the spatial distribution of galaxies, rather than the total matter distribution as traced by lensing. This can introduce systematic differences in our estimates depending on the magnitude cut, as brighter galaxies are expected to be more centrally concentrated, reducing the inferred $r_s$, as discussed in \citet{more+16}. Nevertheless, this effect is not expected to be significant, as \citet{shin+22} showed that changes in the inner cluster profiles due to galaxy selection are confined to radii smaller than $\sim 0.3$ Mpc. In any case, the scale radius is not a critical parameter for the quantities we aim to estimate, and it has a minor impact on the determination of the splashback radius or mass.

In the error estimates for the splashback mass, we consider uncertainties in both $M_{200c}$ and $R_{\text{sp}}$. We assume that these errors are independent and combine them directly by propagating the errors from $M_{200c}$ to $M_{\text{sp}}$ ($\sigma_{M_\text{sp}, M_{200c}}$) in Equation \ref{eq:M_sp}, as well as the error in $M_{\text{sp}}$ given the distribution of $R_{\text{sp}}$ ($\sigma_{M_\text{sp}, R_\text{sp}}$) for each cluster, derived from the posterior sampling in MCMC. Thus, the total error is given by:

\begin{equation}
\sigma_{M_\text{sp}} = \sqrt{ \sigma_{M_\text{sp}, M_{200c}}^2 + \sigma_{M_\text{sp}, R_\text{sp}}^2}.
\end{equation}

Our results indicate that the combined error is predominantly influenced by error propagation from $M_{200c}$.

\section{Results} \label{sec:results}

In this section, we present and discuss the main outcomes of our analysis, ranging from model selection and parameter constraints to the scaling relations between the splashback radius and mass, $R_{\text{sp}}$--$M_{\text{sp}}$. 

\subsection{Model selection} \label{subsec:model_sel}

\begin{figure*} \centering \includegraphics[width=0.85\linewidth]{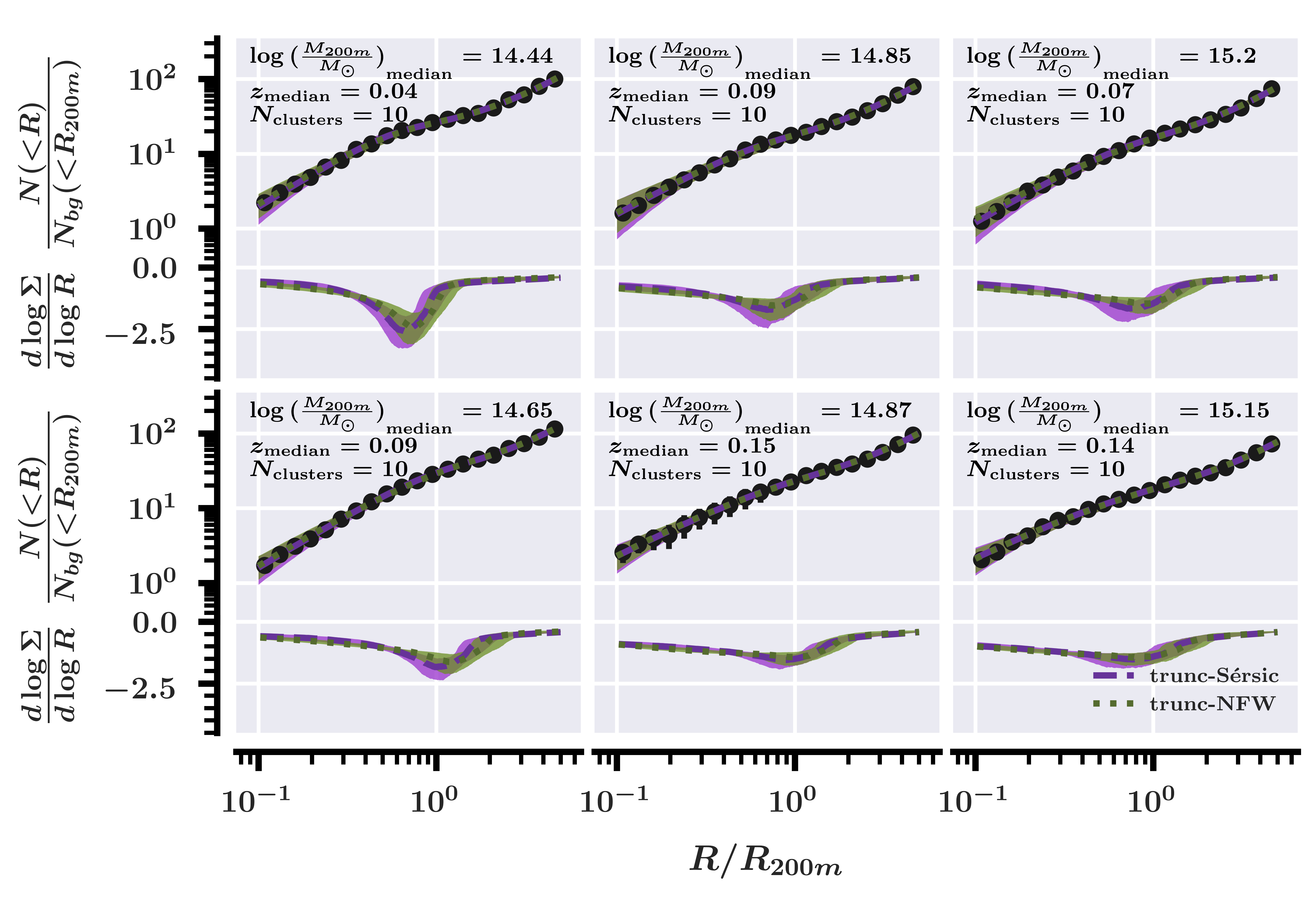} \caption{Model fitting procedure for stacked SDSS clusters. Black points represent the stacked cumulative distributions. Rows represent different redshift bins, and columns represent mass bins. The fitted models are shown as dotted green lines for trunc-NFW and dashed purple lines for trunc-Sérsic. The smaller bottom panels display the logarithmic slope of the surface density in each case.}\label{fig:model_select}
\end{figure*}

\begin{figure} \centering \includegraphics[width=1.0\linewidth]{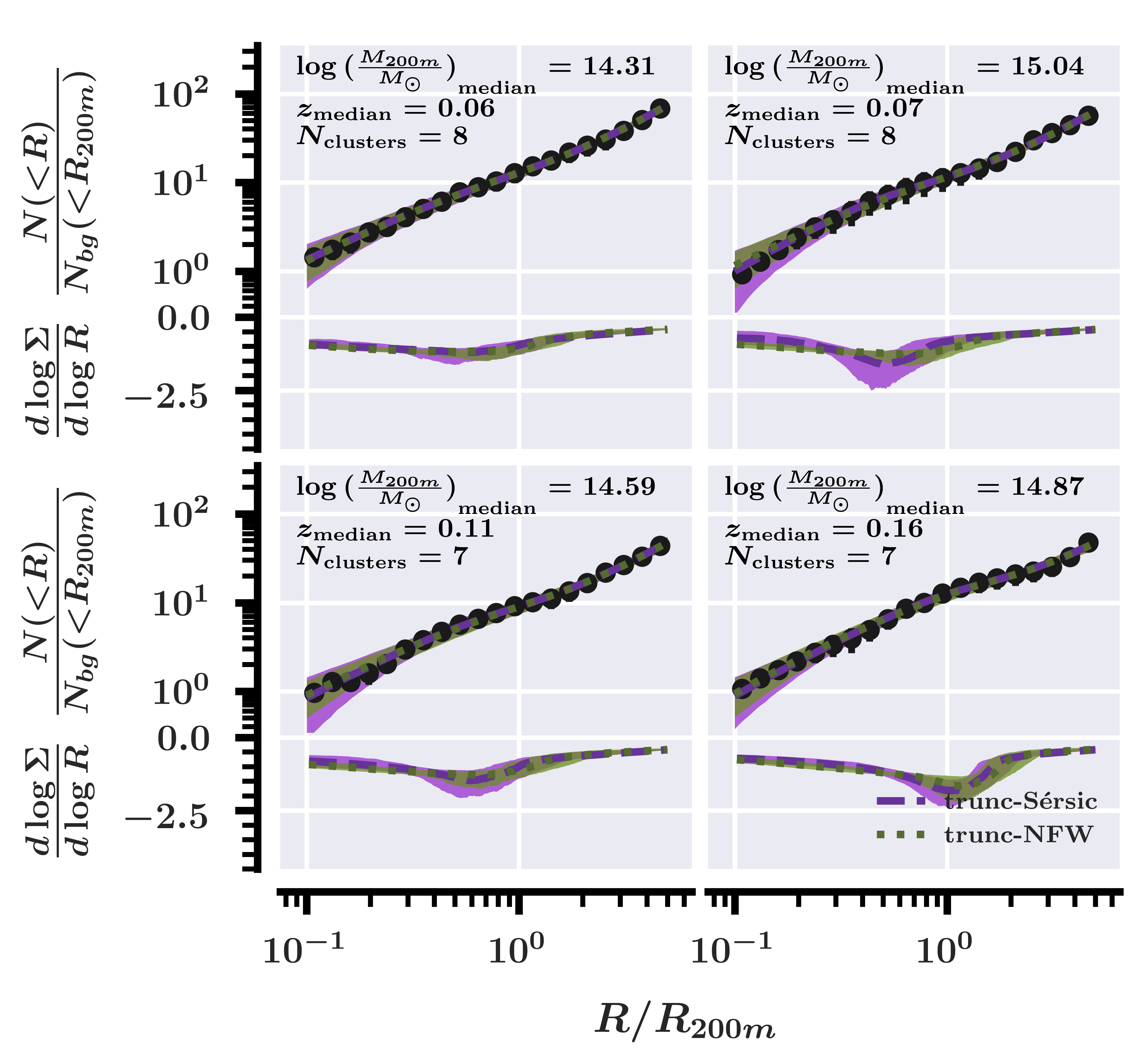} \caption{Model fitting procedure for stacked Mock clusters. Black points represent the stacked cumulative distributions. Rows represent different redshift bins, and columns represent mass bins. The fitted models are shown as dotted green lines for trunc-NFW and dashed purple lines for trunc-Sérsic. The smaller bottom panels display the logarithmic slope of the surface density in each case.}\label{fig:model_select_mock}
\end{figure}

As described in Section \ref{subsubsec:stacked}, we binned our 60 SDSS and 30 Mock clusters separately into intervals of redshift and mass, stacking them to fit the trunc-Sérsic and trunc-NFW models.  To identify the model that best represents our data, we applied statistical methods, a crucial step since our primary objective is to fit individual profiles accurately. The resulting fits for each stacked bin in the SDSS and Mock samples are shown in Figure \ref{fig:model_select} and Figure \ref{fig:model_select_mock}, respectively, where the black points represent the stacked cumulative distribution for binned clusters. Rows correspond to different redshift bins, while columns correspond to mass bins. The fitted models are represented by the dotted green lines for trunc-NFW and dashed purple lines for trunc-Sérsic. The smaller panels below each plot show the logarithmic slope of the surface density for each model.

As shown, the fits for the two models do not display significant visual differences in either the SDSS or Mock stacks, especially regarding the position of the slope profile minima, which act as proxies for the splashback radius. To make a robust decision between the models, we applied three statistical metrics to the cumulative profile fitting, the chi-squared ($\chi^2$), the Akaike Information Criterion \citep[AIC;][]{akaike74}, and the Bayesian Information Criterion \citep[BIC;][]{schwarz78}.

The standard chi-squared statistic is given by:

\begin{equation} \chi^2 = \sum{ \frac{(O - E)^2}{\sigma^2} }, \end{equation}

where $O$ represents the observed values, $E$ the expected values from the model, and $\sigma$ the uncertainties in $O$, which we assume to be the dispersion in the distribution of the stacked profiles. This statistic is useful for discriminating between models, particularly in its reduced form. However, it is important to note that the trunc-Sérsic model has one more degree of freedom (dof) than trunc-NFW, but given that the degrees of freedom are much larger than 1 in both cases, the comparison remains valid.

We also considered the AIC, defined as:

\begin{equation} \text{AIC} = 2k - 2\log(\hat{L}), \end{equation}

where $k$ is the number of model parameters, and $\hat{L}$ is the maximum likelihood, which in our case is Poissonian. AIC penalizes models that do not explain the data well by incorporating the number of parameters used.

The BIC, defined as:

\begin{equation} \text{BIC} = k\log(n) - 2\log(\hat{L}), \end{equation}

where $n$ is the number of data points (100 in all cases), penalizes models with a larger number of free parameters to avoid overfitting.

With $\chi^2$, AIC and BIC, better models have smaller values of these quantites.

The results for these metrics, applied to each binned dataset and sample,  are summarized in Table \ref{tab:model_select}. We compare the normalized splashback radius $R_{\text{sp}}/R_{200\text{m}}$, $\chi^2$, AIC, and BIC for trunc-NFW and trunc-Sérsic. Notably, while the splashback radius is slightly larger for the trunc-NFW model, the values are consistent within $1\sigma$ in all cases. However, it is evident that the trunc-NFW model consistently outperforms the trunc-Sérsic one across all metrics, for both SDSS and Mock samples, which supports the robustness of this model.

Additionally, to verify whether a truncation function is indeed necessary, we present in Table \ref{tab:model_wout_trunc} the same set of metrics for Sérsic and NFW models without a truncation function (see Equation \ref{eq:f_t}), in a similar spirit to \citep{baxter+17}. In this case, the Sérsic model appears to outperform the NFW model; nevertheless, both models perform worse across almost all metrics compared to their truncated counterparts, with the trunc-NFW remaining the best-performing model. Even though in some cases the non-truncated models yield lower AIC and BIC values, such occurrences are rare and show no clear trend with mass or redshift, suggesting these are merely stochastic fluctuations. This demonstrates that including a truncation function $f_t$ significantly improves the modeling of the density profiles. It is important to note that for the non-truncated models we do not estimate a splashback radius, as by definition this quantity arises from models with a truncation.

Based on these findings, we select trunc-NFW as our primary model for individual fits, as discussed in the following sections.

Another interesting aspect is that, while it is difficult to comment on evolution with redshift due to the narrow interval we are analyzing, our results suggest a mass dependence in the $R_{\text{sp}}/R_{200\text{m}}$ ratio, with higher masses exhibiting larger ratios, particularly in the lower redshift bin. This trend has also been observed in previous observational studies \citep[e.g.][]{xu+24} and simulations \citep[e.g.][]{oneil+24}. However, in our higher redshift bin, this trend appears to invert, though again, the narrow interval limits definitive conclusions. As suggested by \citet{towler+24}, this trend could be an artifact of the correlation between cluster mass and mass accretion rate, given that lower-mass clusters tend to exhibit smaller accretion rates, as also discussed in \citet{diemer+14}. Additionally, it is important to note that we treat all these models as more empirical than physical, aiming to identify the best-fitting description of our data rather than strictly adhering to halo model assumptions from cosmology.

\begin{deluxetable*}{ccccccc}
\tabletypesize{\scriptsize}
\tablewidth{0pt} 
\tablecaption{Splashback radius and statistical metric values for each stacked bin of redshift and mass, comparing the trunc-NFW and trunc-Sérsic models for both SDSS and Mock samples.\label{tab:model_select}}
\tablehead{
 \colhead{$z_{\text{med}}$} & \colhead{$\log{(M_{200\text{m}}/M_{\odot})}_{\text{med}}$} & \colhead{Model} &  \colhead{$R_{\text{sp}}/R_{200\text{m}}$} & \colhead{$\chi^2$} & \colhead{AIC} & \colhead{BIC}} 
\startdata
\multicolumn{7}{c}{SDSS sample}\\
\cline{1-7}
0.04 & 14.44 & \begin{tabular}{c}trunc-Sérsic\\ trunc-NFW\end{tabular} & \begin{tabular}{c}$0.71 \pm 0.10$\\ $0.79 \pm 0.12$\end{tabular} & \begin{tabular}{c}136.48\\ 47.66\end{tabular} & \begin{tabular}{c}150.48\\
59.66\end{tabular} & \begin{tabular}{c}168.72\\ 75.29\end{tabular}\\
\cline{3-7}
0.09 & 14.85 & \begin{tabular}{c}trunc-Sérsic\\ trunc-NFW\end{tabular} & \begin{tabular}{c}$0.79 \pm 0.18$\\ $0.87 \pm 0.23$\end{tabular} & \begin{tabular}{c}133.95\\ 20.08\end{tabular} & \begin{tabular}{c}147.95\\
32.08\end{tabular} & \begin{tabular}{c}166.19\\ 47.71\end{tabular}\\
\cline{3-7}
0.07 & 15.20 & \begin{tabular}{c}trunc-Sérsic\\ trunc-NFW\end{tabular} & \begin{tabular}{c}$0.84 \pm 0.20$\\ $0.99 \pm 0.24$\end{tabular} & \begin{tabular}{c}115.59\\ 33.44\end{tabular} & \begin{tabular}{c}129.59\\
45.44\end{tabular} & \begin{tabular}{c}147.82\\ 61.08\end{tabular}\\
\cline{3-7}
0.09 & 14.65 & \begin{tabular}{c}trunc-Sérsic\\ trunc-NFW\end{tabular} & \begin{tabular}{c}$1.13 \pm 0.20$\\ $1.32 \pm 0.24$\end{tabular} & \begin{tabular}{c}232.24\\ 74.54\end{tabular} & \begin{tabular}{c}246.24\\
86.54\end{tabular} & \begin{tabular}{c}264.48\\ 102.17\end{tabular}\\
\cline{3-7}
0.15 & 14.87 & \begin{tabular}{c}trunc-Sérsic\\ trunc-NFW\end{tabular} & \begin{tabular}{c}$0.94 \pm 0.22$\\ $1.12 \pm 0.30$\end{tabular} & \begin{tabular}{c}19.99\\ 9.14\end{tabular} & \begin{tabular}{c}33.99\\
21.14\end{tabular} & \begin{tabular}{c}52.23\\ 36.77\end{tabular}\\
\cline{3-7}
0.14 & 15.15 & \begin{tabular}{c}trunc-Sérsic\\ trunc-NFW\end{tabular} & \begin{tabular}{c}$0.87 \pm 0.28$\\ $0.98 \pm 0.32$\end{tabular} & \begin{tabular}{c}22.40\\ 13.19\end{tabular} & \begin{tabular}{c}36.40\\
25.19\end{tabular} & \begin{tabular}{c}54.64\\ 40.82\end{tabular}\\
\cline{1-7}
\multicolumn{7}{c}{Mock sample}\\
\cline{1-7}
0.06 & 14.31 & \begin{tabular}{c}trunc-Sérsic\\ trunc-NFW\end{tabular} & \begin{tabular}{c}$0.67 \pm 0.30$\\ $0.83 \pm 0.36$\end{tabular} & \begin{tabular}{c}28.56\\ 3.80\end{tabular} & \begin{tabular}{c}42.56\\
15.80\end{tabular} & \begin{tabular}{c}60.80\\ 31.43\end{tabular}\\
\cline{3-7}
0.07 & 15.04 & \begin{tabular}{c}trunc-Sérsic\\ trunc-NFW\end{tabular} & \begin{tabular}{c}$0.58 \pm 0.19$\\ $0.82 \pm 0.34$\end{tabular} & \begin{tabular}{c}14.57\\ 13.70\end{tabular} & \begin{tabular}{c}28.57\\
25.70\end{tabular} & \begin{tabular}{c}46.80\\ 41.33\end{tabular}\\
\cline{3-7}
0.11 & 14.59 & \begin{tabular}{c}trunc-Sérsic\\ trunc-NFW\end{tabular} & \begin{tabular}{c}$0.70 \pm 0.24$\\ $0.82 \pm 0.36$\end{tabular} & \begin{tabular}{c}42.28\\ 23.90\end{tabular} & \begin{tabular}{c}56.28\\
35.90\end{tabular} & \begin{tabular}{c}74.52\\ 51.53\end{tabular}\\
\cline{3-7}
0.16 & 14.87 & \begin{tabular}{c}trunc-Sérsic\\ trunc-NFW\end{tabular} & \begin{tabular}{c}$1.14 \pm 0.24$\\ $1.30 \pm 0.29$\end{tabular} & \begin{tabular}{c}26.71\\ 10.77\end{tabular} & \begin{tabular}{c}40.71\\
22.77\end{tabular} & \begin{tabular}{c}58.94\\ 38.40\end{tabular}\\
\enddata
\end{deluxetable*}

\begin{deluxetable}{cccccc}
\tabletypesize{\scriptsize}
\tablewidth{0pt} 
\tablecaption{Statistical metric values for each stacked bin of redshift and mass, comparing the NFW and Sérsic models (non-truncated) for both SDSS and Mock samples.\label{tab:model_wout_trunc}}
\tablehead{
 \colhead{$z_{\text{med}}$} & \colhead{$\log{(M_{200\text{m}}/M_{\odot})}_{\text{med}}$} & \colhead{Model} & \colhead{$\chi^2$} & \colhead{AIC} & \colhead{BIC}} 
\startdata
\multicolumn{6}{c}{SDSS sample}\\
\cline{1-6}
0.04 & 14.44 & \begin{tabular}{c}Sérsic\\ NFW\end{tabular} & \begin{tabular}{c}102.06\\ 747.31	\end{tabular} & \begin{tabular}{c}112.06\\
755.31\end{tabular} & \begin{tabular}{c}125.09\\ 765.73\end{tabular}\\
\cline{3-6}
0.09 & 14.85 & \begin{tabular}{c}Sérsic\\ NFW\end{tabular} & \begin{tabular}{c}52.64\\ 304.93\end{tabular} & \begin{tabular}{c}42.82\\
312.93\end{tabular} & \begin{tabular}{c}55.84\\ 323.35\end{tabular}\\
\cline{3-6}
0.07 & 15.20 & \begin{tabular}{c}Sérsic\\ NFW\end{tabular} & \begin{tabular}{c}37.84\\ 98.86\end{tabular} & \begin{tabular}{c}47.84\\
106.86\end{tabular} & \begin{tabular}{c}60.86\\ 117.28\end{tabular}\\
\cline{3-6}
0.09 & 14.65 & \begin{tabular}{c}Sérsic\\ NFW\end{tabular} & \begin{tabular}{c}68.78\\ 184.20\end{tabular} & \begin{tabular}{c}78.78\\
192.20\end{tabular} & \begin{tabular}{c}91.81\\ 202.62\end{tabular}\\
\cline{3-6}
0.15 & 14.87 & \begin{tabular}{c}Sérsic\\ NFW\end{tabular} & \begin{tabular}{c}32.82\\ 438.58\end{tabular} & \begin{tabular}{c}42.82\\
446.58\end{tabular} & \begin{tabular}{c}55.84\\ 457.00\end{tabular}\\
\cline{3-6}
0.14 & 15.15 & \begin{tabular}{c}Sérsic\\ NFW\end{tabular} & \begin{tabular}{c}41.32\\ 58.14\end{tabular} & \begin{tabular}{c}51.32\\
66.14\end{tabular} & \begin{tabular}{c}64.34\\ 76.56\end{tabular}\\
\cline{1-6}
\multicolumn{6}{c}{Mock sample}\\
\cline{1-6}
0.06 & 14.31 & \begin{tabular}{c}Sérsic\\ NFW\end{tabular} & \begin{tabular}{c}11.62\\ 18.33\end{tabular} & \begin{tabular}{c}21.62\\
26.33\end{tabular} & \begin{tabular}{c}34.64\\ 36.75\end{tabular}\\
\cline{3-6}
0.07 & 15.04 & \begin{tabular}{c}Sérsic\\ NFW\end{tabular} & \begin{tabular}{c}19.37\\ 19.43\end{tabular} & \begin{tabular}{c}29.37\\
27.43\end{tabular} & \begin{tabular}{c}42.39\\ 37.85\end{tabular}\\
\cline{3-6}
0.11 & 14.59 & \begin{tabular}{c}Sérsic\\ NFW\end{tabular} & \begin{tabular}{c}65.14\\ 67.41\end{tabular} & \begin{tabular}{c}75.14\\
75.41\end{tabular} & \begin{tabular}{c}88.17\\ 85.83\end{tabular}\\
\cline{3-6}
0.16 & 14.87 & \begin{tabular}{c}Sérsic\\ NFW\end{tabular} & \begin{tabular}{c}23.41\\ 17.62\end{tabular} & \begin{tabular}{c}33.41\\
25.62\end{tabular} & \begin{tabular}{c}46.43\\ 36.04\end{tabular}\\
\enddata
\end{deluxetable}

\subsection{Constraints selection}

In this section, we discuss how global properties applied across galaxy clusters can affect the values of the splashback radii obtained. Among these properties, we focus on the center definition, the magnitude limit, and galaxy colors. From this point forward, all determinations are made for individual systems.

\subsubsection{Center definition} \label{subsubsec:center}

Due to the formation processes of galaxy clusters, which occupy the higher end of the mass function, these systems can exist in various dynamical states, ranging from virialized to highly perturbed systems due to large-scale interactions. As the last structures to collapse in a hierarchical scenario, $\Lambda$CDM cosmological simulations and observations suggest that 30\%-50\% of clusters are in non-relaxed states, showing notable substructures in optical and X-ray observations at $z \sim 0$ \citep[e.g.][]{dressler+88, smith+10}. At higher redshifts, this fraction may increase to up to 80\% as clusters are still actively assembling \citep[e.g.][]{fakhouri+10}. This diversity introduces inconsistencies in defining the cluster center. Ideally, the optical center, defined by the Brightest Cluster Galaxy (BCG), should coincide with the minimum of the gravitational potential well of the cluster, which is also traced by the gas distribution (represented by the X-ray emission). However, real clusters almost always show an offset between these two definitions, which can even serve as a proxy for assessing the dynamical state of the cluster \citep[e.g.][]{mann+12, zenteno+20, astudillo+24}, with larger offsets indicating highly perturbed clusters.

We examine whether different cluster center definitions impact the distribution of splashback radii in both SDSS and Mock samples. For our observational clusters with an identified X-ray counterpart (52 clusters), we compare the distributions of splashback radii when using the BCG, X-ray peak, and geometric center. The geometric center, defined as the median galaxy position in the cluster, is less commonly employed but is easily accessible and, in an ideally symmetric cluster, it should also coincide with the potential minimum.  We did not fully conduct this analysis on the Mock catalog due to limitations in extracting X-ray properties from the SAM simulations, limiting the sample to optical and geometric centers only.

It is worth noting that we made efforts to exclude highly perturbed clusters from our sample, as described in our data selection. This means that the differences between center definitions are on the order of $\approx 0.3$ Mpc, which does not pose a significant issue for modeling.

\begin{figure} \centering \includegraphics[width=1.0\linewidth]{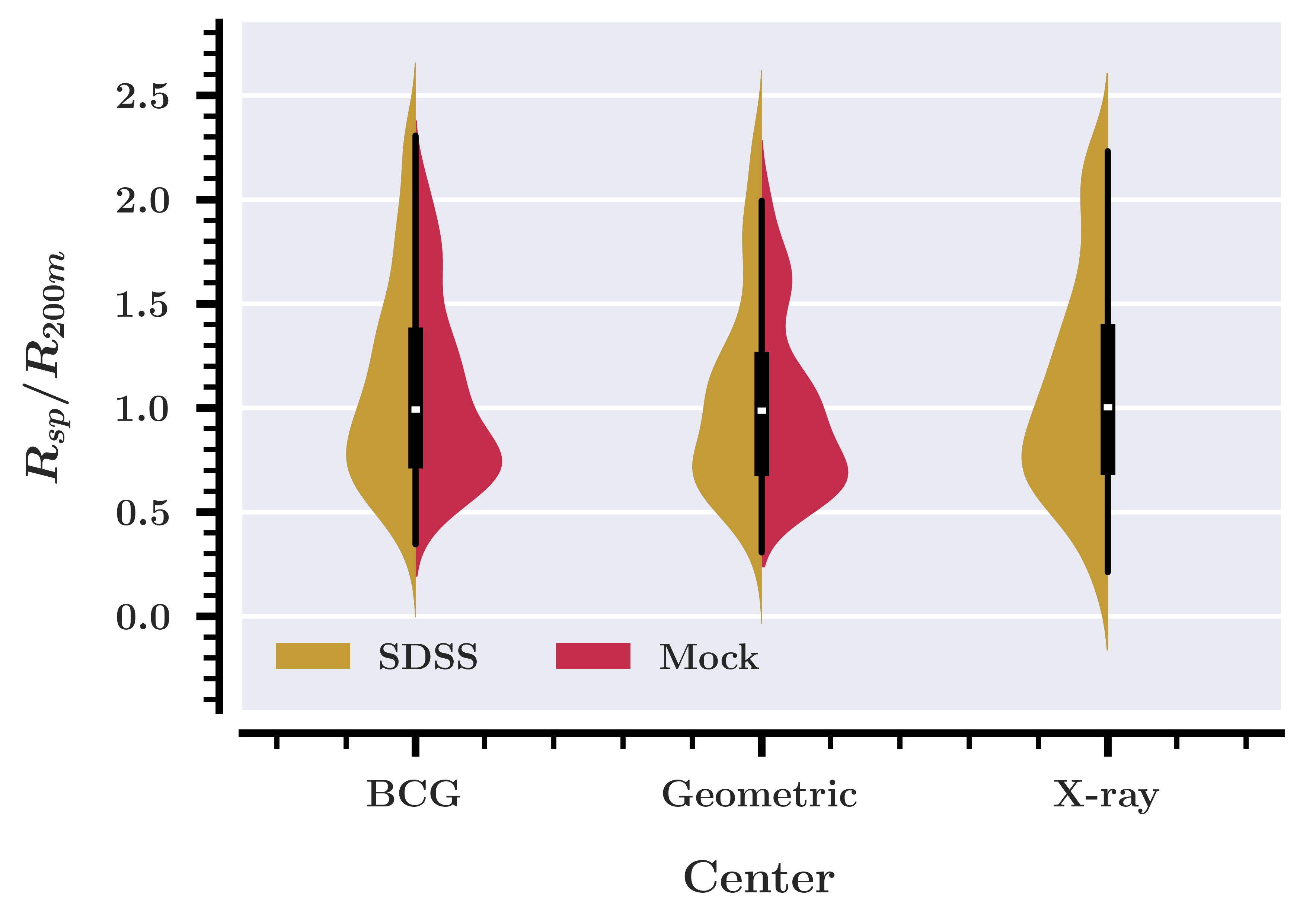} \caption{Normalized distributions of splashback radii for different center definitions (BCG, X-ray peak, and geometric) for the SDSS (yellow) and Mock (pink) clusters. Black box plot represents the distribution for both samples combined.} \label{fig:center_def}
\end{figure}

The results are displayed in Figure \ref{fig:center_def}, which shows the normalized distributions of splashback radii estimated for individual clusters. For completeness, Appendix \ref{app:coma_cluster} includes an example of individual modeling for the Coma Cluster. As illustrated, these distributions remain largely consistent, even across different center definitions for SDSS and Mock samples, shown in yellow and pink, respectively. This consistency supports our expectation that highly perturbed clusters are not present in our sample. The black box plot represents the distribution for both samples combined.

This invariance is significant, given that our model does not account for miscentering, which could otherwise introduce biases. In fact, \citet{more+16} demonstrate that miscentering due to perturbed clusters can overestimate the splashback radius by approximately 20\%. However, miscentering due to systematic errors, such as misidentifying the BCG, generally affects scales smaller than 0.4 Mpc, which is not a concern in our analysis.

Another interesting result is that our distributions are all centered around $R_{\text{sp}}/R_{200m} \approx 1$ for both SDSS and Mock samples, which is generating significant discussions in the literature \citep[e.g.][]{more+16, oneil+21, oneil+22, rana+23, oshea+24}. In fact, given the low accretion rates expected for galaxy clusters, we would expect the splashback radius to be greater than $R_{200m}$, as seen in dark matter simulations \citep{more+15}. However, the majority of studies have not found this trend yet. Several factors have been proposed to explain these discrepancies, one of which is the difficulty in quantifying miscentering in real data. In any case, consistent with the conclusions of \citet{more+16}, this work does not suggest miscentering as a potential cause for this discrepancy.

\subsubsection{Magnitude limit}

Another important factor that can influence the determination of the splashback radius is the magnitude limit of the galaxies within the cluster. This aspect was once identified as a potential cause of discrepancies between dark matter simulations and observations of $R_{\text{sp}}$ \citep[e.g][]{more+16}, largely due to galaxy selection effects and, at a deeper level, the influence of dynamical friction \citep[DF, ][]{chandrasekhar43, vandenbosch+99, adhikari+16}. Dynamical friction is a drag force that acts on objects moving through a background of particles solely because of gravitational attraction, such as a galaxy falling into a galaxy cluster. This force is directly proportional to the mass of the infalling object, potentially decaying the orbits of massive galaxies, which eventually merge with the BCG.

In this context, increasing the magnitude limit for a cluster implies selecting brighter and thus more massive galaxies, which may have experienced significant orbital decay. Since the splashback feature is highly dependent on the apocenter of particle orbits, we would expect smaller values of $R_{\text{sp}}$ as the magnitude limit increases. \citet{chang+18}, for instance, found that high-luminosity galaxies trace smaller splashback radii, although the effect was small compared to their measurement uncertainties.

\begin{figure} \centering \includegraphics[width=1.0\linewidth]{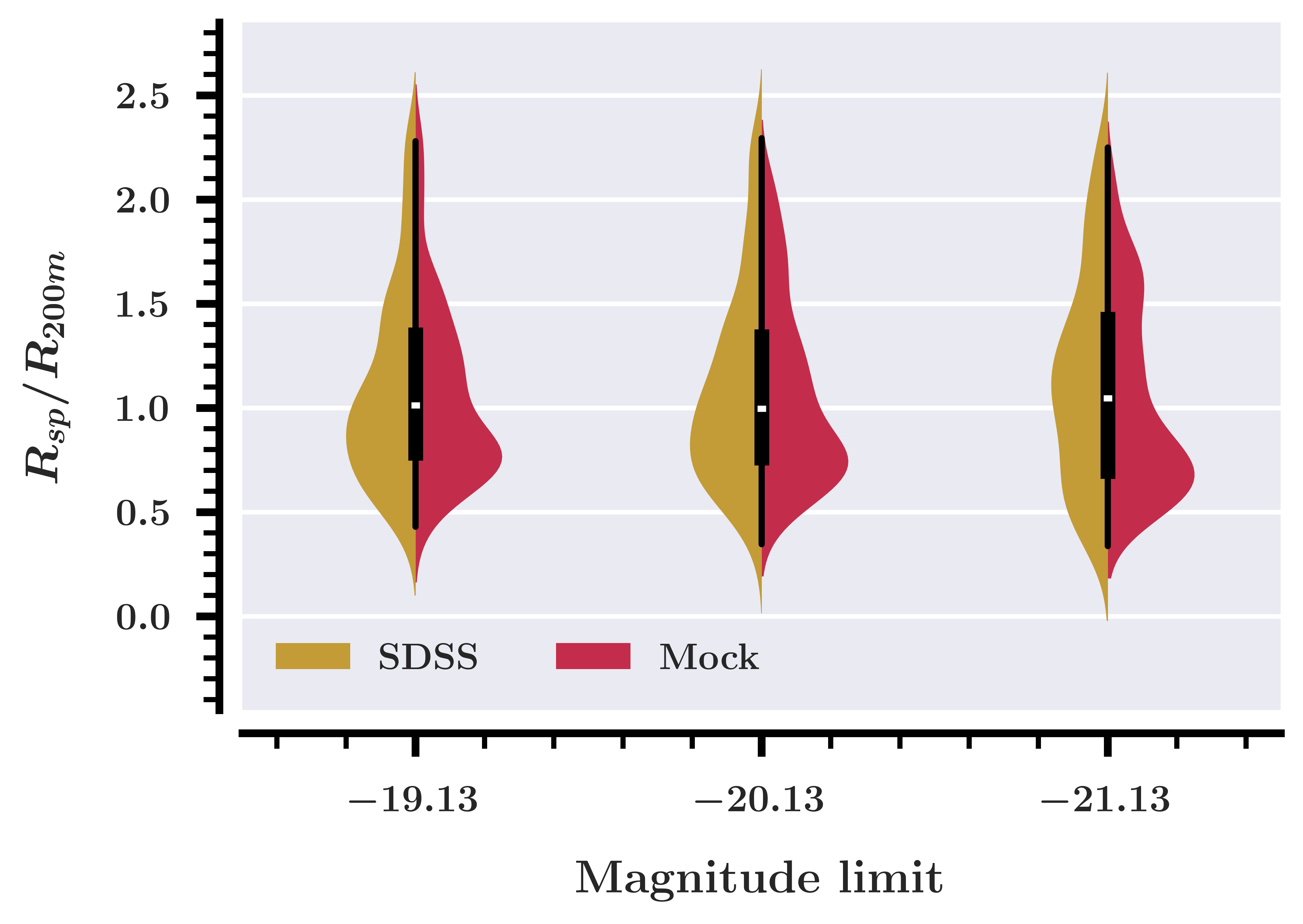} \caption{Normalized distributions of splashback radii for different magnitude limits in the r-band for the SDSS (yellow) and Mock (pink) clusters. Black box plot represents the distribution for both samples combined.} \label{fig:mag_lim}
\end{figure}

However, when we varied the magnitude limits by one order of magnitude around our central value of $-20.13$ in the $r$-band, we observed no significant difference in the distributions of the splashback radius for our cluster samples, as shown in Figure \ref{fig:mag_lim}. Indeed, all distributions of the ratio $R_{\text{sp}}/R_{200m}$ remain centered around $\approx 1$, except for the Mock catalog at the brighter limit of $-21.13$ mag, where we noted a decrease of approximately 9\% in the ratio. Nonetheless, this bright end is strongly affected by selection effects, resulting in a significant loss of galaxies that ultimately distorts our distributions, as also noted for the observational counterpart. In any case, the decreasing trend is not observed for the SDSS sample, although a visual change in the distribution is apparent. This suggests that dynamical friction does not play a significant role in the splashback feature, consistent with most of the literature \citep[e.g.][]{more+16, murata+20, oneil+22, oshea+24}.

Thus, DF cannot explain the discrepancies in the ratio $R_{\text{sp}}/R_{200m}$. Nonetheless, it is also worth noting that our cluster samples are biased toward high masses, with a median value of $M_{200m} \approx 10^{14}\, M_\odot$. In such environments, galaxies would need to be extremely massive to experience appreciable dynamical friction effects, likely more massive than those selected by our magnitude cuts. Thus, we may not be probing the regime where DF becomes relevant.

In any case, this invariance of splashback radii with respect to magnitude is an interesting result on its own, as it suggests that the method can be straightforwardly applied to other galaxy surveys beyond SDSS.

\subsubsection{Galaxy colors}

Galaxy colors, on the other hand, can reveal different aspects of cluster dynamics. As mentioned earlier, \citet{adhikari+21} demonstrated that by relating the splashback radius values for different galaxy populations with their average time of infall, we can use the splashback radius as a ``clock" to understand galaxy quenching.

In general, red galaxies are those that have resided in the cluster for the longest time, approximately 3.3 Gyr, while blue galaxies represent the youngest, most recently infalling population, which, on average, has not yet reached its apocenter. Since these blue galaxies have not reached their maximum clustercentric distance, using them to estimate the splashback radius would result in a smaller value than expected when considering the entire cluster. Indeed, as shown by \citet{oneil+24}, who found similar results studying the splashback radii for galaxy populations with different times of residence (i.e., the time since the galaxy entered the cluster), it is actually a mistake to call the apparent apocenter of these younger populations the splashback radius. In fact, the splashback feature has not yet formed for these galaxies.

\begin{figure} \centering \includegraphics[width=1.0\linewidth]{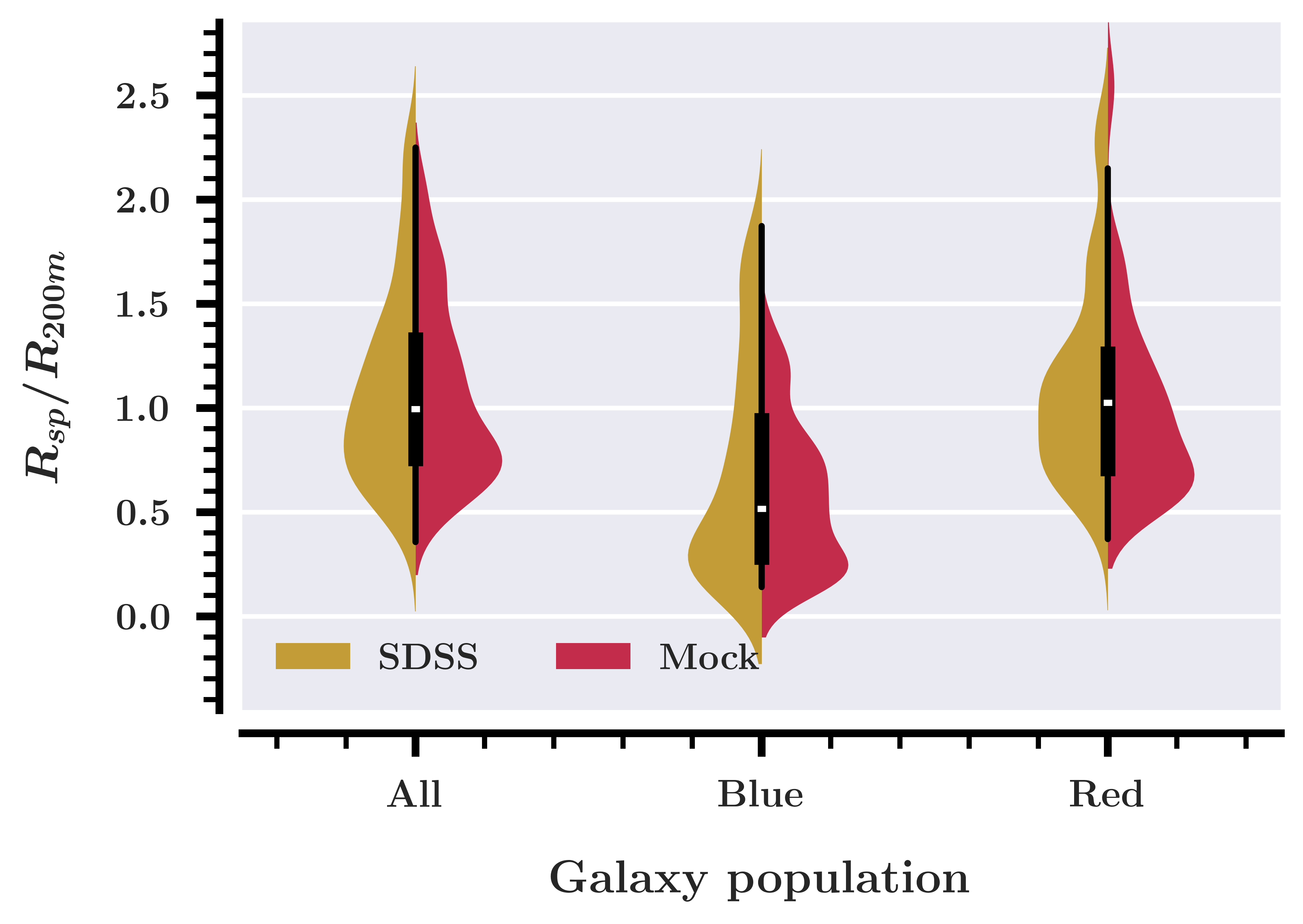} \caption{Normalized distributions of splashback radii for different galaxy populations, using a (u - r) color cut to separate blue and red galaxies in the SDSS (yellow) and Mock (pink) clusters. Black box plot represents the distribution for both samples combined.} \label{fig:colors}
\end{figure}

These expectations are exactly what we observe in Figure \ref{fig:colors}, where we divide our SDSS and Mock clusters into three components: all galaxies, blue galaxies only, and red galaxies only. The selection of galaxies was based on a color cut following the findings of \citet{strateva+01}, who showed that (u - r) = 2.22 effectively separates early (E, S0, and Sa) from late (Sb, Sc, and Irr) morphological types. As expected, the blue population shows a smaller average splashback radius than the red component, with $R_{\text{sp}}/R_{200m}$ ratios around 0.5 for both SDSS and Mock samples. In contrast, the red population, whose distribution aligns closely with the splashback radii for all galaxies within the clusters, shows an $R_{\text{sp}}/R_{200m}$ ratio around 1. This trend is interesting to observe in the context of cluster population behavior, but it may not be recommended as a direct indicator of infall time for a single-cluster measurement.

Additionally, it is important to note that these color cuts may be redshift-dependent. However, since all our clusters are in the local universe $(z < 0.2)$, this effect is unlikely to significantly alter our results.

\subsubsection{Field contamination}

Finally, another important aspect to consider in our modeling and splashback radius determination is the effect of field contamination. A simple cut in recession velocity space, defined as an interval of $\pm \sigma_v$ from the cluster's central velocity, does not guarantee that we are selecting only cluster members. If we assume NFW-like density profiles for galaxy clusters, we would expect a caustic transition between the cluster galaxies and the field, represented by the foreground and background \citep{diaferio99}. Although our modeling accounts for the field contribution via the two-halo term in Equation \ref{eq:2-halo}, it is important to understand the potential effect of missed and misselected galaxies in our determinations.

\begin{figure} \centering \includegraphics[width=1.0\linewidth]{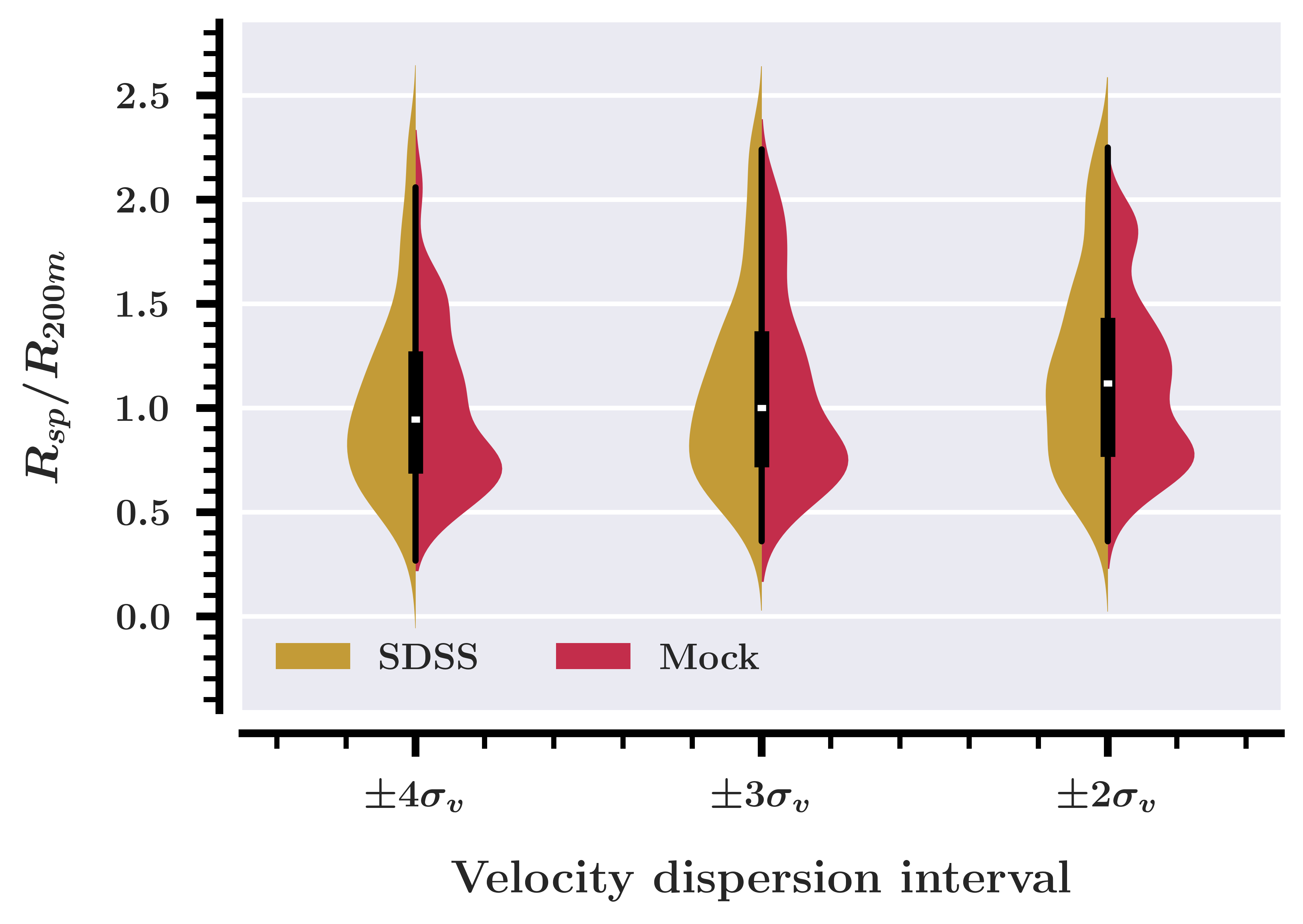} \caption{Normalized distributions of splashback radii for different velocity dispersion intervals in the SDSS (yellow) and Mock (pink) clusters. Black box plot represents the distribution for both samples combined.} \label{fig:vel_disp}
\end{figure}

To address this issue, we varied the recession velocity intervals in terms of the cluster's velocity dispersion by applying cuts within $\pm 4$, $\pm 3$, and $\pm 2$ $\sigma_v$. The results are shown in Figure \ref{fig:vel_disp}, with the same conditions as before. Here, unlike the previous cases, we observe a small trend, with smaller intervals yielding greater $R_{\text{sp}}/R_{200m}$ ratios for both SDSS and Mock samples. In fact, the differences in the central values of each distribution are on the order of 5\%, which is not statistically significant given the sparse distributions we are working with. Nonetheless, this could still represent a real effect and be responsible for the differences observed between dark matter-only simulations and observations.

Working with spectroscopic redshifts, these differences do not pose a significant problem for our modeling, as we could select a fixed interval and calibrate all our results accordingly, as done so far with the $\pm 3\sigma_v$ selection. However, this could become a potential issue when working with photometric redshifts, where the typical uncertainties in redshift determination are one order of magnitude greater than the cluster's velocity dispersion \citep[e.g.][]{beck+16, lima+22}. This is a challenge we aim to address in an upcoming paper.

\subsection{Disturbed clusters}

As discussed in Section \ref{subsubsec:center}, galaxy clusters exhibit diverse dynamical states due to their nature, ranging
from virialized to highly perturbed systems. Keeping this in mind, we employ literature-based proxies to assess the cluster dynamical state using observational features, aiming to compare the splashback radius distributions of disturbed and relaxed clusters. This analysis is of particular interest because a cluster's dynamical state can reflect whether it is a fast-accreting system with a high mass accretion rate or a slow-accreting one with a lower rate, as suggested by \citet{towler+24}. This is especially relevant when cluster disturbances arise from recent mergers, which inject significant mass into the system, momentarily increasing the mass accretion rate and observationally perturbing the cluster. Moreover, as previously noted, there is a well-established anti-correlation between the $R_{\text{sp}}/R_{200m}$ ratio and the mass accretion rate \citep[e.g.][]{diemer+14, more+15}, where higher ratios indicate lower accretion rates, and lower ratios indicate higher rates.

For this analysis, we classify a cluster as disturbed if it satisfies both criteria from \citet{lopes+18}, defined in terms of substructure fraction:

\begin{itemize}
    \item \textbf{Magnitude gap:} The magnitude gap between the brightest and the fourth brightest galaxy within $R_{200c}$ ($\Delta m_{1,4}$):
    \begin{equation}
        \Delta m_{1,4} < 1
    \end{equation}

    \item \textbf{Center offset:} The offset between the optical center and the X-ray centroid peak ($D_{X-BCG}$):
    \begin{equation}
        D_{X-BCG} > 0.02 R_{200c}
    \end{equation}
\end{itemize}

It is important to note that our sample does not include highly perturbed clusters, so the transition between relaxed and disturbed systems based on the above criteria is relatively smooth. Furthermore, this approach serves as a simple method to infer cluster dynamical states, but a complete characterization requires more detailed and complex analyses, encompassing an entire field of study \citep[e.g.][]{sanderson+09, smith+10, dariush+10}. As these criteria rely on X-ray data, this analysis is limited to the SDSS sample, resulting in 31 relaxed clusters and 29 disturbed ones.

\begin{figure}
\centering
\includegraphics[width=1.0\linewidth]{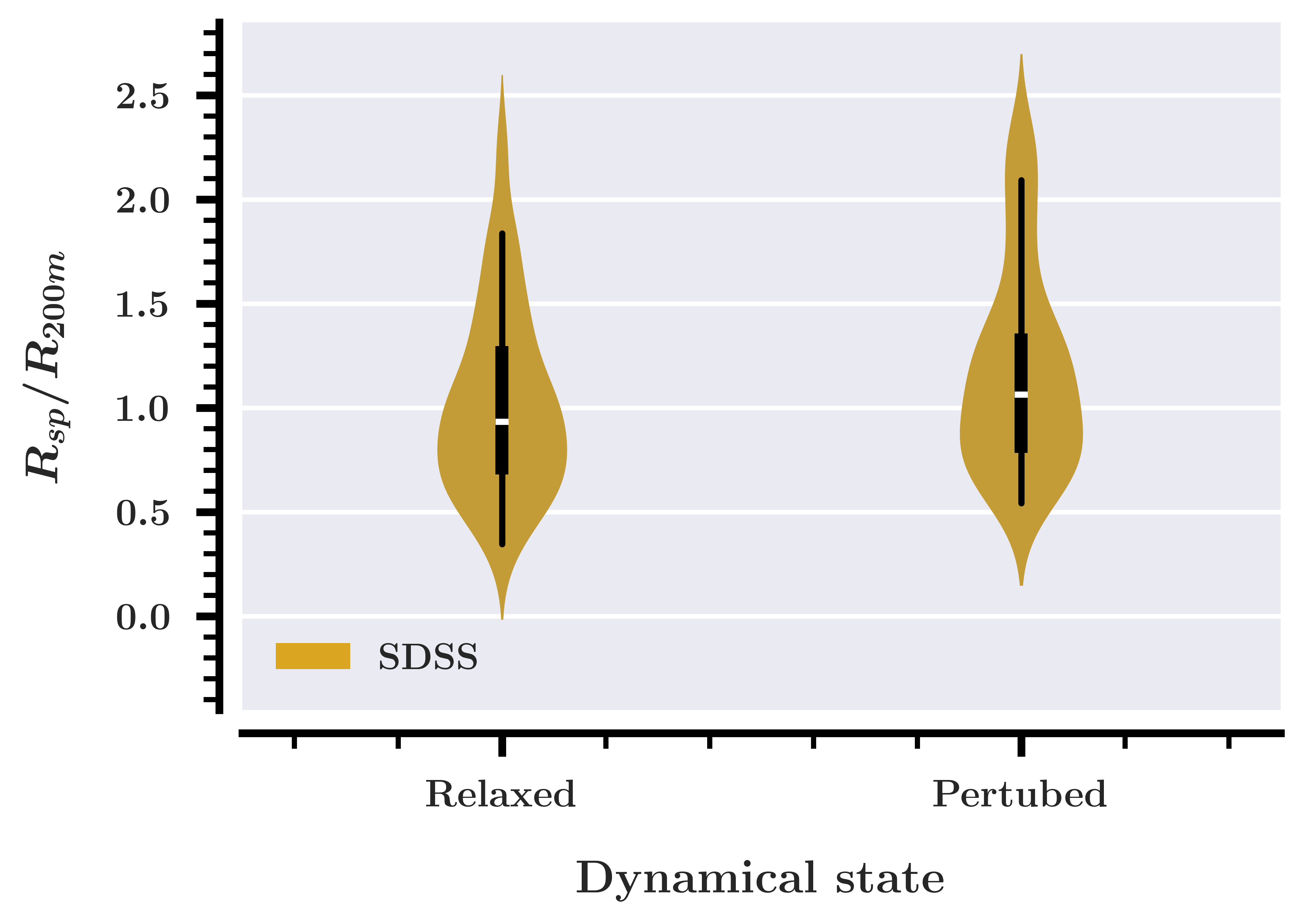}
\caption{Normalized distributions of splashback radii for perturbed and relaxed clusters in the SDSS sample.}
\label{fig:dynamical_state}
\end{figure}

The results are shown in Figure \ref{fig:dynamical_state}, where, although the difference is not substantial, there is a gap between the two distributions. Quantitatively, relaxed clusters exhibit a median $R_{\text{sp}}/R_{200m}$ ratio of $0.93 \pm 0.08$, while disturbed clusters show a ratio of $1.06 \pm 0.09$. This indicates that relaxed clusters are more concentrated, in terms of splashback features, than disturbed ones. Nevertheless, the values are consistent within $1\sigma$. Interestingly, this contrasts with the findings of \citet{towler+24}, where relaxed clusters were reported to have higher $R_{\text{sp}}/R_{200m}$ ratios. Their results align with the expectation that relaxed clusters experience lower mass accretion rates, as predicted by \citet{diemer+14} and \citet{more+15}. However, as we rely on galaxy positions as proxies for splashback radii, a bias between the actual mass profile and the galaxy distribution may play a significant role. Intuitively, disturbed clusters might exhibit a more spread-out galaxy distribution. Additionally, our relaxed clusters might have had
a faster accretion rate in the past. In any case, further statistical analyses are required to better understand this discrepancy.

Furthermore, we would expect that perturbed clusters would have larger dispersions in the individual values of $R_{\text{sp}}$ (i.e., $\sigma_{R_{\text{sp}}}$), as the steepening drop in these systems is supposedly less prominent. Nonetheless, in our data, we do not observe any significant difference in the splashback uncertainties between the two groups, with typical values around $0.2$ Mpc, or approximately $10\%$ in normalized units.

\subsection{Relation $M_{\text{sp}} \textendash R_{\text{sp}}$}

\begin{figure*}
\centering \includegraphics[width=0.85\linewidth]{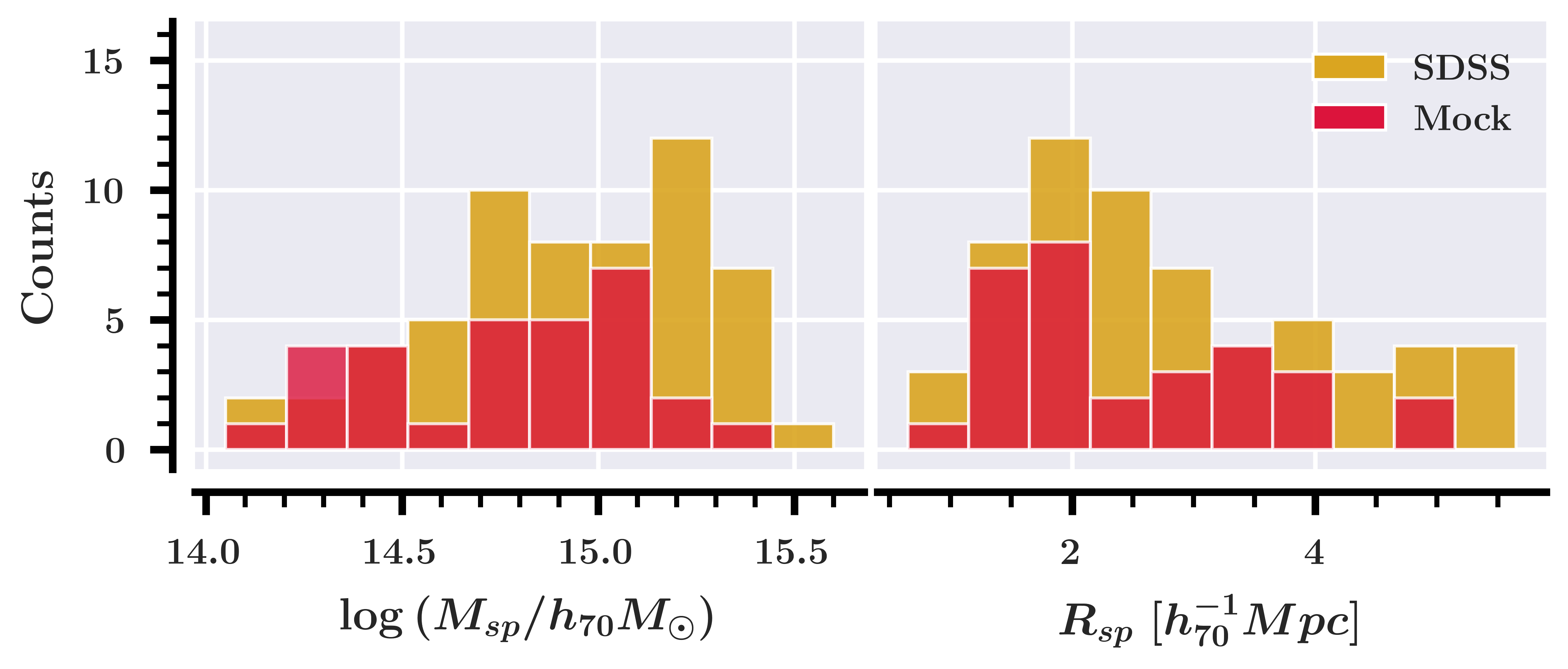}
\caption{Distributions of splashback masses (left panel) and radii (right panel) for our samples. SDSS clusters are shown in yellow, while Mock clusters are in pink.} \label{fig:sp_dist}
\end{figure*}

\begin{figure}
\centering \includegraphics[width=0.85\linewidth]{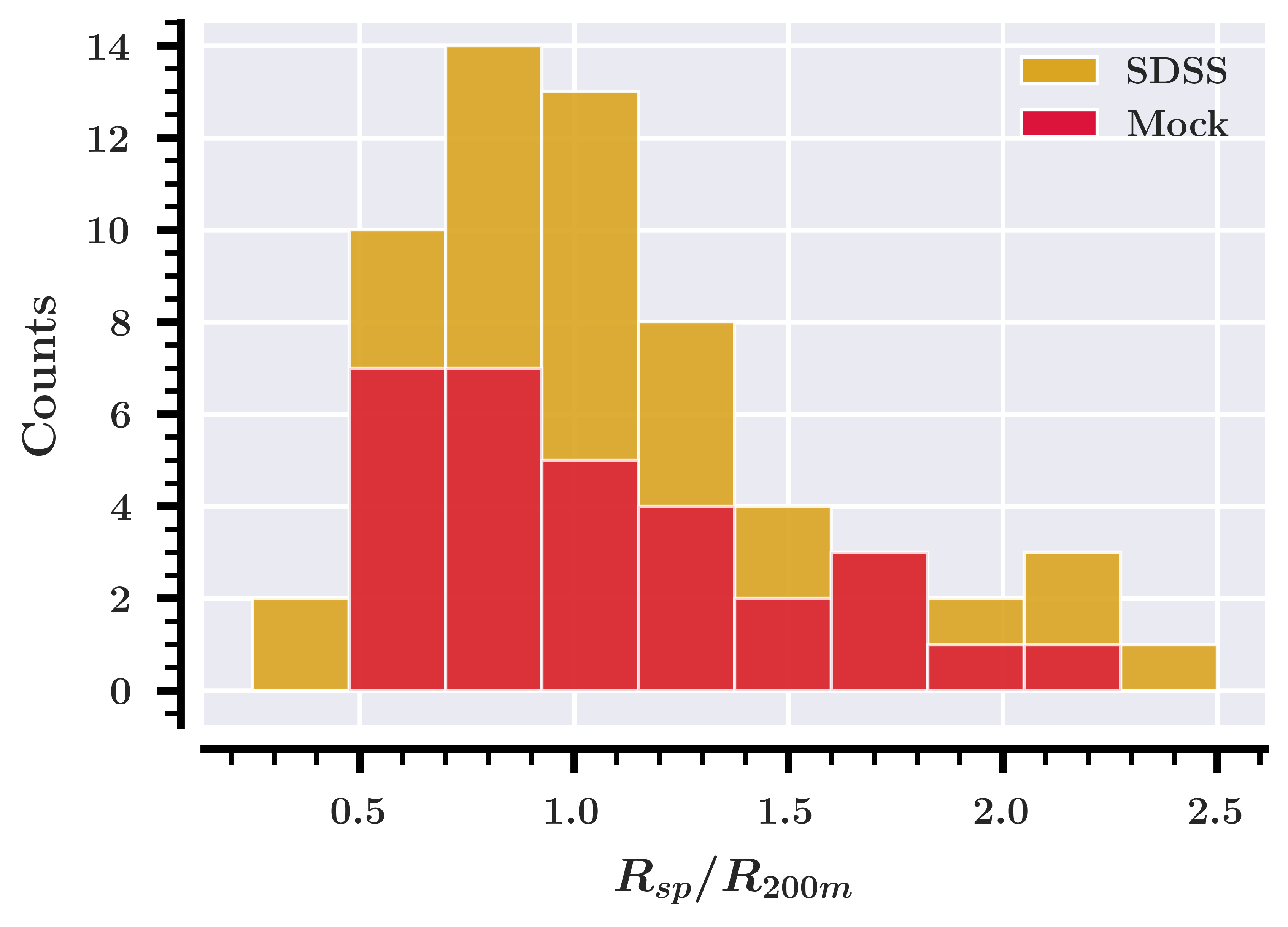}
\caption{Distribution of the ratio $R_{sp}/R_{200m}$ for SDSS (yellow) and Mock (pink) clusters.} \label{fig:delta_sp}
\end{figure}

By modeling each cluster with the trunc-NFW model, we estimate the splashback radii for all objects in the SDSS and Mock samples (see an example of individual modeling in Appendix \ref{app:coma_cluster}). Additionally, using the posterior parameters from the MCMC sampling, we estimate the splashback mass by applying Equation \ref{eq:M_r} and \ref{eq:M_sp}. All uncertainties were considered in the determination of $M_{\text{sp}}$, although the error in $M_{200c}$ dominates, as described in Section \ref{subsec:Msp}. The distributions of the splashback mass and radius for both samples are shown in Figure \ref{fig:sp_dist}, where SDSS clusters are represented in yellow and Mock clusters in pink. Interestingly, despite the smaller number of mock clusters, their distribution trends appear to match those of the SDSS clusters. Figure \ref{fig:delta_sp} shows in more detail the distribution of the ratio $R_{\text{sp}}/R_{200m}$, using the same color coding as above. The values clearly cluster around unity for both samples, in agreement with previous findings in the literature.

The values found for each cluster in the SDSS sample are listed in Table \ref{tab:sp}, where we present the cluster positions relative to the optical center, as well as the splashback masses and radii. The median splashback radius for the SDSS sample is approximately $2.55 \pm 0.18$ Mpc, which is consistent with CDM-only simulations \citep{adhikari+21}, as expected. However, it is important to note that our SDSS sample is biased toward high-mass clusters, with a mean $<M_{200c}> \approx 5 \times 10^{14} M_\odot$. In any sense, the Mock sample similarly yields a median splashback radius of $<R_{\text{sp}}> \approx 2.10 \pm 0.23$ Mpc.

One of the main goals of this work is to establish a scaling relation between $R_{\text{sp}}$ and $M_{\text{sp}}$, given that the splashback radius is easily observable. To achieve this, and based on expectations from self-similar spherical collapse models \citep{gunn+72}, we assume a power-law relation between these quantities, incorporating a redshift evolution term:

\begin{equation}\label{eq:mass_fit}
\left(\frac{M_\text{sp}}{10^{14}M_\odot}\right) = A \left(\frac{R_\text{sp}}{\text{Mpc}}\right)^{B}(1 + z)^{C},
\end{equation}

where $A$ is a proportionality constant, and $B$ and $C$ represent the slopes for the splashback radius and redshift, respectively.

This relation can be expressed in linear form by taking the logarithm of both sides:

\begin{equation}
\log{\left(\frac{M_\text{sp}}{10^{14}M_\odot}\right)} = \log{A} + B\log{\left(\frac{R_\text{sp}}{\text{Mpc}}\right)} + C\log{(1 + z)}.
\end{equation}

In a typical scenario, where clusters have a similar density around $R_\text{sp}$, as described in \citet{tully15}, we would expect a constant “splashback density”:

\begin{equation}
\frac{M_\text{sp}}{R_\text{sp}^3} = \text{constant}.
\end{equation}

\begin{figure}
\centering \includegraphics[width=1.0\linewidth]{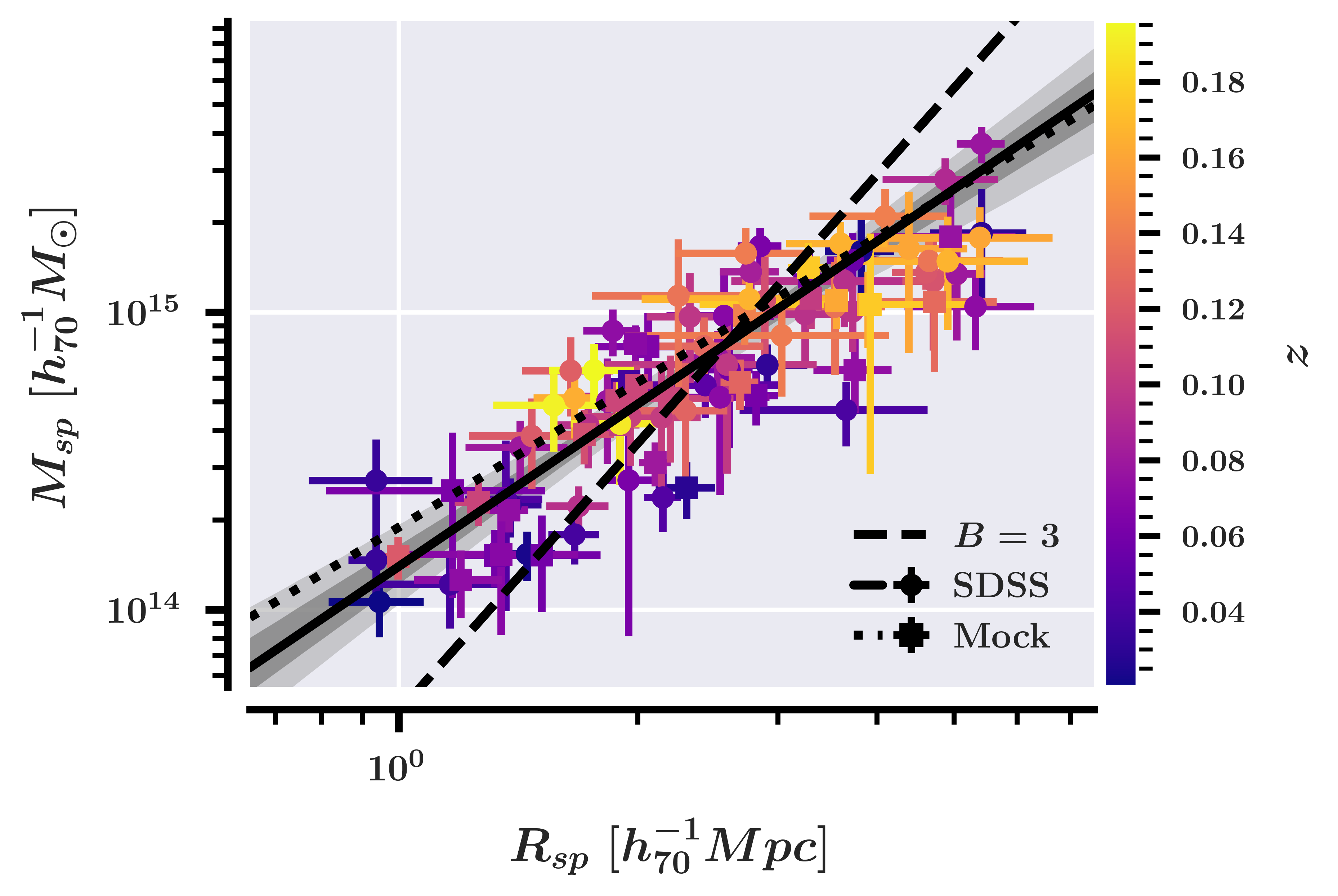} \caption{Relation between $M_\text{sp}$ and $R_\text{sp}$, color-coded by redshift. The black dashed line represents the fit assuming $B = 3$, while the solid black line and the dotted line show the best-fit relations without parameter assumptions for the SDSS and Mock samples, respectively. The gray shaded regions indicate the 68\% and 95\% confidence intervals for the SDSS fit. Circle markers correspond to the SDSS clusters, and square markers correspond to the Mock clusters.} \label{fig:mass_fit}
\end{figure}

Given this constant density, we would expect the B slope in Equation \ref{eq:mass_fit} to be around 3. However, fitting the model to the SDSS data points using an orthogonal distance approach \citep{boggs+90}, which considers both errors in $M_\text{sp}$ and $R_\text{sp}$, we found that $A = 1.48 \pm 0.21$, $B = 1.77 \pm 0.12$, and $C = 2.31 \pm 1.41$. While these results suggest a strong correlation between mass and splashback radius, with a dispersion of about 0.15 dex, they deviate from the constant density scenario. Similar results were obtained for the mock data, with $A = 1.56 \pm 0.33$, $B = 1.62 \pm 0.18$, and $C = 2.10 \pm 2.06$. Despite the smaller sample, these values are consistent within $1\sigma$ with those found for the SDSS sample, supporting the robustness of the scaling relation. Imposing $B = 3$ leads to a strong deviation from the trend in the data points, as shown in Figure \ref{fig:mass_fit}, where we plot $M_{\text{sp}}$ against $R_{\text{sp}}$, color-coded by redshift. The black dashed line represents the fit assuming $B = 3$, while the solid black line and the dotted line show the best-fit relations without parameter assumptions for the SDSS and Mock samples, respectively. The gray shaded regions indicate the 68\% and 95\% confidence intervals for the SDSS fit. Circle markers correspond to the SDSS clusters, and square markers correspond to the Mock clusters. SDSS and Mock fittings were performed separately.

Although the slope of the splashback radius relation is significantly lower than expected in the constant density case, this is not necessarily counterintuitive. Relations for overdensity-defined quantities, such as $R_\Delta \textendash M_\Delta$, typically assume a specific enclosed density. However, the splashback radius represents a physical boundary tied to the cluster's overall gravitational potential rather than a defined overdensity, so the slope of this relation may naturally differ from 3. The splashback radius likely reflects a more complex mass distribution than overdensity-based radii and strongly depends on the cluster's mass accretion rate, which may lead to a lower slope. Additionally, since our splashback radius measurements are, intrinsically, based on 2D projected quantities, projection effects may be present, potentially explaining why the slope approaches 2.

Another important aspect is the strong redshift evolution suggested by our findings, with $C = 2.31 \pm 1.41$, indicating that clusters of the same mass were smaller in the past. This is consistent with hierarchical structure formation in a $\Lambda$CDM universe, where smaller structures merge and grow over cosmic time. However, our sample is limited to low redshifts, making it difficult to fit redshift evolution robustly. Setting $C = 0$ does not alter the other parameters significantly ($A = 1.66 \pm 0.20$ and $B = 1.84 \pm 0.12$, for the SDSS sample), and the dispersion remains around 0.2 dex, suggesting that our sample may not be ideal for studying evolutionary trends. Further investigations with higher redshift samples are needed.

Nonetheless, some redshift evolution is expected given that typical crossing times in galaxy clusters are on the order of 2 Gyr \citep[e.g.][]{boselli+06}. This timescale is crucial for understanding how the splashback radius evolves, as it traces the apocenters of infalling particle orbits reaching maximum clustercentric distance on timescales shorter than the Hubble time ($\approx$ 14 Gyr in $\Lambda$CDM). As substructures continue to fall into clusters, the splashback radius expands, reflecting mass accumulation at the cluster’s outskirts as infalling galaxies complete their orbits and the cluster approaches virialization. Therefore, the evolution of the splashback radius with redshift reflects both structure growth and the cluster’s dynamic state over cosmic history.

Finally, we believe that the scaling relation we established is competitive for mass estimation, as our typical dispersion is around $0.15$ dex. This is comparable to the uncertainties found in mass–richness \citep[e.g.][]{gonzalez+16, simet+16, murata+18} and mass–luminosity relations \citep[e.g.][]{popesso+05, popesso+07, mulroy+17}. These findings could be extended to estimate cluster masses and sizes in large galaxy surveys. As the splashback mass represents a physically motivated cluster mass, its applications could reach far beyond. In future work, we intend to reproduce a cluster mass function using the $M_{\text{sp}}$ values derived from the scaling relations found here.

\begin{deluxetable*}{cccccc}
\tabletypesize{\scriptsize}
\tablewidth{0pt} 
\tablecaption{Measurements splashback features in SDSS clusters. \label{tab:sp}}
\tablehead{
 \colhead{Cluster} & \colhead{R.A.$^{\text{BCG}}$} & \colhead{Dec.$^{\text{BCG}}$} &  \colhead{$z$} & \colhead{$R_{\text{sp}}$} & \colhead{$M_{\text{sp}}$}\\
 \colhead{} & \colhead{(deg)} & \colhead{(deg)} &  \colhead{} & \colhead{($h_{70}^{-1}$Mpc)} & \colhead{($10^{14}$ $h_{70}^{-1} M_{\odot}$)} }
\startdata
A1033 & 157.935 & 35.041 & 0.123 & $1.64 \pm 0.22$ & $7.59 \pm 2.12$\\
A1066 & 159.778 & 5.21 & 0.069 & $2.57 \pm 0.4$ & $11.67 \pm 4.03$\\
A1068 & 160.142 & 40.064 & 0.139 & $3.03 \pm 1.11$ & $10.22 \pm 3.18$\\
A1132 & 164.599 & 56.795 & 0.135 & $4.65 \pm 1.12$ & $16.36 \pm 2.41$\\
A1139 & 164.715 & 1.651 & 0.04 & $1.66 \pm 0.12$ & $2.57 \pm 0.44$\\
A1234 & 170.625 & 21.406 & 0.164 & $1.66 \pm 0.19$ & $6.51 \pm 1.6$\\
A1246 & 171.016 & 21.491 & 0.193 & $1.57 \pm 0.25$ & $5.96 \pm 1.62$\\
A1314 & 173.497 & 49.062 & 0.033 & $1.38 \pm 0.13$ & $3.08 \pm 0.45$\\
A1413 & 178.743 & 23.422 & 0.141 & $4.09 \pm 0.81$ & $26.01 \pm 4.98$\\
A1437 & 180.216 & 3.316 & 0.134 & $3.54 \pm 0.55$ & $14.14 \pm 4.81$\\
A1456 & 180.953 & 4.345 & 0.135 & $2.25 \pm 0.5$ & $13.88 \pm 6.58$\\
A1553 & 187.704 & 10.546 & 0.166 & $4.91 \pm 1.28$ & $17.29 \pm 6.32$\\
A1650 & 194.659 & -1.575 & 0.084 & $2.77 \pm 0.24$ & $17.44 \pm 2.92$\\
A1691 & 197.824 & 39.288 & 0.072 & $5.32 \pm 0.99$ & $13.19 \pm 2.97$\\
A1767 & 203.986 & 59.233 & 0.071 & $2.54 \pm 0.39$ & $6.35 \pm 2.92$\\
A1773 & 205.623 & 2.201 & 0.077 & $2.52 \pm 0.29$ & $9.42 \pm 1.72$\\
A1781 & 206.219 & 29.771 & 0.062 & $1.95 \pm 0.14$ & $3.59 \pm 2.47$\\
A1795 & 207.422 & 26.717 & 0.063 & $2.85 \pm 0.18$ & $21.12 \pm 3.01$\\
A1809 & 208.277 & 5.15 & 0.08 & $1.42 \pm 0.21$ & $4.44 \pm 0.65$\\
A1914 & 216.486 & 37.816 & 0.167 & $2.76 \pm 0.74$ & $12.34 \pm 1.93$\\
A1927 & 217.644 & 25.647 & 0.095 & $3.25 \pm 0.39$ & $12.44 \pm 3.62$\\
A1983 & 223.18 & 16.904 & 0.045 & $2.15 \pm 0.12$ & $3.33 \pm 0.7$\\
A1991 & 223.7 & 18.564 & 0.058 & $2.61 \pm 0.18$ & $8.25 \pm 3.7$\\
A2029 & 227.75 & 5.783 & 0.078 & $5.41 \pm 0.37$ & $44.48 \pm 5.91$\\
A2033 & 227.932 & 6.179 & 0.08 & $5.03 \pm 0.43$ & $16.33 \pm 6.49$\\
A2034 & 227.584 & 33.486 & 0.113 & $2.89 \pm 0.27$ & $14.47 \pm 5.43$\\
A2048 & 228.809 & 4.386 & 0.098 & $2.32 \pm 0.27$ & $12.14 \pm 4.38$\\
A2050 & 229.075 & 0.089 & 0.12 & $1.47 \pm 0.24$ & $4.61 \pm 1.45$\\
A2051 & 229.184 & -0.969 & 0.118 & $4.61 \pm 0.43$ & $17.92 \pm 1.47$\\
A2052 & 229.185 & 7.022 & 0.034 & $0.94 \pm 0.07$ & $1.95 \pm 0.31$\\
A2055 & 229.669 & 6.238 & 0.102 & $2.14 \pm 0.26$ & $5.6 \pm 2.48$\\
A2063 & 230.772 & 8.609 & 0.034 & $0.94 \pm 0.17$ & $3.53 \pm 0.54$\\
A2065 & 230.6 & 27.714 & 0.072 & $1.86 \pm 0.15$ & $11.24 \pm 1.74$\\
A2069 & 231.1 & 30.006 & 0.116 & $4.7 \pm 0.4$ & $15.53 \pm 6.38$\\
A21 & 5.012 & 28.75 & 0.094 & $3.64 \pm 1.01$ & $15.95 \pm 3.78$\\
A2107 & 234.845 & 21.739 & 0.042 & $3.65 \pm 0.97$ & $5.65 \pm 1.02$\\
A2142 & 239.583 & 27.233 & 0.09 & $4.87 \pm 0.81$ & $33.61 \pm 5.0$\\
A2175 & 245.13 & 29.891 & 0.096 & $3.72 \pm 0.32$ & $13.6 \pm 3.12$\\
A2199 & 247.159 & 39.551 & 0.03 & $5.41 \pm 0.75$ & $23.5 \pm 8.09$\\
A2244 & 255.628 & 34.042 & 0.099 & $2.59 \pm 0.77$ & $8.48 \pm 3.95$\\
A2259 & 260.045 & 27.704 & 0.16 & $5.38 \pm 1.27$ & $21.66 \pm 4.76$\\
A2627 & 354.212 & 23.924 & 0.125 & $2.3 \pm 0.3$ & $5.93 \pm 2.59$\\
A2634 & 355.118 & 27.109 & 0.032 & $2.91 \pm 0.36$ & $8.82 \pm 1.14$\\
A2700 & 0.89 & 2.138 & 0.094 & $1.68 \pm 0.15$ & $3.07 \pm 0.35$\\
A291 & 30.43 & -2.197 & 0.196 & $1.76 \pm 0.22$ & $8.22 \pm 1.59$\\
A383 & 42.014 & -3.529 & 0.189 & $1.9 \pm 0.25$ & $5.17 \pm 1.9$\\
A671 & 127.132 & 30.431 & 0.05 & $2.43 \pm 0.6$ & $6.84 \pm 1.04$\\
A7 & 2.939 & 32.416 & 0.104 & $1.96 \pm 0.29$ & $5.25 \pm 1.61$\\
A779 & 140.151 & 33.651 & 0.023 & $1.45 \pm 0.12$ & $2.18 \pm 0.31$\\
A795 & 141.064 & 14.128 & 0.139 & $2.73 \pm 0.5$ & $19.81 \pm 3.42$\\
A85 & 10.459 & -9.43 & 0.055 & $3.72 \pm 0.27$ & $18.66 \pm 4.26$\\
A954 & 153.496 & -0.082 & 0.094 & $1.73 \pm 0.16$ & $5.55 \pm 1.32$\\
A961 & 154.095 & 33.638 & 0.127 & $2.42 \pm 0.45$ & $8.85 \pm 2.07$\\
Coma-Cluster & 195.034 & 27.977 & 0.024 & $3.82 \pm 0.38$ & $18.72 \pm 5.03$\\
MKW-4 & 181.113 & 1.896 & 0.02 & $0.95 \pm 0.13$ & $1.4 \pm 0.19$\\
MKW-9 & 233.133 & 4.681 & 0.039 & $1.16 \pm 0.23$ & $1.52 \pm 0.35$\\
MKW3S & 230.466 & 7.709 & 0.045 & $1.36 \pm 0.15$ & $2.8 \pm 1.6$\\
MS0906 & 137.303 & 10.975 & 0.166 & $3.6 \pm 0.53$ & $21.32 \pm 3.14$\\
RX-J1720.1+2638 & 260.042 & 26.626 & 0.161 & $4.38 \pm 0.81$ & $21.22 \pm 9.58$\\
ZWCL1215 & 184.421 & 3.656 & 0.077 & $1.83 \pm 0.18$ & $6.22 \pm 2.35$\\
\enddata
\tablecomments{Cluster positions given in terms of optical center.}
\end{deluxetable*}

\section{Conclusions}\label{sec:conclusion}

Galaxy cluster properties are important keys to understanding and discriminating between models of structure formation in our universe. While typical values of mass and radius for these systems, defined in terms of spherical overdensities, may undergo pseudo-evolution \citep{diemer+13}, which biases our understanding of growth, the splashback feature offers a solution for defining the physical size and mass of clusters. Unlike traditional definitions, the splashback radius does not suffer from pseudo-evolution and represents the physical boundary of dark matter halos, separating virialized from infalling components.

In this work, we model the cumulative number profile of individual galaxy clusters from both observations and mock catalogs to estimate the splashback radius, identifying it as the location of the minimum in the logarithmic slope of the surface density. We also indirectly estimate the splashback mass by applying analytical models to weak-lensing $M_{200c}$ measurements. Using SDSS spectroscopic data ensures reliable cluster membership determination, limited by magnitude and maximum clustercentric distance. We test two different density models, which we call trunc-NFW and trunc-Sérsic, stacking profiles by bins of mass and redshift, and using statistical metrics to assess which model best fits the data.

We vary cluster properties such as center definition, magnitude limits, galaxy colors and recession velocity interval to evaluate their impact on the estimation of the splashback radius. Additionally, we explore the potential to establish a scaling relation between $R_{\text{sp}}$ and $M_{\text{sp}}$, enabling direct mass estimates from splashback radius measurements, an easily observable quantity. Our main findings are:

\begin{itemize}

\item Both the trunc-NFW and trunc-Sérsic models perform well, with good agreement in the determination of the splashback radius within $1\sigma$. However, trunc-Sérsic systematically yields lower $R_{\text{sp}}$ values (Figures \ref{fig:model_select} and \ref{fig:model_select_mock}). Nevertheless, trunc-NFW outperforms trunc-Sérsic in terms of $\chi^2$, AIC, and BIC in nearly all mass and redshift bins in both SDSS and Mock samples (Table \ref{tab:model_select}).

\item Our values of $R_{\text{sp}}$ in terms of the ratio $R_{\text{sp}}/R_{200m}$ assemble around 1 (Figures \ref{fig:center_def}, \ref{fig:mag_lim}, \ref{fig:colors} and \ref{fig:vel_disp}), which is considered counterintuitive according to expectations from dark matter-only simulations \citep[e.g.][]{more+16, oneil+21, oneil+22, rana+23, oshea+24}. The question remains open.

\item Center definition does not play a significant role in determining the splashback radius (Figure \ref{fig:center_def}). This suggests that the choice between optical (BCG), X-ray, or geometric centers is not critical, as systematic miscentering only affects scales smaller than 0.4 Mpc \citep{more+16}, which we exclude from the modeling. However, we do not account for highly perturbed clusters, where the center definition could significantly impact the results or even render determination difficult due to prominent substructures.

\item Varying the magnitude limit also does not affect the distribution of splashback radii, with consistent values regardless of galaxy brightness (Figure \ref{fig:mag_lim}). This robustness is important when applying our approach to other galaxy surveys and has deeper astrophysical implications. Specifically, selecting only bright galaxies could be expected to lower splashback radii due to dynamical friction, which decays orbits of massive galaxies. However, we find no evidence of this effect, reinforcing findings from other studies \citep[e.g.][]{more+16, murata+20, oshea+24}.

\item In contrast, galaxy colors have a significant impact on splashback radius determination. The $R_{\text{sp}}$ distributions for red galaxies are consistent with the overall cluster population, but those for blue galaxies are notably smaller (Figure \ref{fig:colors}). \citet{adhikari+21} proposed that blue galaxies have smaller splashback radii because they are younger infallings and have not yet reached their apocenters. This explanation aligns with our results, though we do not explore this feature in depth.

\item Another important feature is the impact of varying the recession velocity interval for each cluster, specifically in terms of the velocity dispersion. By changing the membership selection interval, and consequently the level of potential field contamination, we observe a slight trend in our measurements. Reducing the velocity interval results in an approximate 5\% increase in the ratio $R_{\text{sp}}/R_{200m}$. However, these differences are not statistically significant given the dispersion in our distributions (Figure \ref{fig:vel_disp}), and since our objective is to calibrate our determinations within a fixed interval, we chose to limit this to within $\pm 3 \sigma_v$ of the cluster's central velocity. Nonetheless, contamination in this type of analysis could contribute to the discrepancies observed between dark matter-only simulations and observations.

\item Relaxed clusters appear to have a smaller $R_{\text{sp}}/R_{200m}$ ratio compared to perturbed ones. Although this difference is consistent within the $1 \sigma$ interval, it contrasts with previous findings in the literature. A potential bias between the actual mass profile and the galaxy distribution within clusters may contribute to this discrepancy. Further statistical analyses are necessary to fully understand this result.

\item Splashback masses and radii show a strong correlation, with both SDSS and Mock samples following the same trend (Figure \ref{fig:mass_fit}). However, the relation we find deviates from the expectation of constant densities around the splashback radius in galaxy clusters, as proposed by \citet{tully15}. Some projection effects could influence the observed relation, although we believe our result is robust. This correlation also shows significant redshift evolution, consistent with a hierarchical growth scenario. Still, the narrow redshift range of our data makes it difficult to confidently confirm this trend, suggesting that further exploration with higher redshift clusters is needed.

\end{itemize}

We find that splashback radii of individual galaxy clusters can be reliably estimated from cumulative number profile modeling, yielding robust results compared to surface density features. By indirectly estimating the splashback mass, we demonstrate that the strong correlation between $R_{\text{sp}}$ and $M_{\text{sp}}$ enables the establishment of a scaling relation in the form of a power law, allowing mass estimates from splashback radii with a dispersion of approximately 0.15 dex. This is comparable to other mass-observable scaling relations, such as mass-richness \citep[e.g.][]{gonzalez+16, simet+16, murata+18} and mass-luminosity \citep[e.g.][]{popesso+05, popesso+07, mulroy+17}, indicating that our approach is competitive.

In future applications, we aim to extend this method to other galaxy surveys, enabling the reproduction of a galaxy cluster mass function using splashback mass as the key observable. This will test our determinations and provide insights into cosmology by fitting analytical approximations to these distributions.

Our overarching goal is to calibrate these scaling relations using splashback measurements derived solely from photometric data (photo-zs), enabling cluster mass and size estimates from low-cost, deep photometric surveys, such as the upcoming Legacy Survey of Space and Time \citep[LSST,][]{ivezic+19}. This approach will be developed in future work, but we argue that our current findings open new opportunities for observational cosmology.


\section*{Acknowledgements}
\begin{acknowledgments}
LG-S was financed by the São Paulo Research Foundation (FAPESP), Brasil (2024/03449-3). LSJ ackowlegdes the support from CNPq (308994/2021-3) and FAPESP (2011/51680-6).
\end{acknowledgments}

\vspace{5mm}
\facilities{Sloan}

\software{Astropy \citep{astropy:2013, astropy:2018, astropy:2022},
          COLOSSUS \citep{colossus},
          emcee \citep{emcee},
          NumPy \citep{numpy},
          SciPy \citep{scipy}}
          
\appendix

\section{Individual Modeling: An Example for the Coma Cluster} \label{app:coma_cluster}

In this section, we present an example of individual modeling for the Coma Cluster of galaxies, selected for its historical and scientific significance. We obtain the cumulative profile of the Coma Cluster by following the procedure outlined in Section \ref{subsubsec:individual}, including only galaxies whose recession velocities fall within $\pm 3\sigma_v$ of the cluster's central velocity (with a velocity dispersion of 975.326 km/s and a central velocity of 7195.02 km/s, corresponding to z = 0.024). We also limit the sample to galaxies brighter than -20.13 mag in the r-band and estimate the cumulative profile over 100 bins spanning a clustercentric comoving distance range of 0.3 to 10 cMpc.

The observed cumulative profile is then modeled using the trunc-NFW model described in Section \ref{subsec:model}. We sample the six free parameters via MCMC with a Poisson likelihood, employing the priors listed in Table \ref{tab:priors}. From the resulting posterior distributions, we approximate the splashback radius as the minimum of the logarithmic density slope, as proposed in previous studies \citep{diemer+14, adhikari+14, more+15}.

\begin{figure*} \centering
\includegraphics[width=.55\linewidth]{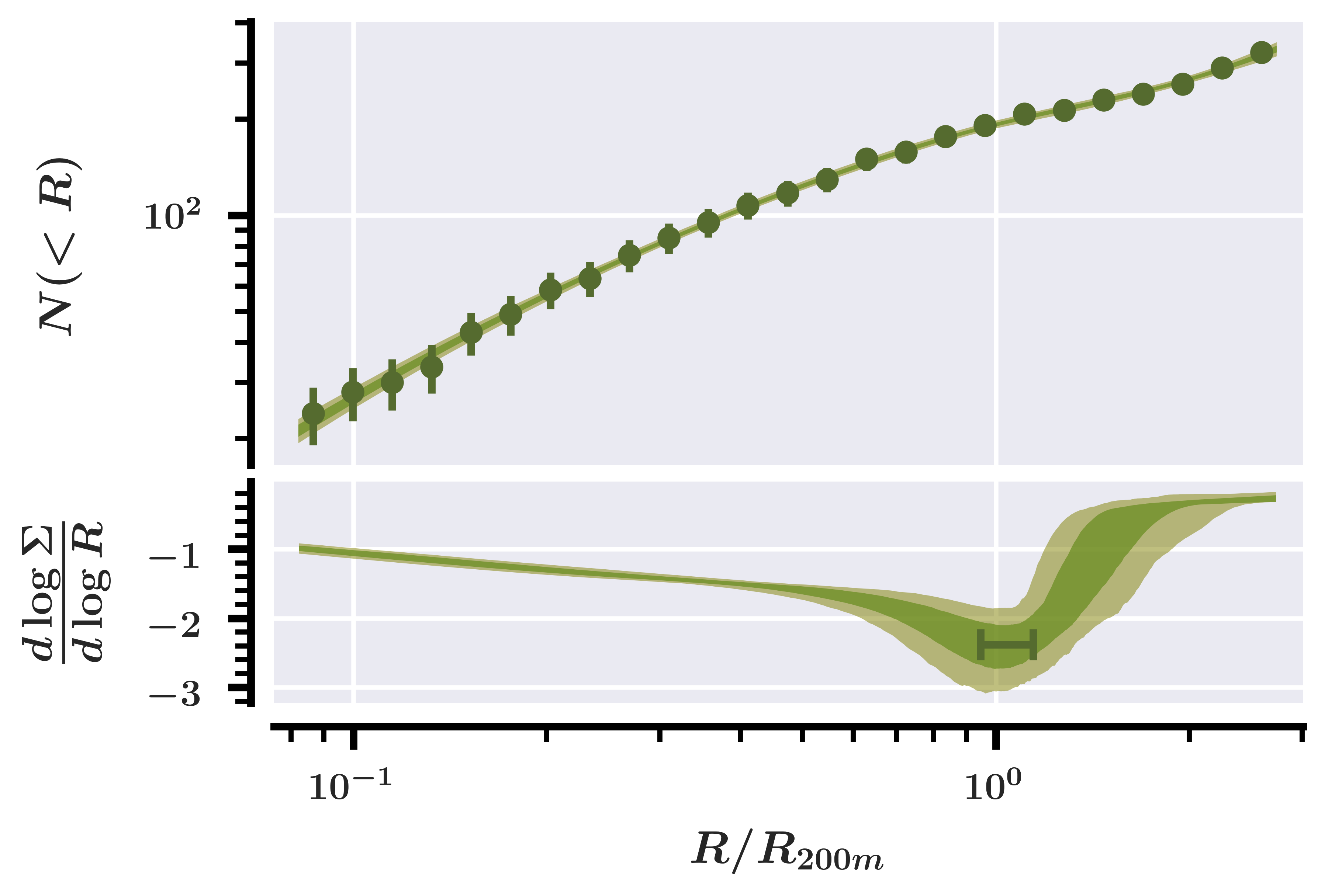}
\includegraphics[width=.44\linewidth]{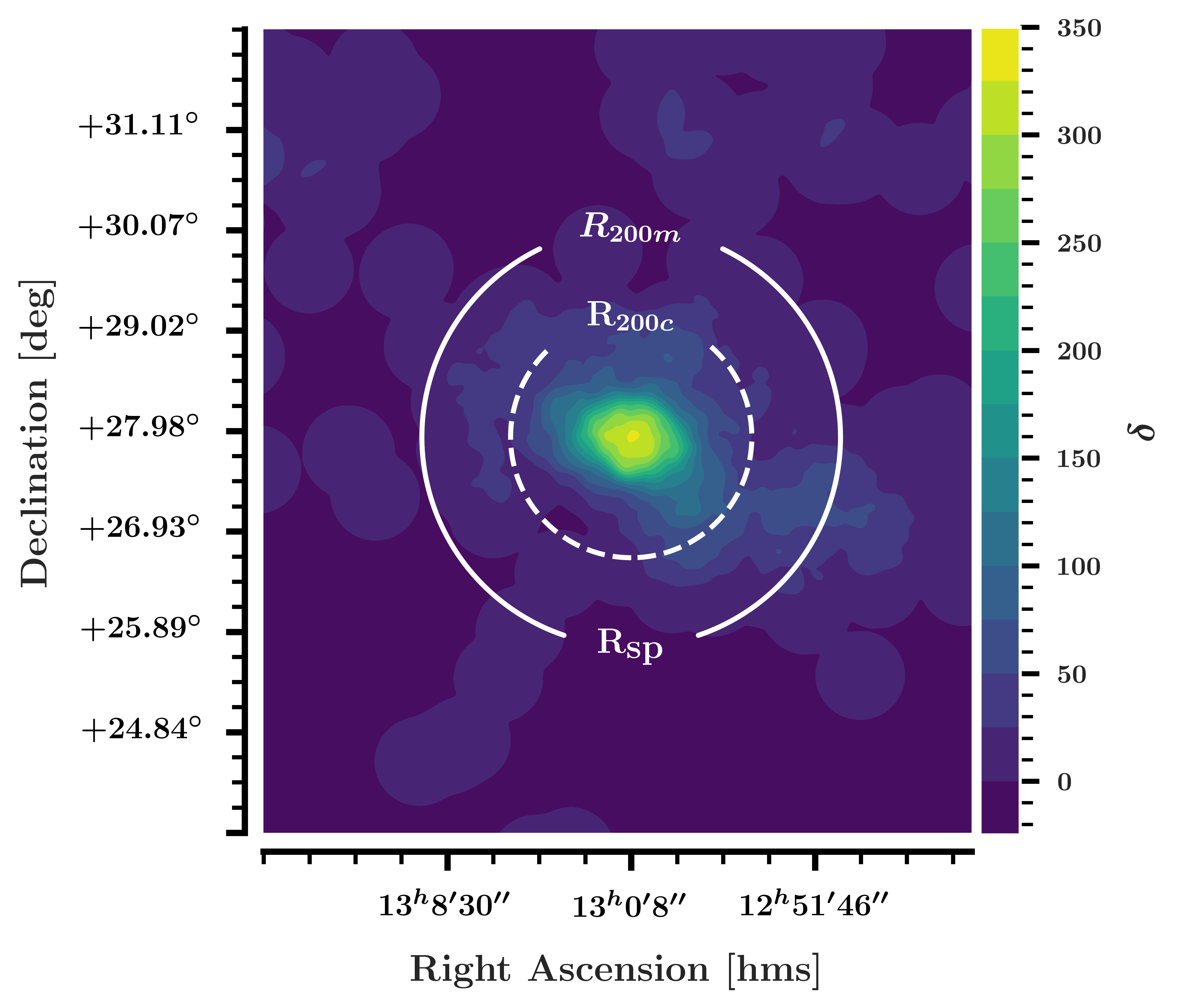}
\includegraphics[width=0.95\linewidth]{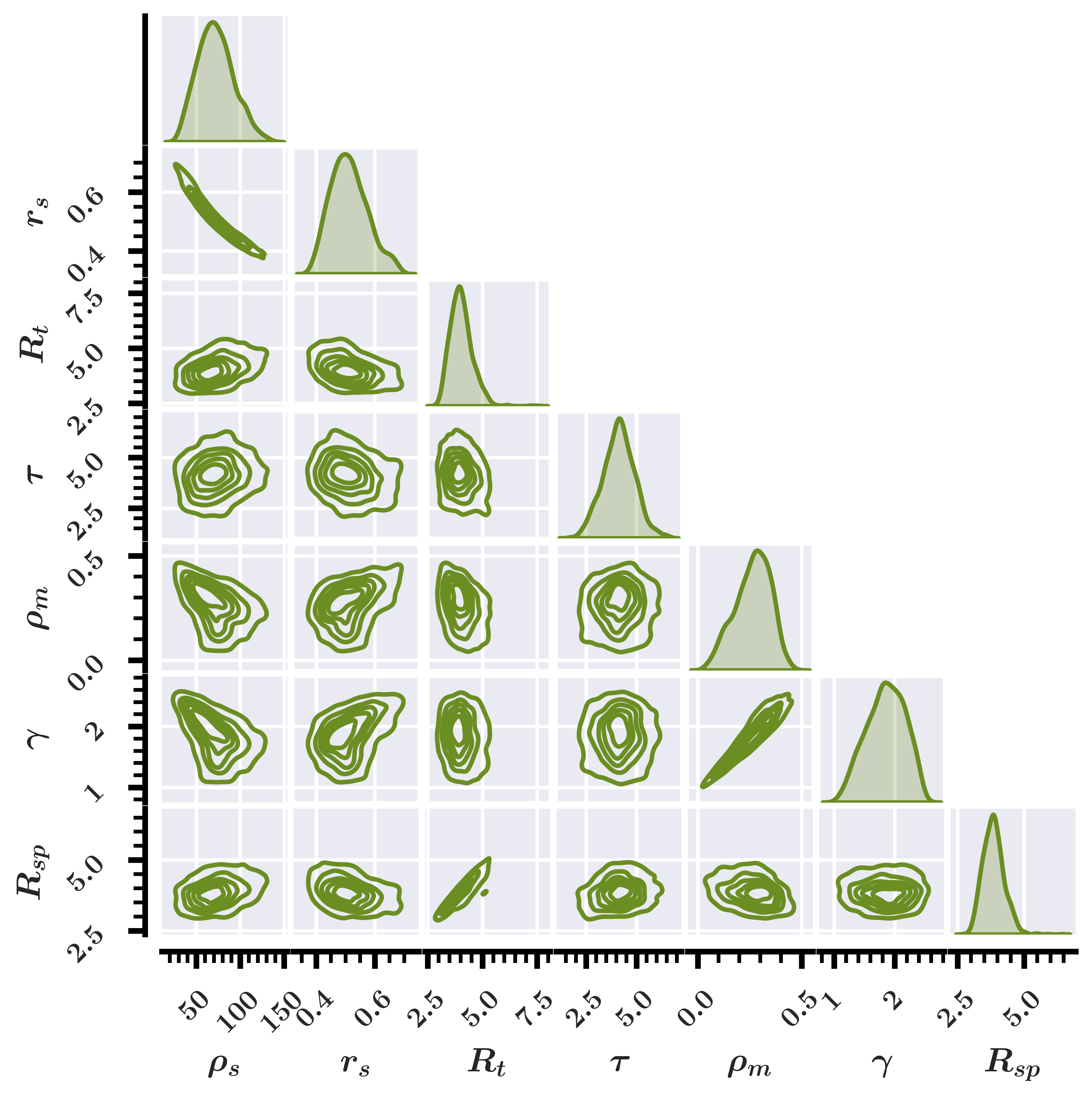}
\caption{Top-left panel: observed cumulative distribution for the Coma Cluster and model fit. Top-right panel: spatial distribution of galaxies, highlighting radii. Bottom panel: MCMC posterior distributions.}\label{fig:coma_example}
\end{figure*}

Figure \ref{fig:coma_example} shows the results for the Coma Cluster. In the top-left panel, we display the observed cumulative distribution as green points, with the 68\% and 95\% confidence intervals shaded in green, representing the posterior uncertainties for the modeled profile. The lower left panel presents the logarithmic density slope, with the error bar marking the $1\sigma$ interval for the splashback radius, which is remarkably close to the cluster's $R_{200m}$ (3.66 Mpc), yielding $R_\text{sp} = 3.82 \pm 0.38$ Mpc. The right panel illustrates the spatial distribution of galaxies in the Coma Cluster, highlighting $R_{200c}$, $R_{200m}$, and $R_{sp}$, with a color map representing the numerical overdensity:

\begin{equation}
\delta = \frac{\Sigma_g - \Sigma_\text{bg}}{\Sigma_\text{bg}},
\end{equation}

where $\Sigma_g$ denotes the point density, and $\Sigma_\text{bg}$ is the background density, estimated as described in Section \ref{subsubsec:stacked}. The larger bottom panel shows the posterior distributions for each parameter from the MCMC sampling, highlighting parameter correlations.

This procedure was applied to each cluster in both the SDSS and Mock samples.

\bibliography{sample631}{}

\begin{thebibliography}{}
\expandafter\ifx\csname natexlab\endcsname\relax\def\natexlab#1{#1}\fi
\providecommand{\url}[1]{\href{#1}{#1}}
\providecommand{\dodoi}[1]{doi:~\href{http://doi.org/#1}{\nolinkurl{#1}}}
\providecommand{\doeprint}[1]{\href{http://ascl.net/#1}{\nolinkurl{http://ascl.net/#1}}}
\providecommand{\doarXiv}[1]{\href{https://arxiv.org/abs/#1}{\nolinkurl{https://arxiv.org/abs/#1}}}

\bibitem[{Ade {et~al.}(2014)Ade, Aghanim, Armitage-Caplan, Arnaud, Ashdown, Atrio-Barandela, Aumont, Baccigalupi, Banday, Barreiro, Bartlett, Battaner, Benabed, Benoît, Benoit-Lévy, Bernard, Bersanelli, Bielewicz, Bobin, Bock, Bonaldi, Bond, Borrill, Bouchet, Bridges, Bucher, Burigana, Butler, Calabrese, Cappellini, Cardoso, Catalano, Challinor, Chamballu, Chary, Chen, Chiang, Chiang, Christensen, Church, Clements, Colombi, Colombo, Couchot, Coulais, Crill, Curto, Cuttaia, Danese, Davies, Davis, de~Bernardis, de~Rosa, de~Zotti, Delabrouille, Delouis, Désert, Dickinson, Diego, Dolag, Dole, Donzelli, Doré, Douspis, Dunkley, Dupac, Efstathiou, Elsner, Enßlin, Eriksen, Finelli, Forni, Frailis, Fraisse, Franceschi, Gaier, Galeotta, Galli, Ganga, Giard, Giardino, Giraud-Héraud, Gjerløw, González-Nuevo, Górski, Gratton, Gregorio, Gruppuso, Gudmundsson, Haissinski, Hamann, Hansen, Hanson, Harrison, Henrot-Versillé, Hernández-Monteagudo, Herranz, Hildebrandt, Hivon, Hobson, Holmes, Hornstrup, Hou, Hovest,
  Huffenberger, Jaffe, Jaffe, Jewell, Jones, Juvela, Keihänen, Keskitalo, Kisner, Kneissl, Knoche, Knox, Kunz, Kurki-Suonio, Lagache, Lähteenmäki, Lamarre, Lasenby, Lattanzi, Laureijs, Lawrence, Leach, Leahy, Leonardi, León-Tavares, Lesgourgues, Lewis, Liguori, Lilje, Linden-Vørnle, López-Caniego, Lubin, Macías-Pérez, Maffei, Maino, Mandolesi, Maris, Marshall, Martin, Martínez-González, Masi, Massardi, Matarrese, Matthai, Mazzotta, Meinhold, Melchiorri, Melin, Mendes, Menegoni, Mennella, Migliaccio, Millea, Mitra, Miville-Deschênes, Moneti, Montier, Morgante, Mortlock, Moss, Munshi, Murphy, Naselsky, Nati, Natoli, Netterfield, Nørgaard-Nielsen, Noviello, Novikov, Novikov, O’Dwyer, Osborne, Oxborrow, Paci, Pagano, Pajot, Paladini, Paoletti, Partridge, Pasian, Patanchon, Pearson, Pearson, Peiris, Perdereau, Perotto, Perrotta, Pettorino, Piacentini, Piat, Pierpaoli, Pietrobon, Plaszczynski, Platania, Pointecouteau, Polenta, Ponthieu, Popa, Poutanen, Pratt, Prézeau, Prunet, Puget, Rachen, Reach,
  Rebolo, Reinecke, Remazeilles, Renault, Ricciardi, Riller, Ristorcelli, Rocha, Rosset, Roudier, Rowan-Robinson, Rubiño-Martín, Rusholme, Sandri, Santos, Savelainen, Savini, Scott, Seiffert, Shellard, Spencer, Starck, Stolyarov, Stompor, Sudiwala, Sunyaev, Sureau, Sutton, Suur-Uski, Sygnet, Tauber, Tavagnacco, Terenzi, Toffolatti, Tomasi, Tristram, Tucci, Tuovinen, Türler, Umana, Valenziano, Valiviita, Van~Tent, Vielva, Villa, Vittorio, Wade, Wandelt, Wehus, White, White, Wilkinson, Yvon, Zacchei, \& Zonca}]{planck14}
Ade, P. A.~R., Aghanim, N., Armitage-Caplan, C., {et~al.} 2014, Astronomy \& Astrophysics, 571, A16, \dodoi{10.1051/0004-6361/201321591}

\bibitem[{Adhikari {et~al.}(2014)Adhikari, Dalal, \& Chamberlain}]{adhikari+14}
Adhikari, S., Dalal, N., \& Chamberlain, R.~T. 2014, Journal of Cosmology and Astroparticle Physics, 2014, 019–019, \dodoi{10.1088/1475-7516/2014/11/019}

\bibitem[{Adhikari {et~al.}(2016)Adhikari, Dalal, \& Clampitt}]{adhikari+16}
Adhikari, S., Dalal, N., \& Clampitt, J. 2016, Journal of Cosmology and Astroparticle Physics, 2016, 022, \dodoi{10.1088/1475-7516/2016/07/022}

\bibitem[{Adhikari {et~al.}(2018)Adhikari, Sakstein, Jain, Dalal, \& Li}]{adhikari+18}
Adhikari, S., Sakstein, J., Jain, B., Dalal, N., \& Li, B. 2018, Journal of Cosmology and Astroparticle Physics, 2018, 033, \dodoi{10.1088/1475-7516/2018/11/033}

\bibitem[{Adhikari {et~al.}(2021)Adhikari, hyeon Shin, Jain, Hilton, Baxter, Chang, Wechsler, Battaglia, Bond, Bocquet, Choi, DeRose, Devlin, Dunkley, Evrard, Ferraro, Hill, Hughes, Gallardo, Lokken, MacInnis, Madhavacheril, McMahon, Nati, Newburgh, Niemack, Page, Palmese, Partridge, Rozo, Rykoff, Salatino, Schillaci, Sehgal, Sifón, To, Wollack, Wu, Xu, Aguena, Allam, Amon, Annis, Avila, Bacon, Bertin, Bhargava, Brooks, Burke, Rosell, Kind, Carretero, Castander, Choi, Costanzi, da~Costa, Vicente, Desai, Diehl, Doel, Everett, Ferrero, Ferté, Flaugher, Fosalba, Frieman, García-Bellido, Gaztanaga, Gruen, Gruendl, Gschwend, Gutierrez, Hartley, Hinton, Hollowood, Honscheid, James, Jeltema, Kuehn, Kuropatkin, Lahav, Lima, Maia, Marshall, Martini, Melchior, Menanteau, Miquel, Morgan, Ogando, Paz-Chinchón, Malagón, Sanchez, Santiago, Scarpine, Serrano, Sevilla-Noarbe, Smith, Soares-Santos, Suchyta, Swanson, Varga, Wilkinson, Zhang, Austermann, Beall, Becker, Denison, Duff, Hilton, Hubmayr, Ullom, Lanen, Vale,
  Collaboration), \& Collaboration)}]{adhikari+21}
Adhikari, S., hyeon Shin, T., Jain, B., {et~al.} 2021, The Astrophysical Journal, 923, 37, \dodoi{10.3847/1538-4357/ac0bbc}

\bibitem[{Akaike(1974)}]{akaike74}
Akaike, H. 1974, IEEE Transactions on Automatic Control, 19, 716, \dodoi{10.1109/TAC.1974.1100705}

\bibitem[{Allen {et~al.}(2011)Allen, Evrard, \& Mantz}]{allen+11}
Allen, S.~W., Evrard, A.~E., \& Mantz, A.~B. 2011, Annual Review of Astronomy and Astrophysics, 49, 409, \dodoi{https://doi.org/10.1146/annurev-astro-081710-102514}

\bibitem[{{Allen} {et~al.}(2004)}]{allen+04}
{Allen}, S.~W., {et~al.} 2004, Monthly Notices of the Royal Astronomical Society, 353, 457

\bibitem[{Araya-Araya {et~al.}(2021)Araya-Araya, Vicentin, Sodré, Overzier, \& Cuevas}]{araya-araya21}
Araya-Araya, P., Vicentin, M.~C., Sodré, Laerte, J., Overzier, R.~A., \& Cuevas, H. 2021, Monthly Notices of the Royal Astronomical Society, 504, 5054, \dodoi{10.1093/mnras/stab1133}

\bibitem[{{Arnaud} {et~al.}(2005)}]{arnaud+05}
{Arnaud}, M., {et~al.} 2005, Astronomy \& Astrophysics, 441, 893

\bibitem[{{Astropy Collaboration} {et~al.}(2013){Astropy Collaboration}, {Robitaille}, {Tollerud}, {Greenfield}, {Droettboom}, {Bray}, {Aldcroft}, {Davis}, {Ginsburg}, {Price-Whelan}, {Kerzendorf}, {Conley}, {Crighton}, {Barbary}, {Muna}, {Ferguson}, {Grollier}, {Parikh}, {Nair}, {Unther}, {Deil}, {Woillez}, {Conseil}, {Kramer}, {Turner}, {Singer}, {Fox}, {Weaver}, {Zabalza}, {Edwards}, {Azalee Bostroem}, {Burke}, {Casey}, {Crawford}, {Dencheva}, {Ely}, {Jenness}, {Labrie}, {Lim}, {Pierfederici}, {Pontzen}, {Ptak}, {Refsdal}, {Servillat}, \& {Streicher}}]{astropy:2013}
{Astropy Collaboration}, {Robitaille}, T.~P., {Tollerud}, E.~J., {et~al.} 2013, \aap, 558, A33, \dodoi{10.1051/0004-6361/201322068}

\bibitem[{{Astropy Collaboration} {et~al.}(2018){Astropy Collaboration}, {Price-Whelan}, {Sip{\H{o}}cz}, {G{\"u}nther}, {Lim}, {Crawford}, {Conseil}, {Shupe}, {Craig}, {Dencheva}, {Ginsburg}, {Vand erPlas}, {Bradley}, {P{\'e}rez-Su{\'a}rez}, {de Val-Borro}, {Aldcroft}, {Cruz}, {Robitaille}, {Tollerud}, {Ardelean}, {Babej}, {Bach}, {Bachetti}, {Bakanov}, {Bamford}, {Barentsen}, {Barmby}, {Baumbach}, {Berry}, {Biscani}, {Boquien}, {Bostroem}, {Bouma}, {Brammer}, {Bray}, {Breytenbach}, {Buddelmeijer}, {Burke}, {Calderone}, {Cano Rodr{\'\i}guez}, {Cara}, {Cardoso}, {Cheedella}, {Copin}, {Corrales}, {Crichton}, {D'Avella}, {Deil}, {Depagne}, {Dietrich}, {Donath}, {Droettboom}, {Earl}, {Erben}, {Fabbro}, {Ferreira}, {Finethy}, {Fox}, {Garrison}, {Gibbons}, {Goldstein}, {Gommers}, {Greco}, {Greenfield}, {Groener}, {Grollier}, {Hagen}, {Hirst}, {Homeier}, {Horton}, {Hosseinzadeh}, {Hu}, {Hunkeler}, {Ivezi{\'c}}, {Jain}, {Jenness}, {Kanarek}, {Kendrew}, {Kern}, {Kerzendorf}, {Khvalko}, {King}, {Kirkby}, {Kulkarni},
  {Kumar}, {Lee}, {Lenz}, {Littlefair}, {Ma}, {Macleod}, {Mastropietro}, {McCully}, {Montagnac}, {Morris}, {Mueller}, {Mumford}, {Muna}, {Murphy}, {Nelson}, {Nguyen}, {Ninan}, {N{\"o}the}, {Ogaz}, {Oh}, {Parejko}, {Parley}, {Pascual}, {Patil}, {Patil}, {Plunkett}, {Prochaska}, {Rastogi}, {Reddy Janga}, {Sabater}, {Sakurikar}, {Seifert}, {Sherbert}, {Sherwood-Taylor}, {Shih}, {Sick}, {Silbiger}, {Singanamalla}, {Singer}, {Sladen}, {Sooley}, {Sornarajah}, {Streicher}, {Teuben}, {Thomas}, {Tremblay}, {Turner}, {Terr{\'o}n}, {van Kerkwijk}, {de la Vega}, {Watkins}, {Weaver}, {Whitmore}, {Woillez}, {Zabalza}, \& {Astropy Contributors}}]{astropy:2018}
{Astropy Collaboration}, {Price-Whelan}, A.~M., {Sip{\H{o}}cz}, B.~M., {et~al.} 2018, \aj, 156, 123, \dodoi{10.3847/1538-3881/aabc4f}

\bibitem[{{Astropy Collaboration} {et~al.}(2022){Astropy Collaboration}, {Price-Whelan}, {Lim}, {Earl}, {Starkman}, {Bradley}, {Shupe}, {Patil}, {Corrales}, {Brasseur}, {N{"o}the}, {Donath}, {Tollerud}, {Morris}, {Ginsburg}, {Vaher}, {Weaver}, {Tocknell}, {Jamieson}, {van Kerkwijk}, {Robitaille}, {Merry}, {Bachetti}, {G{"u}nther}, {Aldcroft}, {Alvarado-Montes}, {Archibald}, {B{'o}di}, {Bapat}, {Barentsen}, {Baz{'a}n}, {Biswas}, {Boquien}, {Burke}, {Cara}, {Cara}, {Conroy}, {Conseil}, {Craig}, {Cross}, {Cruz}, {D'Eugenio}, {Dencheva}, {Devillepoix}, {Dietrich}, {Eigenbrot}, {Erben}, {Ferreira}, {Foreman-Mackey}, {Fox}, {Freij}, {Garg}, {Geda}, {Glattly}, {Gondhalekar}, {Gordon}, {Grant}, {Greenfield}, {Groener}, {Guest}, {Gurovich}, {Handberg}, {Hart}, {Hatfield-Dodds}, {Homeier}, {Hosseinzadeh}, {Jenness}, {Jones}, {Joseph}, {Kalmbach}, {Karamehmetoglu}, {Ka{l}uszy{'n}ski}, {Kelley}, {Kern}, {Kerzendorf}, {Koch}, {Kulumani}, {Lee}, {Ly}, {Ma}, {MacBride}, {Maljaars}, {Muna}, {Murphy}, {Norman}, {O'Steen},
  {Oman}, {Pacifici}, {Pascual}, {Pascual-Granado}, {Patil}, {Perren}, {Pickering}, {Rastogi}, {Roulston}, {Ryan}, {Rykoff}, {Sabater}, {Sakurikar}, {Salgado}, {Sanghi}, {Saunders}, {Savchenko}, {Schwardt}, {Seifert-Eckert}, {Shih}, {Jain}, {Shukla}, {Sick}, {Simpson}, {Singanamalla}, {Singer}, {Singhal}, {Sinha}, {Sip{H{o}}cz}, {Spitler}, {Stansby}, {Streicher}, {{{S}}umak}, {Swinbank}, {Taranu}, {Tewary}, {Tremblay}, {Val-Borro}, {Van Kooten}, {Vasovi{'c}}, {Verma}, {de Miranda Cardoso}, {Williams}, {Wilson}, {Winkel}, {Wood-Vasey}, {Xue}, {Yoachim}, {Zhang}, {Zonca}, \& {Astropy Project Contributors}}]{astropy:2022}
{Astropy Collaboration}, {Price-Whelan}, A.~M., {Lim}, P.~L., {et~al.} 2022, \apj, 935, 167, \dodoi{10.3847/1538-4357/ac7c74}

\bibitem[{Astudillo {et~al.}(2024)Astudillo, Carrasco, Castellón, Zenteno, \& Cuevas}]{astudillo+24}
Astudillo, S.~V., Carrasco, E.~R., Castellón, J. L.~N., Zenteno, A., \& Cuevas, H. 2024, The effect of dynamical states on galaxy clusters populations. I. Classification of dynamical states.
\newblock \doarXiv{2408.02519}

\bibitem[{Baxter {et~al.}(2017)Baxter, Chang, Jain, Adhikari, Dalal, Kravtsov, More, Rozo, Rykoff, \& Sheth}]{baxter+17}
Baxter, E., Chang, C., Jain, B., {et~al.} 2017, The Astrophysical Journal, 841, 18, \dodoi{10.3847/1538-4357/aa6ff0}

\bibitem[{Beck {et~al.}(2016)Beck, Dobos, Budavári, Szalay, \& Csabai}]{beck+16}
Beck, R., Dobos, L., Budavári, T., Szalay, A.~S., \& Csabai, I. 2016, Monthly Notices of the Royal Astronomical Society, 460, 1371, \dodoi{10.1093/mnras/stw1009}

\bibitem[{Behroozi {et~al.}(2014)Behroozi, Wechsler, Lu, Hahn, Busha, Klypin, \& Primack}]{behroozi+14}
Behroozi, P.~S., Wechsler, R.~H., Lu, Y., {et~al.} 2014, The Astrophysical Journal, 787, 156, \dodoi{10.1088/0004-637x/787/2/156}

\bibitem[{{Bertschinger}(1985)}]{bertschinger85}
{Bertschinger}, E. 1985, Astrophysical Journal Supplement Series, 58, 39

\bibitem[{Boggs \& Rogers(1990)}]{boggs+90}
Boggs, P.~T., \& Rogers, J.~E. 1990, Contemporary Mathematics, 112, 186

\bibitem[{Boselli \& Gavazzi(2006)}]{boselli+06}
Boselli, A., \& Gavazzi, G. 2006, Publications of the Astronomical Society of the Pacific, 118, 517, \dodoi{10.1086/500691}

\bibitem[{{Carlberg} {et~al.}(1997)}]{carlberg+97}
{Carlberg}, G.~R., {et~al.} 1997, The Astrophysical Journal, 478, 462

\bibitem[{Chandrasekhar(1943)}]{chandrasekhar43}
Chandrasekhar, S. 1943, Dynamical Friction. I. General Considerations: the Coefficient of Dynamical Friction.

\bibitem[{Chang {et~al.}(2018)Chang, Baxter, Jain, Sánchez, Adhikari, Varga, Fang, Rozo, Rykoff, Kravtsov, Gruen, Hartley, Huff, Jarvis, Kim, Prat, MacCrann, McClintock, Palmese, Rapetti, Rollins, Samuroff, Sheldon, Troxel, Wechsler, Zhang, Zuntz, Abbott, Abdalla, Allam, Annis, Bechtol, Benoit-Lévy, Bernstein, Brooks, Buckley-Geer, Rosell, Kind, Carretero, D’Andrea, da~Costa, Davis, Desai, Diehl, Dietrich, Drlica-Wagner, Eifler, Flaugher, Fosalba, Frieman, García-Bellido, Gaztanaga, Gerdes, Gruendl, Gschwend, Gutierrez, Honscheid, James, Jeltema, Krause, Kuehn, Lahav, Lima, March, Marshall, Martini, Melchior, Menanteau, Miquel, Mohr, Nord, Ogando, Plazas, Sanchez, Scarpine, Schindler, Schubnell, Sevilla-Noarbe, Smith, Smith, Soares-Santos, Sobreira, Suchyta, Swanson, Tarle, Weller, \& Collaboration)}]{chang+18}
Chang, C., Baxter, E., Jain, B., {et~al.} 2018, The Astrophysical Journal, 864, 83, \dodoi{10.3847/1538-4357/aad5e7}

\bibitem[{Chilingarian {et~al.}(2010)Chilingarian, Melchior, \& Zolotukhin}]{chilingarian10}
Chilingarian, I.~V., Melchior, A.-L., \& Zolotukhin, I.~Y. 2010, Monthly Notices of the Royal Astronomical Society, 405, 1409, \dodoi{10.1111/j.1365-2966.2010.16506.x}

\bibitem[{Chilingarian \& Zolotukhin(2011)}]{chilingarian11}
Chilingarian, I.~V., \& Zolotukhin, I.~Y. 2011, Monthly Notices of the Royal Astronomical Society, 419, 1727, \dodoi{10.1111/j.1365-2966.2011.19837.x}

\bibitem[{Ciotti \& Bertin(1999)}]{ciotti+99}
Ciotti, L., \& Bertin, G. 1999, Astronomy and Astrophysics, 352, 447, \dodoi{10.48550/arXiv.astro-ph/9911078}

\bibitem[{Contigiani {et~al.}(2019)Contigiani, Vardanyan, \& Silvestri}]{contigiani+19}
Contigiani, O., Vardanyan, V., \& Silvestri, A. 2019, Phys. Rev. D, 99, 064030, \dodoi{10.1103/PhysRevD.99.064030}

\bibitem[{Cypriano {et~al.}(2004)Cypriano, Laerte~Sodré, Kneib, \& Campusano}]{cypriano+04}
Cypriano, E.~S., Laerte~Sodré, J., Kneib, J.-P., \& Campusano, L.~E. 2004, The Astrophysical Journal, 613, 95, \dodoi{10.1086/422896}

\bibitem[{Dariush {et~al.}(2010)Dariush, Raychaudhury, Ponman, Khosroshahi, Benson, Bower, \& Pearce}]{dariush+10}
Dariush, A.~A., Raychaudhury, S., Ponman, T.~J., {et~al.} 2010, Monthly Notices of the Royal Astronomical Society, 405, 1873, \dodoi{10.1111/j.1365-2966.2010.16569.x}

\bibitem[{Davis \& Peebles(1983)}]{davis+83}
Davis, M., \& Peebles, P. J.~E. 1983, Astrophysical Journal, 267, 465, \dodoi{10.1086/160884}

\bibitem[{Diaferio(1999)}]{diaferio99}
Diaferio, A. 1999, Monthly Notices of the Royal Astronomical Society, 309, 610, \dodoi{10.1046/j.1365-8711.1999.02864.x}

\bibitem[{Diemer(2018)}]{colossus}
Diemer, B. 2018, The Astrophysical Journal Supplement Series, 239, 35, \dodoi{10.3847/1538-4365/aaee8c}

\bibitem[{Diemer(2020)}]{diemer20}
---. 2020, The Astrophysical Journal, 903, 87, \dodoi{10.3847/1538-4357/abbf52}

\bibitem[{Diemer(2022)}]{diemer22}
---. 2022, Monthly Notices of the Royal Astronomical Society, 519, 3292, \dodoi{10.1093/mnras/stac3778}

\bibitem[{Diemer \& Joyce(2019)}]{diemer19}
Diemer, B., \& Joyce, M. 2019, The Astrophysical Journal, 871, 168, \dodoi{10.3847/1538-4357/aafad6}

\bibitem[{Diemer \& Kravtsov(2014)}]{diemer+14}
Diemer, B., \& Kravtsov, A.~V. 2014, The Astrophysical Journal, 789, 1, \dodoi{10.1088/0004-637x/789/1/1}

\bibitem[{Diemer {et~al.}(2013)Diemer, More, \& Kravtsov}]{diemer+13}
Diemer, B., More, S., \& Kravtsov, A.~V. 2013, The Astrophysical Journal, 766, 25, \dodoi{10.1088/0004-637x/766/1/25}

\bibitem[{{Dietrich, J. P.} {et~al.}(2009){Dietrich, J. P.}, {Biviano, A.}, {Popesso, P.}, {Zhang, Y.-Y.}, {Lombardi, M.}, \& {Böhringer, H.}}]{dietrich+09}
{Dietrich, J. P.}, {Biviano, A.}, {Popesso, P.}, {et~al.} 2009, A\&A, 499, 669, \dodoi{10.1051/0004-6361/200811433}

\bibitem[{Dressler \& Shectman(1988)}]{dressler+88}
Dressler, A., \& Shectman, S.~A. 1988, Astronomical Journal, 95, 985

\bibitem[{{Ettori, S.} {et~al.}(2011){Ettori, S.}, {Gastaldello, F.}, {Leccardi, A.}, {Molendi, S.}, {Rossetti, M.}, {Buote, D.}, \& {Meneghetti, M.}}]{ettori+11}
{Ettori, S.}, {Gastaldello, F.}, {Leccardi, A.}, {et~al.} 2011, A\&A, 526, C1, \dodoi{10.1051/0004-6361/201015271e}

\bibitem[{Fakhouri {et~al.}(2010)Fakhouri, Ma, \& Boylan-Kolchin}]{fakhouri+10}
Fakhouri, O., Ma, C.-P., \& Boylan-Kolchin, M. 2010, Monthly Notices of the Royal Astronomical Society, 406, 2267, \dodoi{10.1111/j.1365-2966.2010.16859.x}

\bibitem[{{Fillmore} \& {Goldreich}(1984)}]{fillmore+84}
{Fillmore}, J.~A., \& {Goldreich}, P. 1984, \apj, 281, 1

\bibitem[{Foreman-Mackey {et~al.}(2013)Foreman-Mackey, Hogg, Lang, \& Goodman}]{emcee}
Foreman-Mackey, D., Hogg, D.~W., Lang, D., \& Goodman, J. 2013, Publications of the Astronomical Society of the Pacific, 125, 306, \dodoi{10.1086/670067}

\bibitem[{{Gavazzi, R.} {et~al.}(2009){Gavazzi, R.}, {Adami, C.}, {Durret, F.}, {Cuillandre, J.-C.}, {Ilbert, O.}, {Mazure, A.}, {Pelló, R.}, \& {Ulmer, M. P.}}]{gavazzi+09}
{Gavazzi, R.}, {Adami, C.}, {Durret, F.}, {et~al.} 2009, A\&A, 498, L33, \dodoi{10.1051/0004-6361/200911841}

\bibitem[{Giocoli {et~al.}(2024)Giocoli, Palmucci, Lesci, Moscardini, Despali, Marulli, Maturi, Radovich, Sereno, Bardelli, Castignani, Covone, Ingoglia, Romanello, Roncarelli, \& Puddu}]{giocoli+24}
Giocoli, C., Palmucci, L., Lesci, G.~F., {et~al.} 2024, A\&A, 687, A79, \dodoi{10.1051/0004-6361/202449561}

\bibitem[{Gonzalez {et~al.}(2016)Gonzalez, Rodriguez, García~Lambas, Merchán, Foëx, \& Chalela}]{gonzalez+16}
Gonzalez, E.~J., Rodriguez, F., García~Lambas, D., {et~al.} 2016, Monthly Notices of the Royal Astronomical Society, 465, 1348, \dodoi{10.1093/mnras/stw2803}

\bibitem[{Gunn \& Gott(1972)}]{gunn+72}
Gunn, J.~E., \& Gott, J.~R. 1972, Astrophysical Journal, 176, 1, \dodoi{10.1086/151605}

\bibitem[{Haggar {et~al.}(2024)Haggar, Amoura, Mpetha, Taylor, Walker, \& Power}]{haggar+24}
Haggar, R., Amoura, Y., Mpetha, C.~T., {et~al.} 2024, The Astrophysical Journal, 972, 28, \dodoi{10.3847/1538-4357/ad5cee}

\bibitem[{{Haiman} {et~al.}(2001)}]{haiman+01}
{Haiman}, Z., {et~al.} 2001, The Astrophysical Journal, 553, 545

\bibitem[{Harris {et~al.}(2020)Harris, Millman, van~der Walt, Gommers, Virtanen, Cournapeau, Wieser, Taylor, Berg, Smith, Kern, Picus, Hoyer, van Kerkwijk, Brett, Haldane, del R{\'{i}}o, Wiebe, Peterson, G{\'{e}}rard-Marchant, Sheppard, Reddy, Weckesser, Abbasi, Gohlke, \& Oliphant}]{numpy}
Harris, C.~R., Millman, K.~J., van~der Walt, S.~J., {et~al.} 2020, Nature, 585, 357, \dodoi{10.1038/s41586-020-2649-2}

\bibitem[{Henriques {et~al.}(2015)Henriques, White, Thomas, Angulo, Guo, Lemson, Springel, \& Overzier}]{henriques15}
Henriques, B. M.~B., White, S. D.~M., Thomas, P.~A., {et~al.} 2015, Monthly Notices of the Royal Astronomical Society, 451, 2663, \dodoi{10.1093/mnras/stv705}

\bibitem[{Herbonnet {et~al.}(2020)Herbonnet, Sifón, Hoekstra, Bahé, van der Burg, Melin, von der Linden, Sand, Kay, \& Barnes}]{herbonnet20}
Herbonnet, R., Sifón, C., Hoekstra, H., {et~al.} 2020, Monthly Notices of the Royal Astronomical Society, 497, 4684, \dodoi{10.1093/mnras/staa2303}

\bibitem[{Jones {et~al.}(2001--)Jones, Oliphant, Peterson, {et~al.}}]{scipy}
Jones, E., Oliphant, T., Peterson, P., {et~al.} 2001--, {SciPy}: Open source scientific tools for {Python}.
\newblock \url{http://www.scipy.org/}

\bibitem[{Kopylova \& Kopylov(2022)}]{kopylova+20}
Kopylova, F.~G., \& Kopylov, A.~I. 2022, Astrophysical Bulletin, 77, 347–360

\bibitem[{Kravtsov \& Borgani(2012)}]{kravtsov+12}
Kravtsov, A.~V., \& Borgani, S. 2012, Annual Review of Astronomy and Astrophysics, 50, 353, \dodoi{https://doi.org/10.1146/annurev-astro-081811-125502}

\bibitem[{Kubo {et~al.}(2009)Kubo, Annis, Hardin, Kubik, Lawhorn, Lin, Nicklaus, Nelson, Reis, Seo, Soares-Santos, Stebbins, \& Yunker}]{kubo+09}
Kubo, J.~M., Annis, J., Hardin, F.~M., {et~al.} 2009, The Astrophysical Journal, 702, L110, \dodoi{10.1088/0004-637X/702/2/L110}

\bibitem[{Lacey \& Cole(1993)}]{lacey+93}
Lacey, C., \& Cole, S. 1993, Monthly Notices of the Royal Astronomical Society, 262, 627, \dodoi{10.1093/mnras/262.3.627}

\bibitem[{Lebeau {et~al.}(2024)Lebeau, Ettori, Aghanim, \& Sorce}]{lebeau+24}
Lebeau, T., Ettori, S., Aghanim, N., \& Sorce, J.~G. 2024, A\&A, 689, A19, \dodoi{10.1051/0004-6361/202450146}

\bibitem[{Lima {et~al.}(2022)Lima, Sodré, Bom, Teixeira, Nakazono, Buzzo, Queiroz, Herpich, Castellon, Dantas, Dors, de~Souza, Akras, Jiménez-Teja, Kanaan, Ribeiro, \& Schoennell}]{lima+22}
Lima, E., Sodré, L., Bom, C., {et~al.} 2022, Astronomy and Computing, 38, 100510, \dodoi{https://doi.org/10.1016/j.ascom.2021.100510}

\bibitem[{Lopes {et~al.}(2018)Lopes, Trevisan, Laganá, Durret, Ribeiro, \& Rembold}]{lopes+18}
Lopes, P. A.~A., Trevisan, M., Laganá, T.~F., {et~al.} 2018, Monthly Notices of the Royal Astronomical Society, 478, 5473, \dodoi{10.1093/mnras/sty1374}

\bibitem[{Mann \& Ebeling(2012)}]{mann+12}
Mann, A.~W., \& Ebeling, H. 2012, Monthly Notices of the Royal Astronomical Society, 420, 2120, \dodoi{10.1111/j.1365-2966.2011.20170.x}

\bibitem[{Merritt {et~al.}(2005)Merritt, Navarro, Ludlow, \& Jenkins}]{merritt+05}
Merritt, D., Navarro, J.~F., Ludlow, A., \& Jenkins, A. 2005, The Astrophysical Journal, 624, L85, \dodoi{10.1086/430636}

\bibitem[{More {et~al.}(2015)More, Diemer, \& Kravtsov}]{more+15}
More, S., Diemer, B., \& Kravtsov, A.~V. 2015, The Astrophysical Journal, 810, 36, \dodoi{10.1088/0004-637x/810/1/36}

\bibitem[{More {et~al.}(2016)More, Miyatake, Takada, Diemer, Kravtsov, Dalal, More, Murata, Mandelbaum, Rozo, Rykoff, Oguri, \& Spergel}]{more+16}
More, S., Miyatake, H., Takada, M., {et~al.} 2016, The Astrophysical Journal, 825, 39, \dodoi{10.3847/0004-637x/825/1/39}

\bibitem[{Mulroy {et~al.}(2017)Mulroy, McGee, Gillman, Smith, Haines, Démoclès, Okabe, \& Egami}]{mulroy+17}
Mulroy, S.~L., McGee, S.~L., Gillman, S., {et~al.} 2017, Monthly Notices of the Royal Astronomical Society, 472, 3246

\bibitem[{Murata {et~al.}(2018)Murata, Nishimichi, Takada, Miyatake, Shirasaki, More, Takahashi, \& Osato}]{murata+18}
Murata, R., Nishimichi, T., Takada, M., {et~al.} 2018, The Astrophysical Journal, 854, 120, \dodoi{10.3847/1538-4357/aaaab8}

\bibitem[{Murata {et~al.}(2020)Murata, Sunayama, Oguri, More, Nishizawa, Nishimichi, \& Osato}]{murata+20}
Murata, R., Sunayama, T., Oguri, M., {et~al.} 2020, Publications of the Astronomical Society of Japan, 72, 64, \dodoi{10.1093/pasj/psaa041}

\bibitem[{Navarro {et~al.}(1996)Navarro, Frenk, \& White}]{navarro+96}
Navarro, J.~F., Frenk, C.~S., \& White, S. D.~M. 1996, The Astrophysical Journal, 462, 563, \dodoi{10.1086/177173}

\bibitem[{Okabe {et~al.}(2010)Okabe, Takada, Umetsu, Futamase, \& Smith}]{okabe+10}
Okabe, N., Takada, M., Umetsu, K., Futamase, T., \& Smith, G.~P. 2010, Publications of the Astronomical Society of Japan, 62, 811, \dodoi{10.1093/pasj/62.3.811}

\bibitem[{Okabe \& Umetsu(2008)}]{okabe+08}
Okabe, N., \& Umetsu, K. 2008, Publications of the Astronomical Society of Japan, 60, 345, \dodoi{10.1093/pasj/60.2.345}

\bibitem[{O'Shea {et~al.}(2024)O'Shea, Borrow, O'Neil, \& Vogelsberger}]{oshea+24}
O'Shea, T.~M., Borrow, J., O'Neil, S., \& Vogelsberger, M. 2024, Dynamical friction and measurements of the splashback radius in galaxy clusters.
\newblock \doarXiv{2405.18468}

\bibitem[{O’Neil {et~al.}(2021)O’Neil, Barnes, Vogelsberger, \& Diemer}]{oneil+21}
O’Neil, S., Barnes, D.~J., Vogelsberger, M., \& Diemer, B. 2021, Monthly Notices of the Royal Astronomical Society, 504, 4649, \dodoi{10.1093/mnras/stab1221}

\bibitem[{O’Neil {et~al.}(2022)O’Neil, Borrow, Vogelsberger, \& Diemer}]{oneil+22}
O’Neil, S., Borrow, J., Vogelsberger, M., \& Diemer, B. 2022, Monthly Notices of the Royal Astronomical Society, 513, 835, \dodoi{10.1093/mnras/stac850}

\bibitem[{O’Neil {et~al.}(2024)O’Neil, Borrow, Vogelsberger, Zhao, \& Wang}]{oneil+24}
O’Neil, S., Borrow, J., Vogelsberger, M., Zhao, H., \& Wang, B. 2024, Monthly Notices of the Royal Astronomical Society, 530, 3310, \dodoi{10.1093/mnras/stae990}

\bibitem[{Pedersen \& Dahle(2007)}]{pedersen+07}
Pedersen, K., \& Dahle, H. 2007, The Astrophysical Journal, 667, 26, \dodoi{10.1086/520945}

\bibitem[{Peebles(1980)}]{peebles80}
Peebles, P. J.~E. 1980, The Large-Scale Structure of the Universe (Princeton University Press).
\newblock \url{http://www.jstor.org/stable/j.ctvxrpz4n}

\bibitem[{{Pointecouteau, E.} {et~al.}(2005){Pointecouteau, E.}, {Arnaud, M.}, \& {Pratt, G. W.}}]{pointecouteau+05}
{Pointecouteau, E.}, {Arnaud, M.}, \& {Pratt, G. W.} 2005, A\&A, 435, 1, \dodoi{10.1051/0004-6361:20042569}

\bibitem[{Popesso {et~al.}(2007)Popesso, Biviano, Böhringer, \& Romaniello}]{popesso+07}
Popesso, P., Biviano, A., Böhringer, H., \& Romaniello, M. 2007, A\&A, 464, 451, \dodoi{10.1051/0004-6361:20054708}

\bibitem[{Popesso {et~al.}(2005)Popesso, Biviano, Böhringer, Romaniello, \& Voges}]{popesso+05}
Popesso, P., Biviano, A., Böhringer, H., Romaniello, M., \& Voges, W. 2005, A\&A, 433, 431, \dodoi{10.1051/0004-6361:20041915}

\bibitem[{{Pratt} {et~al.}(2019)}]{pratt+19}
{Pratt}, G.~W., {et~al.} 2019, Space Science Reviews, 215, 25

\bibitem[{Rana {et~al.}(2023)Rana, More, Miyatake, Grandis, Klein, Bulbul, Chiu, Miyazaki, \& Bahcall}]{rana+23}
Rana, D., More, S., Miyatake, H., {et~al.} 2023, Monthly Notices of the Royal Astronomical Society, 522, 4181, \dodoi{10.1093/mnras/stad1239}

\bibitem[{Ryu \& Lee(2021)}]{ryu+21}
Ryu, S., \& Lee, J. 2021, The Astrophysical Journal, 917, 98, \dodoi{10.3847/1538-4357/ac0c14}

\bibitem[{Sanderson {et~al.}(2009)Sanderson, Edge, \& Smith}]{sanderson+09}
Sanderson, A. J.~R., Edge, A.~C., \& Smith, G.~P. 2009, Monthly Notices of the Royal Astronomical Society, 398, 1698, \dodoi{10.1111/j.1365-2966.2009.15214.x}

\bibitem[{Schwarz(1978)}]{schwarz78}
Schwarz, G. 1978, The Annals of Statistics, 6, 461 , \dodoi{10.1214/aos/1176344136}

\bibitem[{Sereno(2015)}]{sereno15}
Sereno, M. 2015, Monthly Notices of the Royal Astronomical Society, 450, 3665, \dodoi{10.1093/mnras/stu2505}

\bibitem[{Shi {et~al.}(2016)Shi, Yang, Wang, Zhang, Mo, van~den Bosch, Li, Liu, Lu, Tweed, \& Yang}]{shi+16}
Shi, F., Yang, X., Wang, H., {et~al.} 2016, The Astrophysical Journal, 833, 241, \dodoi{10.3847/1538-4357/833/2/241}

\bibitem[{Shi(2016)}]{shi16}
Shi, X. 2016, Monthly Notices of the Royal Astronomical Society, 459, 3711, \dodoi{10.1093/mnras/stw925}

\bibitem[{Shin {et~al.}(2019)Shin, Adhikari, Baxter, Chang, Jain, Battaglia, Bleem, Bocquet, DeRose, Gruen, Hilton, Kravtsov, McClintock, Rozo, Rykoff, Varga, Wechsler, Wu, Zhang, Aiola, Allam, Bechtol, Benson, Bertin, Bond, Brodwin, Brooks, Buckley-Geer, Burke, Carlstrom, Carnero Rosell, Carrasco Kind, Carretero, Castander, Choi, Cunha, Crawford, da Costa, De Vicente, Desai, Devlin, Dietrich, Doel, Dunkley, Eifler, Evrard, Flaugher, Fosalba, Gallardo, García-Bellido, Gaztanaga, Gerdes, Gralla, Gruendl, Gschwend, Gupta, Gutierrez, Hartley, Hill, Ho, Hollowood, Honscheid, Hoyle, Huffenberger, Hughes, James, Jeltema, Kim, Krause, Kuehn, Lahav, Lima, Madhavacheril, Maia, Marshall, Maurin, McMahon, Menanteau, Miller, Miquel, Mohr, Naess, Nati, Newburgh, Niemack, Ogando, Page, Partridge, Patil, Plazas, Rapetti, Reichardt, Romer, Sanchez, Scarpine, Schindler, Serrano, Smith, Smith, Soares-Santos, Sobreira, Staggs, Stark, Stein, Suchyta, Swanson, Tarle, Thomas, van Engelen, Wollack, \& Xu}]{shin+19}
Shin, T., Adhikari, S., Baxter, E.~J., {et~al.} 2019, Monthly Notices of the Royal Astronomical Society, 487, 2900, \dodoi{10.1093/mnras/stz1434}

\bibitem[{Shin {et~al.}(2021)Shin, Jain, Adhikari, Baxter, Chang, Pandey, Salcedo, Weinberg, Amsellem, Battaglia, Belyakov, Dacunha, Goldstein, Kravtsov, Varga, Abbott, Aguena, Alarcon, Allam, Amon, Andrade-Oliveira, Annis, Bacon, Bechtol, Becker, Bernstein, Bertin, Bocquet, Bond, Brooks, Buckley-Geer, Burke, Campos, Rosell, Kind, Carretero, Chen, Choi, Costanzi, da Costa, DeRose, Desai, De Vicente, Devlin, Diehl, Dietrich, Dodelson, Doel, Doux, Drlica-Wagner, Eckert, Elvin-Poole, Everett, Ferraro, Ferrero, Ferté, Flaugher, Frieman, Gallardo, Gatti, Gaztanaga, Gerdes, Gruen, Gruendl, Gutierrez, Harrison, Hartley, Hill, Hilton, Hinton, Hollowood, Hughes, James, Jarvis, Jeltema, Koopman, Krause, Kuehn, Kuropatkin, Lahav, Lima, Lokken, MacCrann, Madhavacheril, Maia, McCullough, McMahon, Melchior, Menanteau, Miquel, Mohr, Moodley, Morgan, Myles, Nati, Navarro-Alsina, Niemack, Ogando, Page, Palmese, Partridge, Paz-Chinchón, Pereira, Pieres, Malagón, Prat, Raveri, Rodriguez-Monroy, Rollins, Romer, Rykoff,
  Salatino, Sánchez, Sanchez, Santiago, Scarpine, Schillaci, Secco, Serrano, Sevilla-Noarbe, Sheldon, Sherwin, Sifón, Smith, Soares-Santos, Staggs, Suchyta, Swanson, Tarle, Thomas, To, Troxel, Tutusaus, Vavagiakis, Weller, Wollack, Yanny, Yin, \& Zhang}]{shin+22}
Shin, T., Jain, B., Adhikari, S., {et~al.} 2021, Monthly Notices of the Royal Astronomical Society, 507, 5758, \dodoi{10.1093/mnras/stab2505}

\bibitem[{Simet {et~al.}(2016)Simet, McClintock, Mandelbaum, Rozo, Rykoff, Sheldon, \& Wechsler}]{simet+16}
Simet, M., McClintock, T., Mandelbaum, R., {et~al.} 2016, Monthly Notices of the Royal Astronomical Society, 466, 3103, \dodoi{10.1093/mnras/stw3250}

\bibitem[{Smith {et~al.}(2010)Smith, Khosroshahi, Dariush, Sanderson, Ponman, Stott, Haines, Egami, \& Stark}]{smith+10}
Smith, G.~P., Khosroshahi, H.~G., Dariush, A., {et~al.} 2010, Monthly Notices of the Royal Astronomical Society, 409, 169, \dodoi{10.1111/j.1365-2966.2010.17311.x}

\bibitem[{Springel {et~al.}(2005)Springel, White, Jenkins, Frenk, Yoshida, Gao, Navarro, Thacker, Croton, Helly, Peacock, Cole, Thomas, Couchman, Evrard, Colberg, \& Pearce}]{springel05}
Springel, V., White, S. D.~M., Jenkins, A., {et~al.} 2005, Nature, 435, 629–636, \dodoi{10.1038/nature03597}

\bibitem[{Strateva {et~al.}(2001)Strateva, Željko Ivezić, Knapp, Narayanan, Strauss, Gunn, Lupton, Schlegel, Bahcall, Brinkmann, Brunner, Budavári, Csabai, Castander, Doi, Fukugita, Győry, Hamabe, Hennessy, Ichikawa, Kunszt, Lamb, McKay, Okamura, Racusin, Sekiguchi, Schneider, Shimasaku, \& York}]{strateva+01}
Strateva, I., Željko Ivezić, Knapp, G.~R., {et~al.} 2001, The Astronomical Journal, 122, 1861, \dodoi{10.1086/323301}

\bibitem[{Strauss {et~al.}(2002)Strauss, Weinberg, Lupton, Narayanan, Annis, Bernardi, Blanton, Burles, Connolly, Dalcanton, Doi, Eisenstein, Frieman, Fukugita, Gunn, Ivezić, Kent, Kim, Knapp, Kron, Munn, Newberg, Nichol, Okamura, Quinn, Richmond, Schlegel, Shimasaku, SubbaRao, Szalay, Vanden~Berk, Vogeley, Yanny, Yasuda, York, \& Zehavi}]{strauss02}
Strauss, M.~A., Weinberg, D.~H., Lupton, R.~H., {et~al.} 2002, The Astronomical Journal, 124, 1810–1824, \dodoi{10.1086/342343}

\bibitem[{Sunyaev \& Zeldovich(1972)}]{sunyaev+72}
Sunyaev, R.~A., \& Zeldovich, Y.~B. 1972, Comments on Astrophysics and Space Physics, 4, 173

\bibitem[{Sérsic(1963)}]{sersic63}
Sérsic, J.~L. 1963, Boletín de la Asociación Argentina de Astronomía, 6, 41

\bibitem[{Towler {et~al.}(2024)Towler, Kay, Schaye, Kugel, Schaller, Braspenning, Elbers, Frenk, Kwan, Salcido, van Daalen, Vandenbroucke, \& Altamura}]{towler+24}
Towler, I., Kay, S.~T., Schaye, J., {et~al.} 2024, Monthly Notices of the Royal Astronomical Society, 529, 2017, \dodoi{10.1093/mnras/stae654}

\bibitem[{Tully(2015)}]{tully15}
Tully, R.~B. 2015, The Astronomical Journal, 149, 54, \dodoi{10.1088/0004-6256/149/2/54}

\bibitem[{Umetsu \& Diemer(2017)}]{umetsu17}
Umetsu, K., \& Diemer, B. 2017, The Astrophysical Journal, 836, 231, \dodoi{10.3847/1538-4357/aa5c90}

\bibitem[{{Umetsu} {et~al.}(2020)}]{umetsu+20}
{Umetsu}, K., {et~al.} 2020, The Astronomy and Astrophysics Review, 28, 7

\bibitem[{Vallés-Pérez {et~al.}(2020)Vallés-Pérez, Planelles, \& Quilis}]{valles-parez+20}
Vallés-Pérez, D., Planelles, S., \& Quilis, V. 2020, Monthly Notices of the Royal Astronomical Society, 499, 2303, \dodoi{10.1093/mnras/staa3035}

\bibitem[{van~den Bosch {et~al.}(1999)van~den Bosch, Lewis, Lake, \& Stadel}]{vandenbosch+99}
van~den Bosch, F.~C., Lewis, G.~F., Lake, G., \& Stadel, J. 1999, The Astrophysical Journal, 515, 50, \dodoi{10.1086/307023}

\bibitem[{{Vikhlinin} {et~al.}(2009)}]{vikhlinin+09}
{Vikhlinin}, A., {et~al.} 2009, The Astrophysical Journal, 692, 1060

\bibitem[{Wang {et~al.}(2014)Wang, Yang, Shen, Mo, van~den Bosch, Luo, Wang, Lau, Wang, Kang, \& Li}]{wang14}
Wang, L., Yang, X., Shen, S., {et~al.} 2014, Monthly Notices of the Royal Astronomical Society, 439, 611, \dodoi{10.1093/mnras/stt2481}

\bibitem[{Wang {et~al.}(2013)Wang, Brunner, \& Dolence}]{wang+13}
Wang, Y., Brunner, R.~J., \& Dolence, J.~C. 2013, Monthly Notices of the Royal Astronomical Society, 432, 1961, \dodoi{10.1093/mnras/stt450}

\bibitem[{Wetzel {et~al.}(2014)Wetzel, Tinker, Conroy, \& Bosch}]{wetzel+14}
Wetzel, A.~R., Tinker, J.~L., Conroy, C., \& Bosch, F. C. v.~d. 2014, Monthly Notices of the Royal Astronomical Society, 439, 2687, \dodoi{10.1093/mnras/stu122}

\bibitem[{Wojtak \& Łokas(2010)}]{wojtak+10}
Wojtak, R., \& Łokas, E.~L. 2010, Monthly Notices of the Royal Astronomical Society, 408, 2442, \dodoi{10.1111/j.1365-2966.2010.17297.x}

\bibitem[{Xu {et~al.}(2024)Xu, Shan, Li, Yao, Wang, Li, \& Zhang}]{xu+24}
Xu, W., Shan, H., Li, R., {et~al.} 2024, The Astrophysical Journal, 971, 157, \dodoi{10.3847/1538-4357/ad57c7}

\bibitem[{Zenteno {et~al.}(2020)Zenteno, Hernández-Lang, Klein, Vergara Cervantes, Hollowood, Bhargava, Palmese, Strazzullo, Romer, Mohr, Jeltema, Saro, Lidman, Gruen, Ojeda, Katzenberger, Aguena, Allam, Avila, Bayliss, Bertin, Brooks, Buckley-Geer, Burke, Capasso, Carnero Rosell, Carrasco Kind, Carretero, Castander, Costanzi, da Costa, De Vicente, Desai, Diehl, Doel, Eifler, Evrard, Flaugher, Floyd, Fosalba, Frieman, García-Bellido, Gerdes, Gonzalez, Gruendl, Gschwend, Gutierrez, Hartley, Hinton, Honscheid, James, Kuehn, Lahav, Lima, McDonald, Maia, March, Melchior, Menanteau, Miquel, Ogando, Paz-Chinchón, Plazas, Roodman, Rykoff, Sanchez, Scarpine, Schubnell, Serrano, Sevilla-Noarbe, Smith, Soares-Santos, Suchyta, Swanson, Tarle, Thomas, Varga, Walker, Wilkinson, \& Collaboration)}]{zenteno+20}
Zenteno, A., Hernández-Lang, D., Klein, M., {et~al.} 2020, Monthly Notices of the Royal Astronomical Society, 495, 705, \dodoi{10.1093/mnras/staa1157}

\bibitem[{Zürcher \& More(2019)}]{zurcher+19}
Zürcher, D., \& More, S. 2019, The Astrophysical Journal, 874, 184, \dodoi{10.3847/1538-4357/ab08e8}

\bibitem[{Łokas \& Mamon(2001)}]{lokas+01}
Łokas, E.~L., \& Mamon, G.~A. 2001, Monthly Notices of the Royal Astronomical Society, 321, 155, \dodoi{10.1046/j.1365-8711.2001.04007.x}

\bibitem[{Željko Ivezić {et~al.}(2019)Željko Ivezić, Kahn, Tyson, Abel, Acosta, Allsman, Alonso, AlSayyad, Anderson, Andrew, Angel, Angeli, Ansari, Antilogus, Araujo, Armstrong, Arndt, Astier, Éric Aubourg, Auza, Axelrod, Bard, Barr, Barrau, Bartlett, Bauer, Bauman, Baumont, Bechtol, Bechtol, Becker, Becla, Beldica, Bellavia, Bianco, Biswas, Blanc, Blazek, Blandford, Bloom, Bogart, Bond, Booth, Borgland, Borne, Bosch, Boutigny, Brackett, Bradshaw, Brandt, Brown, Bullock, Burchat, Burke, Cagnoli, Calabrese, Callahan, Callen, Carlin, Carlson, Chandrasekharan, Charles-Emerson, Chesley, Cheu, Chiang, Chiang, Chirino, Chow, Ciardi, Claver, Cohen-Tanugi, Cockrum, Coles, Connolly, Cook, Cooray, Covey, Cribbs, Cui, Cutri, Daly, Daniel, Daruich, Daubard, Daues, Dawson, Delgado, Dellapenna, de~Peyster, de~Val-Borro, Digel, Doherty, Dubois, Dubois-Felsmann, Durech, Economou, Eifler, Eracleous, Emmons, Neto, Ferguson, Figueroa, Fisher-Levine, Focke, Foss, Frank, Freemon, Gangler, Gawiser, Geary, Gee, Geha, Gessner,
  Gibson, Gilmore, Glanzman, Glick, Goldina, Goldstein, Goodenow, Graham, Gressler, Gris, Guy, Guyonnet, Haller, Harris, Hascall, Haupt, Hernandez, Herrmann, Hileman, Hoblitt, Hodgson, Hogan, Howard, Huang, Huffer, Ingraham, Innes, Jacoby, Jain, Jammes, Jee, Jenness, Jernigan, Jevremović, Johns, Johnson, Johnson, Jones, Juramy-Gilles, Jurić, Kalirai, Kallivayalil, Kalmbach, Kantor, Karst, Kasliwal, Kelly, Kessler, Kinnison, Kirkby, Knox, Kotov, Krabbendam, Krughoff, Kubánek, Kuczewski, Kulkarni, Ku, Kurita, Lage, Lambert, Lange, Langton, Guillou, Levine, Liang, Lim, Lintott, Long, Lopez, Lotz, Lupton, Lust, MacArthur, Mahabal, Mandelbaum, Markiewicz, Marsh, Marshall, Marshall, May, McKercher, McQueen, Meyers, Migliore, Miller, Mills, Miraval, Moeyens, Moolekamp, Monet, Moniez, Monkewitz, Montgomery, Morrison, Mueller, Muller, Arancibia, Neill, Newbry, Nief, Nomerotski, Nordby, O’Connor, Oliver, Olivier, Olsen, O’Mullane, Ortiz, Osier, Owen, Pain, Palecek, Parejko, Parsons, Pease, Peterson, Peterson,
  Petravick, Petrick, Petry, Pierfederici, Pietrowicz, Pike, Pinto, Plante, Plate, Plutchak, Price, Prouza, Radeka, Rajagopal, Rasmussen, Regnault, Reil, Reiss, Reuter, Ridgway, Riot, Ritz, Robinson, Roby, Roodman, Rosing, Roucelle, Rumore, Russo, Saha, Sassolas, Schalk, Schellart, Schindler, Schmidt, Schneider, Schneider, Schoening, Schumacher, Schwamb, Sebag, Selvy, Sembroski, Seppala, Serio, Serrano, Shaw, Shipsey, Sick, Silvestri, Slater, Smith, Smith, Sobhani, Soldahl, Storrie-Lombardi, Stover, Strauss, Street, Stubbs, Sullivan, Sweeney, Swinbank, Szalay, Takacs, Tether, Thaler, Thayer, Thomas, Thornton, Thukral, Tice, Trilling, Turri, Berg, Berk, Vetter, Virieux, Vucina, Wahl, Walkowicz, Walsh, Walter, Wang, Wang, Warner, Wiecha, Willman, Winters, Wittman, Wolff, Wood-Vasey, Wu, Xin, Yoachim, \& Zhan}]{ivezic+19}
Željko Ivezić, Kahn, S.~M., Tyson, J.~A., {et~al.} 2019, The Astrophysical Journal, 873, 111, \dodoi{10.3847/1538-4357/ab042c}

\end{thebibliography}
\bibliographystyle{aasjournal}



\end{document}